\documentclass[aps,prb,10pt,reprint,amsmath,amssymb,amsfonts,showkeys]{revtex4-1}
\pdfoutput=1 
\usepackage{graphicx,graphics}
\usepackage{latexsym,verbatim}
\usepackage{bm}
\usepackage{natbib}
\usepackage[breaklinks=true,colorlinks,citecolor=blue,linkcolor=blue,urlcolor=blue]{hyperref}
\usepackage{siunitx}
\usepackage[utf8]{inputenc}
\usepackage{physics}
\usepackage[percent]{overpic}
\usepackage{tikz}
\usepackage{bbold}

\graphicspath{{figures/}}

\begin{document}

\author{Thomas Benjamin Smith}
\email{tommy.smith023@gmail.com}
\affiliation{School of Physics and Astronomy, University of Manchester, Oxford Road, Manchester M13 9PY, United Kingdom}
\author{Alessandro Principi}
\affiliation{School of Physics and Astronomy, University of Manchester, Oxford Road, Manchester M13 9PY, United Kingdom}

\title{A Bipartite Kronig-Penney Model with Dirac Potential Scatterers}

\begin{abstract}
Here we present a simple extension to the age-old Kronig-Penney model, which is made to be bipartite by varying either the scatterer separations or the potential heights. In doing so, chiral (sublattice) symmetry can be introduced. When such a symmetry is present, topologically protected edge states are seen to exist. 
The solution proceeds through the conventional scattering formalism used to study the Kronig-Penney model, which does not require further tight-binding approximations or mapping into a Su-Schrieffer-Heeger model. 
The topological invariant for this specific system is found to be the winding of the reflection coefficient, ultimately linked to the system wavefunction. 
The solution of such a simple and illustrative 1D problem, whose topological content is extracted without requiring further tight-binding approximations, represents the novel aspect of our paper.
The cases in which chiral symmetry is absent are then seen to not host topologically protected edge states, as verified by the behaviour of the reflection coefficient and the absence of winding.
\end{abstract}


\maketitle

\section{Introduction}

For many years now, the subjects of topological protection and topological order have been hot topics within the theoretical condensed matter physics community. Ever since the seminal papers of Thouless {\it et al.}\cite{TKNN,Thouless:1994}  wherein the quantisation of Hall conductance was shown to be of topological origin, the quest has been on to discover and delve all systems in which the topological character of the bulk system manifests itself within some observable of the finite bounded system;\cite{Kohmoto:1985,Novoselov:2006,Zhang:2015,Avron:2003,Hatsugai:1997,Gurarie:2013,Aoki:1986,Watson:1996} a phenomenon termed bulk-boundary correspondence.

Within this paper we take inspiration from the one-dimensional Su-Schrieffer-Heeger (SSH) model\cite{SSH:1979,SSH:1980,SSH:1988} that is known to host zero-energy topologically protected edge states.\cite{Asboth:2016,Kane:2013} We thus consider an extended, closed Kronig-Penney model\cite{KP:1931,Grosso:2000} with Dirac potentials that is allowed to become bipartite (or dimerised) by the suitable variation of certain physical variables; namely the scatterer separations or the scatterer potential heights.

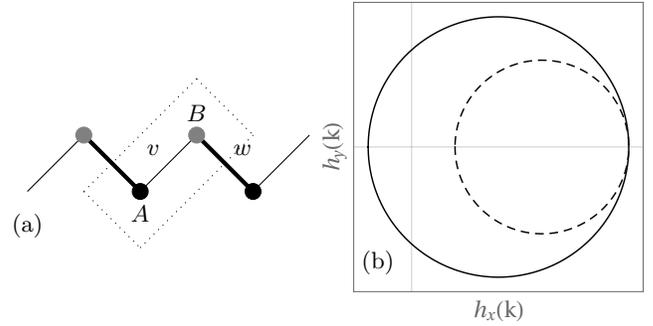
\begin{figure}
\centering
\begin{minipage}{.5\linewidth}
\begin{tikzpicture}[scale=0.75]
\draw (6.5,0.5) node[draw=none, anchor=south east] {$v$};
\draw (7.5,0.5) node[draw=none, anchor=south west] {$w$};
\draw node at (6,-0.1) [anchor=north] {$A$};
\draw node at (7,1.1) [anchor=south] {$B$};
\draw (4,0) -- (5,1);
\draw[ultra thick] (5,1) -- (6,0);
\draw (6,0) -- (7,1);
\draw[ultra thick](7,1) -- (8,0);
\draw (8,0) -- (9,1);
\filldraw[gray] (5,1) circle (4pt);
\filldraw[black] (6,0) circle (4pt);
\filldraw[gray] (7,1) circle (4pt);
\filldraw[black] (8,0) circle (4pt);
\draw[dotted] (5,0) -- (7,2);
\draw[dotted] (6,-1) -- (8,1);
\draw[dotted] (5,0) -- (6,-1);
\draw[dotted] (7,2) -- (8,1);
\draw node at (4,-0.25) [anchor=north] {(a)};
\end{tikzpicture}
\end{minipage}%
\begin{minipage}{.5\linewidth}
\centering
\begin{overpic}[width=\linewidth]{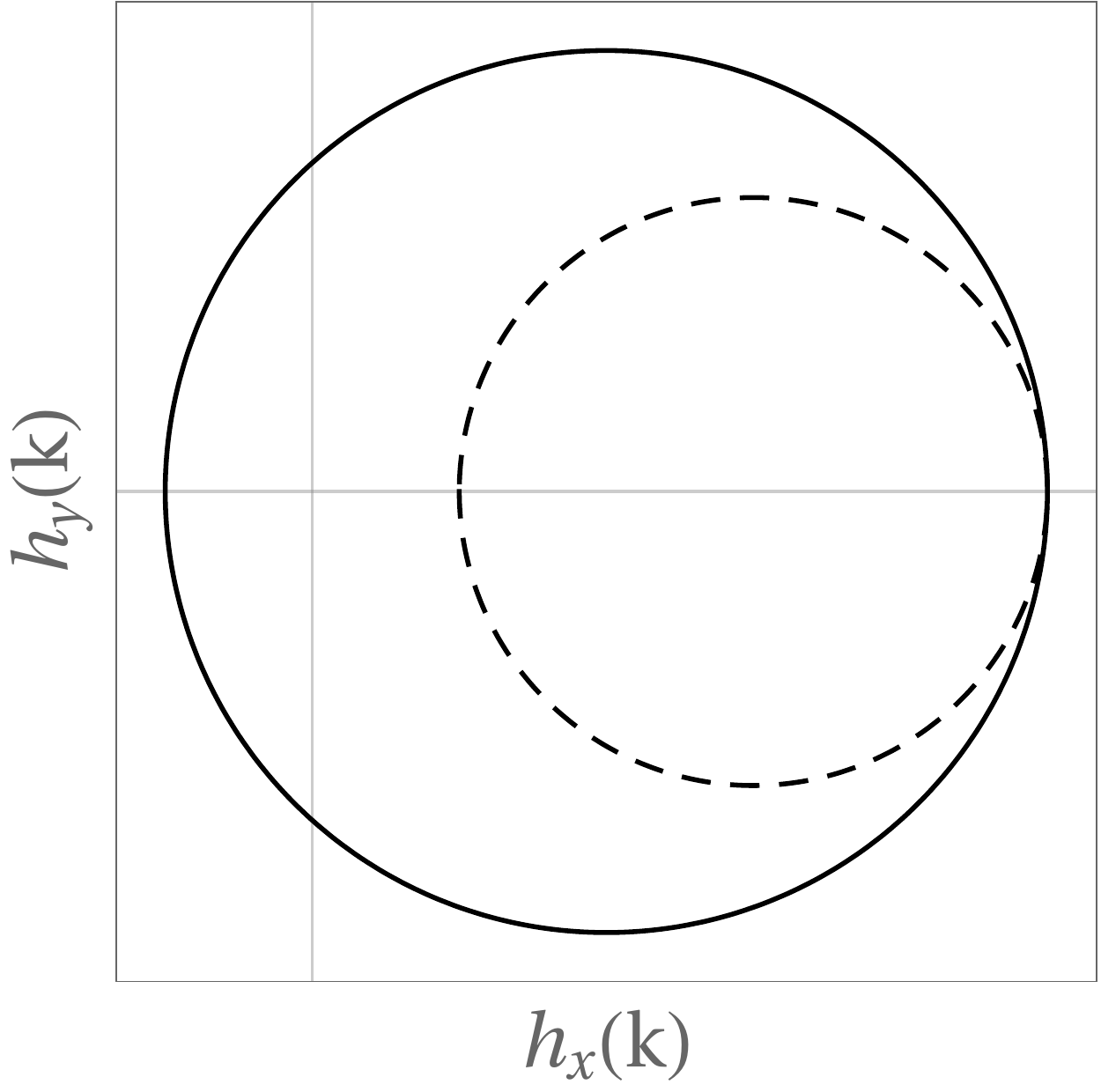}\put(13,17){(b)}
\end{overpic}
\end{minipage}%
\caption{Panel (a): A diagram of the unit cell of the SSH model lattice with hopping parameters $v$ and $w$ between sublattices $A$ (grey) and $B$ (black). Panel (b): The parametric winding of $h(k)=h_x(k)+ih_y(k)=v+we^{-ik}$ through $k$ with $w=1-v$ for $v=.4$ (solid) and $v=.6$ (dashed).}
\label{fig:SSHmodel}
\end{figure}

Within the simple SSH model, the bipartite nature of the system is generated by the hopping parameters $v$ and $w$ between neighbouring lattice sites that alternate along the chain, as may be seen in Fig.~\ref{fig:SSHmodel}(a). This generates two distinct sublattices that are identical save for their different environments; sublattice $A$ has $v$ to the left and $w$ to the right whilst sublattice $B$ has $w$ to the left and $v$ to the right. This is the origin of chiral/sublattice symmetry within the system, which is crucial for the presence of topologically protected edge states.\cite{Asboth:2016} 

Alternatively, bipartition can be achieved by distinguishing the two sublattices with, e.g., an on-site potential even in the presence of equal hoppings ({\it i.e.} $v=w$). 
The two bipartition strategies lead to very different results for what concerns the edge states of the finite system. In the former case their origin is purely topological, whereas in the latter they are as a consequence of a change of symmetry (if they even exist at all).

As is laid out more clearly in Appendix \ref{appB}, the single-particle Hamiltonian of a general SSH model within the bulk/thermodynamic limit, {\it i.e.} under periodic boundary conditions, that possesses nearest-neighbour hoppings, a constant on-site potential of $V$ and sublattice dependent potential $U$ is given by:
\begin{equation}
{\cal H}(k)=
\begin{pmatrix}
V+U&h(k)
\\
h^*(k)&V-U
\end{pmatrix}=\bm{d}(k)\cdot\bm{\sigma},
\end{equation}
where $h(k)=v+we^{-ik}$ with $k$ as the Bloch wavevector. The finite SSH model with $U=0$ is known to host topologically protected edge states. This is a fact guaranteed by the presence of both chiral symmetry and a quantised invariant within the bulk Hamiltonian. This is the concept of bulk-boundary correspondence.

Chiral symmetry requires that:\cite{Asboth:2016}
\begin{equation}
{\cal H}(k)=-\Gamma^\dagger(k){\cal H}(k)\Gamma(k)=-\sigma_z{\cal H}(k)\sigma_z,
\end{equation}
since $\Gamma(k)=\sigma_z$ as it must be an anti-commuting unitary operator. Given that this condition is met, the invariant belongs to $\mathbb{Z}$, {\it i.e.} an integer, and may be found to reside in not only the winding of $h(k)$ about the origin in the $\Re[h(k)]$-$\Im[h(k)]$, or $d_x-d_y$, plane but also the so-called Zak or 1D-Berry phase.\cite{Zak:1989,Berry:1984} Such windings may be seen in Fig.~\ref{fig:SSHmodel}(b) for an SSH model with $V=U=0$ and $w=1-v$. The solid circle corresponds to $v=0.4$ and $w=0.6$, whereas the dashed circle to $v=0.6$ and $w=0.4$. This winding, which is invariant against adiabatic deformations of the Hamiltonian that preserve the winding number, corresponds to the number of protected edge states of the finite system.

Chiral symmetry, however, is absent when the sublattices are not energetically identical. By which it is meant that the presence of a term proportional to $\sigma_z$ in the Hamiltonian, such as the $U\sigma_z$ one here, destroys the symmetry since it energetically distinguishes the sublattices. Such a term may arise as a sublattice dependent on-site potential term or from next-nearest neighbour hoppings between the same sublattice. Regardless, when ${\cal H}(k)$ contains terms proportional to $\sigma_z$, ${\cal H}(k)\neq-\Gamma^\dagger(k)H(k)\Gamma(k)$. The lack of the symmetry causes the winding number to be zero for all $v$ as the $d_z\sigma_z$ term pushes the circle trivially out of the $d_x-d_y$ plane. Hence, $\bm{d}(k)$ fails to wind the origin and any edge state present in the system is a conventional Shockley,\cite{Shockley:1939} or Tamm,\cite{Tamm:1932} state. Furthermore, the Zak phase too loses its quantisation and topological behaviour as a result.

The significance of topologically protected edge states lies in their resistance and robustness against adiabatic lattice deformations and perturbations, to say nothing about their fundamental theoretical interest. Provided that a system is in a topologically non-trivial phase then any adiabatic deformation ({\it i.e.} a change of its parameters) that leaves it within the same topological phase will not affect the existence of the protected edge states.\cite{Asboth:2016,Kane:2013}

As such, it is expected that this behaviour should be present in a bipartite Kronig-Penney model where the widths between the scatterers act in the same way as the hopping parameters within the SSH model.

In relation to the Kronig-Penney model, there has been much study undertaken into the finite system that possesses open boundary conditions. In such cases, charge quanta may be pumped through the chain by a suitable adiabatic deformation of parameters and the quantisation is of a topological origin.\cite{Gasparian:2005,Wang:2013}

In the present case, we impose hard wall boundary conditions such that all states must exist within the chain itself and dimerise the system, {\it i.e.} make it bipartite. Then the parameter space is the first Brillouin zone.

This paper is divided into four parts.

Firstly, the general bulk system is solved by considering a periodic geometry of the chain such that the boundary conditions are periodic. Then we need only consider a single unit cell whilst making use of Bloch's theorem. A pseudo scattering matrix for the unit cell is found in terms of a real eigenvalue problem involving the wavefunction coefficients. The reflection coefficient is then shown to exhibit a topological character akin to $h(k)$.

Secondly, the general finite system is presented and explained. Due to the non-exact nature of the energies within the model, the final calculation must be performed numerically.

Thirdly, four different cases are considered. We consider systems in which we not only fix the potential heights to a constant and vary the separations between the potential scatterers but also  fix the separations to be equal and vary the potential heights instead. In both of these cases, we consider systems in which the potentials have negative and positive strengths. In the former case, we search for states that are bound to the potentials which have negative energy and thus imaginary wavevector. In the latter case, the states propagate within the wells with positive energy and real wavevector.

Finally, conclusions are drawn and the results are discussed in the context of physical systems for which this theory may be applied.

\section{The Bulk Solution}

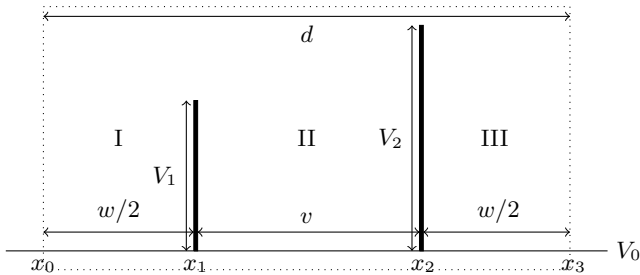
\begin{figure}
\centering
\begin{tikzpicture}
\draw (1,0)--(9,0);
\filldraw (3.5,0)--(3.5,2)--(3.55,2)--(3.55,0)--(3.5,0);
\filldraw (6.5,0)--(6.5,3)--(6.55,3)--(6.55,0)--(6.5,0);
%\filldraw (8.5,0)--(8.5,2)--(8.6,2)--(8.6,0)--(8.5,0);
\draw[dotted] (1.5,3.25)--(8.5,3.25)--(8.5,-.25)--(1.5,-.25)--(1.5,3.25);
\draw[<->] (1.5,3.12)--(8.5,3.12) node at (5,3.12) [anchor=north] {$d$};
\draw[<->] (3.55,.25)--(6.5,.25) node at (5,.25) [anchor=south] {$v$};
\draw[<->] (6.55,.25)--(8.5,.25) node at (7.55,.25) [anchor=south] {$w/2$};
\draw[<->] (1.5,.25)--(3.5,.25) node at (2.5,.25) [anchor=south] {$w/2$};
\draw[<->] (3.4,0)--(3.4,2) node at (3.4,1) [anchor=east] {$V_1$};
\draw[<->] (6.4,0)--(6.4,3) node at (6.4,1.5) [anchor=east] {$V_2$};
\draw node at (1.5,0) [anchor=north] {$x_0$};
\draw node at (3.525,0) [anchor=north] {$x_1$};
\draw node at (6.55,0) [anchor=north] {$x_2$};
\draw node at (8.55,0) [anchor=north] {$x_3$};
\draw node at (9,0) [anchor=west] {$V_0$};
\draw node at (2.5,1.5) {I};
\draw node at (5,1.5) {II};
\draw node at (7.5,1.5) {III};
\end{tikzpicture}
\caption{A diagram of the unit cell of the considered bipartite Kronig-Penney model with Dirac scatterers. The bipartite-ness may be realised by not only varying $v$ and $w$ but also varying $V_1$ and $V_2$.}
\label{fig:eKPmodel}
\end{figure}

We solve the time independent Schr{\"o}dinger equation as given by:
\begin{equation} \label{eq:schroedinger_def}
\left[-\frac{\hbar^2}{2m}\frac{d^2}{dx^2}+V(x)\right]\Psi_k(x)=E(k)\Psi_k(x),
\end{equation}
for the system as shown in Fig.~\ref{fig:eKPmodel} with potential scatterers that have Dirac-delta profiles. The potential $V(x)$ within the unit cell is then given by:
\begin{equation}
V(x)=V_1\delta(x-x_1)+V_2\delta(x-x_2).
\end{equation}
However, rather than either mapping this differential equation into an effective Hamiltonian or using a tight-binding approximation, we solve the system in the standard scattering paradigm. The objective of which is to maintain a certain transparency to the analysis so that any conclusions drawn are clearer as a result. 
Analyses based on either an effective Hamiltonian or tight-binding would introduce unnecessary approximations.

In passing, we note that the topological features of edge states have been also studied, with a similar scattering matrix approach, in Ref.~\onlinecite{Fulga:2012}. There, however, the focus is on lattice ({\it i.e.} tight-binding) models. Here we adopt instead a ``wave-mechanics" description of the simplest possible problem, suitable for generalisation to 1D problems with more complicated potentials, and we clearly identify the quantities that bear topological significance.

Following the conventional scattering paradigm, we solve the problem posed by Eq.~(\ref{eq:schroedinger_def}) in the regions in which the potential vanishes, and we then match such solutions by using the standard boundary conditions for Dirac potentials. These are that, at the position of each scatterer, the wavefunction is continuous and that the difference of the derivatives of the wavefunction yields the product of the Dirac potential height and some constant. Finally, periodic boundary conditions are imposed upon the chain such that Bloch's theorem may be used. This introduces the Bloch wavevector, which we will denote $k$, through $\Psi(x+d)=\Psi(x)e^{ikd}$.

To solve the problem, we first make the {\it Ansatz} that the wavefunction within each unit cell is simply the linear combination of right and left moving waves with coefficients that differ between wells. In other words:
\begin{equation} \label{eqn:bulkWFs}
\Psi_k(x)=\mathcal{N}_c(k)\sum_{n=1}^3\Theta_n(x)\psi_{n,k}(x),
\end{equation}
where $\Theta_n(x)=\theta(x-x_{n-1})\theta(x_n-x)$, $\psi_{n,k}(x)=C_ne^{iq_kx}+D_ne^{-iq_kx}$ and the wavevector $q_k$ is related to the energy of the wavefunction by $E(k)=\hbar^2q_k^2/(2m)$. The normalisation constant is found in the standard way through:
\begin{equation}
\int_{x_0}^{x_3}\frac{dx}{x_3-x_0}|\Psi_k(x)|^2=1.
\end{equation}
(The definition of $x_0, x_3$ as the edges of the unit cell may be found in Fig.~\ref{fig:eKPmodel}.)

The boundary conditions for the scatterers are then formulated within the unit cell with the first and second scatterers located at $x_1$ and $x_2$ respectively. The edges of the unit cell are then $x_0=-d/2$ and $x_3=d/2$, thus we have that:
\begin{align} \label{eqn:bulkBCs}
\psi_{1,k}(x_1)=\psi_{2,k}(x_1),&
\quad~
\psi_{2,k}(x_2)=\psi_{3,k}(x_2),
\\
V_1[\psi_{1,k}(x_1)+\psi_{2,k}(x_1)]&=\psi'_{2,k}(x_1)-\psi'_{1,k}(x_1),
\\%\frac{d\psi_2}{dx}\bigg|_{0}-\frac{d\psi_1}{dx}\bigg|_{d}e^{-ikd}
V_2[\psi_{2,k}(x_2)+\psi_{3,k}(x_2)]&=\psi'_{3,k}(x_2)-\psi'_{2,k}(x_2),
%\frac{d\psi_2}{dx}\bigg|_{v}-\frac{d\psi_1}{dx}\bigg|_{v}
\\
\psi_{3,k}(x)&=\psi_{1,k}(x-d)e^{ikd}.
\end{align}

Keeping within the scattering paradigm, we build a real eigenvalue equation in terms of an unitary (non-Hermitian) matrix in the case of real (imaginary) wavevector and positive (negative) potential heights. In the positive-potential case, such a matrix is a scattering matrix. This is not strictly true in the negative-potential case, but we will continue to call it as such to simplify the discussion. The scatterings across the two potentials within the unit cell are solved from which the following real eigenvalue equation is found:
\begin{align} \label{eqn:SmatEq}
\begin{pmatrix}
D_1(k)
\\
C_2(k)
\end{pmatrix}=S(k)
\begin{pmatrix}
D_1(k)
\\
C_2(k)
\end{pmatrix},
\end{align}
the details of which are laid out in Appendix \ref{appA}. This $S$-matrix has a general form given by:
\begin{equation}\label{eqn:Smat}
S(k)=
\begin{pmatrix}
r(k) & t(k)
\\
-t^*(k)e^{i\phi_k} & r^*(k)e^{i\phi_k}
\end{pmatrix}.
\end{equation}
The bulk SSH model Hamiltonian, as may be seen in Eq.~(\ref{eqn:SSHhams}), is in the basis of sublattice sites $A$ and $B$ whilst the present eigenvalue equation relates the equivalent quantities $D_1,C_2$ (or $D_2,C_1$) in a similar fashion. Hence, $r(k)$ and/or $t(k)$ are likely to reveal topological properties akin to $h(k)=v+we^{-ik}$ in the SSH case. 

In fact, the scattering coefficients may be ultimately derived from the bulk Green's function, and therefore inherit its topological properties\cite{Essin:2011} (akin to those of the wavefunction, from which it can be in principle constructed). Given the non-interacting nature of the problem at hand, a Green's function approach here would be unnecessarily complicated and less transparent. 
%Note, however, that it is generally simpler to determine $r(k)$ and $t(k)$ than the full wavefunction.

%{\bf 
%Within the field, most (if not all) one-dimensional topologically insulating systems have been investigated and classified. Generally, the topological invariant of 1D tight-binding two-level Hamiltonians may be classified by their symmetry properties.\cite{Altland:1997,Kitaev:2009,Chiu:2016} The invariant may then be found to reside within the bulk Green's function\cite{Essin:2011} and/or Zak phase,\cite{Zak:1989} both of which must be determined through the wavefunction. Since the wavefunction may only be determined numerically here and because the scattering coefficients may be derived ultimately from the bulk Green's function\cite{Essin:2011} (thereby inheriting its topological properties) we shall, for simplicity's sake, analytically solve for the scattering coefficients and compare their behaviours with the numerical Zak phase.}

Note that, within the present context, left-eigenvectors of the eigenvalue problem presented have no physical meaning. This is as a result of solving a set of boundary conditions within the matrix formalism.

The quantities $r(k)$ and $t(k)$ are pseudo reflection and transmission coefficients for the entire unit cell whilst $\phi_k$ is the phase of the scattering matrix and are given by:
\begin{align}
r(k)&=e^{iq_kd}\left[\frac{V_1V_2e^{iq_k[d+2(x_1-x_2)]}-q_k^2e^{ikd}}{(V_1-iq_k)(V_2-iq_k)}\right],\label{eqn:rcoeff}
\\
t(k)&=iq_k\left[\frac{V_1e^{iq_k(d+2x_1)}e^{-ikd}+V_2e^{2iq_kx_2}}{(V_1-iq_k)(V_2-iq_k)}\right],\label{eqn:tcoeff}
\\
e^{i\phi_k}&=e^{2iq_kd}\left[\frac{(V_1+iq_k)(V_2+iq_k)}{(V_1-iq_k)(V_2-iq_k)}\right].\label{eqn:ph}
\end{align}
An interesting note may be observed now: the transmission coefficient $t(k)$ is not invariant with the choice of unit cell. This is because it depends on $x_1$ and $x_2$ in their own rights whereas $r(k)$ depends only on $x_1-x_2$, which is always $v$, regardless of all else. Furthermore, if the eigenvalue equation is generated in the basis of $(C_1(k),D_2(k))$ then $t(k)$ has yet another different expression whilst $r(k)$ remains invariant.

However, $t(k)$ can be made invariant by redefining $C_2(k)$ with respect to an appropriate phase factor, which renders the Zak phase unchanged. Yet, as will be seen, the winding of $t(k)$ does not correspond to the protection of edge states. As such, the behaviour of $t(k)$ ought to be ignored in favour of $r(k)$.

The transcendental equation that defines the energy bands of any states within the system is found by solving the eigenvalue problem as given in Eq.~(\ref{eqn:SmatEq}). When done so, it is found that:
\begin{multline}
\cos(kd)=\left(1-\frac{V_1V_2}{q_k^2}\right)\cos(q_kd)+\frac{V_1+V_2}{q_k}\sin(q_kd)\\+\frac{V_1V_2}{q_k^2}\cos\left\{q_k[d+2(x_1-x_2)]\right\},
\end{multline}
where the specification of $x_1$ and $x_2$ is, again, arbitrary since this equation depends on their difference and the two scatterers are always positioned a distance of $v$ apart within the unit cell. When the right-hand side has a value greater (lesser) than $+1~(-1)$ there are no real solutions and thus the band gaps are defined. Since this equation relates $k$ to $q_k$ with $q_k\propto\sqrt{E(k)}$ it may only be solved using numerical root finding methods.

%As one might expect, this bears a close resemblance to the band defining equation as seen in the standard Kronig-Penney model. Indeed, when the two scatterers are both situated at $x_1=x_2=0$ with heights $V_1=V_2=V$ then we obtain the standard Kronig-Penney transcendental equation:\cite{KP:1931}
%\begin{equation}
%\cos(kd)=\cos(q_kd)+\frac{2V}{q_k}\sin(q_kd).
%\end{equation}

Finally, as a confirmation of the existence of the topological character within the reflection coefficient, the winding of $r(k)$ will be plotted and compared to the Zak phase as calculated from the unit-cell periodic wavefunctions. The Zak phase is defined by:\cite{Zak:1989,Delplace:2011}
\begin{equation}
\theta_{\cal Z}=i\int_{-\pi/d}^{+\pi/d}dk\bra{u_k}\ket{\partial_ku_k},
\end{equation}
where the inner product signifies to take an integral in the dimension $x$ over the unit cell and  $u_k(x)=e^{-ikx}\Psi_k(x)$ is the unit-cell periodic wavefunction. As such, for this system, the Zak phase is calculated explicitly as:
\begin{multline}
\theta_{\cal Z}=i\int_{-\pi/d}^{+\pi/d}dk|{\cal N}_c(k)|^2\sum_{n=1}^3\int_{x_{n-1}}^{x_n}dx
\\
\times\left[\left(-ix+\frac{\partial_k{\cal N}_c}{{\cal N}_c(k)}\right)|\psi_{n,k}(x)|^2+\psi_{n,k}^*(x)\partial_k\psi_{n,k}(x)\right].
\end{multline}
%The Zak phase takes values of zero or $\pi$ in the topologically trivial and non-trivial regimes respectively. Thus the topological invariant be defined as $W=\theta_{\cal Z}/\pi$.
As made clear by Zak in his seminal work,\cite{Zak:1989} $\theta_{\cal Z}$ is well-defined and quantised into units of $\pi$ if, and only if, the unit-cell density, $|\Psi_k(x)|^2$, is centro-symmetric.

As such, the unit cells in each case must be constructed in such a way that this condition is met. As mentioned, changing the unit cell does not alter the expressions for the reflection coefficient nor the transcendental equation. These expressions apply to the entire unit cell as a whole and not the microscopic detail whilst the wavefunctions are entirely determined from the unit cell detail.

The result of choosing a symmetric $|\Psi_k(x)|^2$ is that the contribution to the Zak phase of the polarisation term (proportional to $x$) vanishes. Furthermore, the normalisation contribution must also vanish as it is symmetric over the Brillouin zone. Thus, the Zak phase is given only in terms of the curvature contribution of the wavefunctions:
\begin{multline}
\theta_{\cal Z}=i\int_{-\pi/d}^{+\pi/d}dk|{\cal N}_c(k)|^2\\\times\sum_{n=1}^3\int_{x_{n-1}}^{x_n}dx\left[\psi^*_{n,k}(x)\partial_k\psi_{n,k}(x)\right].
\end{multline}
Then the Zak phase takes the distinct values of $0$ and $\pi$ when within the topologically trivial and non-trivial phases, respectively. The integer invariant is then defined as ${\cal W}_{\cal Z}=\theta_{\cal Z}/\pi$, which may also be observed as the winding number of the reflection coefficient. This may be calculated explicitly using:
\begin{equation}
{\cal W}_r=\frac{1}{2\pi i}\int_{-\pi/d}^{+\pi/d} dk\{\partial_k\ln\left[r(k)\right]\}.
\end{equation}

\section{The Finite Solution}

\begin{figure}
\centering
\begin{tikzpicture}[scale=0.715, every node/.style={scale=0.715}]
\draw (-9,0)--(-4,0);
\draw[dotted] (-4,0)--(0,0);
\draw (0,0)--(2,0);
\draw[->] (-9,0)--(-9,5) node at (-9,5) [anchor=south] {$+\infty$};
\draw[->] (2,0)--(2,5) node at (2,5) [anchor=south] {$+\infty$};
\draw node at (-9,-0) [anchor=north] {$-L/2$};
\draw node at (2,-0) [anchor=north] {$+L/2$};
\filldraw (-8,0)--(-8,2.5)--(-7.9,2.5)--(-7.9,0)--(-8,0);
\filldraw (-5.9,0)--(-5.9,3.5)--(-5.8,3.5)--(-5.8,0)--(-5.9,0);
\filldraw (-4.9,0)--(-4.9,2.5)--(-4.8,2.5)--(-4.8,0)--(-4.9,0);
\filldraw (0.9,0)--(0.9,2.5)--(1,2.5)--(1,0)--(0.9,0);
\draw[<->] (-9,.25)--(-8,.25) node at (-8.5,.25) [anchor=south] {$v$};
\draw[<->] (-7.9,.25)--(-5.9,.25) node at (-6.9,.25) [anchor=south] {$w$};
\draw[->] (-9,1)--(-8,1) node at (-9,1) [anchor=east] {$x_1$};
\draw[->] (-9,1.5)--(-5.9,1.5) node at (-9,1.5) [anchor=east] {$x_2$};
\draw[->] (-9,2)--(-4.9,2) node at (-9,2) [anchor=east] {$x_3$};
\draw[<->] (-4.6,0)--(-4.6,2.5) node at (-4.6,2) [anchor=west] {$V_{\cal O}$};
\draw[<->] (-5.6,0)--(-5.6,3.5) node at (-5.6,3)[anchor=west] {$V_{\cal E}$};
\draw node at (-7.95,2.5) [anchor=south] {$1$};
\draw node at (-5.85,3.5) [anchor=south] {$2$};
\draw node at (-4.85,2.5) [anchor=south] {$3$};
\draw node at (1,2.5) [anchor=south] {$N$};
\end{tikzpicture}
\caption{A diagram of the most general finite bipartite Kronig-Penney model with positive Dirac scatterers. If $N$ is odd (even) the final scatterer has the same (different) potential height as the first scatterer and the final well has the different (same) width as the first well.}
\label{fig:fdKPmodel}
\end{figure}
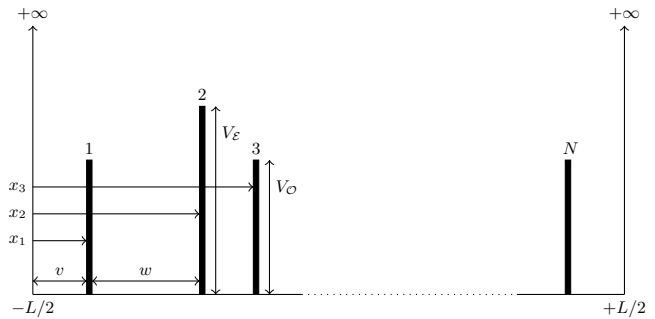

The analysis of the finite system differs since periodic boundary conditions no longer apply. Instead, closed boundary conditions are imposed such that the wavefunction vanishes outside of the chain of scatterers. In other words, we terminate the chain with hard walls that possess infinite potential heights.

The {\it Ansatz} for the wavefunction is identical to the bulk case and takes the form of a superposition of all the well wavefunctions provided that they are suitably confined to their wells with appropriate step functions. Explicitly:
\begin{equation}
\Psi_q(x)=\mathcal{N}_c(q)\sum_{n=1}^{N+1}\Theta_n(x)\psi_{n,q}(x),
\end{equation}
where the well wavefunction is given by $\psi_{n,q}(x)=C_n(q)e^{iqx}+D_n(q)e^{-iqx}$, $q$ is the quasi-momentum for the entire wavefunction, $\Theta_n(x)=\theta(x-x_{n-1})\theta(x_n-x)$ again and $\mathcal{N}_c(q)$ is the normalisation constant over the entire chain. As in the bulk case, the coefficients are dependent upon the wavevector, $q$. This is assumed from now on so their explicit dependence is dropped for brevity and clarity. As a result, the energy of the state is again given simply by $E=\hbar^2q^2/(2m)$.

The ends of the chain are defined as $x_0=-L/2$ and $x_{N+1}=+L/2$ (see Fig.~\ref{fig:fdKPmodel}) and the hard walls impose that the wavefunction vanishes when $x<-L/2$ and $x>L/2$, {\it i.e.} $C_0=D_0=C_{N+2}=D_{N+2}=0$.

The method of solution then proceeds similarly to the bulk case. However, rather than build an eigenvalue scattering matrix equation we now use the boundary conditions to generate a zero-eigenvalue equation:
\begin{equation}
{\rm M}\cdot\bm{v}=\bm{0},
\end{equation}
where ${\rm M}$ is a $2N\times2N$ square matrix and \\$\bm{v}=(C_1,D_1,\cdots,C_{N+1},D_{N+1})^{\rm T}$ is the vector of wavefunction coefficients.

The non-trivial solution ($\bm{v}\neq\bm{0}$) occurs when the determinant of the matrix is equal to zero: ${\det}~{\rm M}=0$. From this condition, the energy bands are found numerically and the non-trivial vector $\bm{v}$ is determined using the Singular Valued Decomposition technique.\cite{Klema:1980,Kalman:1996}

The boundary conditions at the hard walls are given by:
\begin{align}
\psi_{1,q}(x_0)=0,\quad\psi_{N+1,q}(x_{N+1})=0,
\end{align}
whilst the boundary conditions of the $n$th scatterer manifest themselves in this context as:
\begin{align}
\psi_{n,q}(x_n)&=\psi_{n+1,q}(x_n),
\\
V_n[\psi_{n,q}(x_n)+\psi_{n+1,q}(x_n)]&=\psi'_{n+1,q}(x_n)-\psi'_{n,q}(x),
%\frac{d\psi_{n+1}}{dx}\bigg|_{x_n}-\frac{d\psi_n}{dx}\bigg|_{x_n}
\end{align}
where $V_n=V_{\cal E}$ and $V_n=V_{\cal O}$ for even and odd $n$ respectively and natural units are still assumed. As measured from $-L/2$, the positions of the odd scatterers are given by $x_n^{\cal O}=v+(n-1)(v+w)/2$ and the even scatterers by $x_n^{\cal E}=n(v+w)/2$.

Since $x_0=-L/2$ and $x_{N+1}=+L/2$, by definition, the length of the chain is $x_{N+1}-x_0=L$, which takes different values for odd or even $N$. For both cases, $x_0^{\cal O}=x_0^{\cal E}=-L/2$ by construction. When $N$ is odd, $N+1$ is even so the position of the right wall is $x^{\cal E}_{N+1}=+L/2=(N+1)(v+w)/2$. When $N$ is even, $N+1$ is odd so the position of the right wall is $x^{\cal O}_{N+1}=+L/2=v+N(v+w)/2$. Finally, for consistency, the width of the second well is always chosen to be $w=d-v$, where $d$ is the size of the unit cell in the bulk, from which the length of the chain changes appropriately. Thus:
\begin{equation}
L_{\cal O}(v)=v+\frac{d}{2}N,\quad L_{\cal E}(v)=\frac{d}{2}(N+1).
\end{equation}

All that remains is to normalise the wavefunction appropriately by finding ${\cal N}_c(q)$ such that its probability density over the entire chain is equal to one:
\begin{equation}
\int_{-L/2}^{+L/2}\frac{dx}{L}|\Psi_q(x)|^2=1.
\end{equation}

\section{Results}

In total we solve four different cases. The two situations of negative and positive potential scatterers are considered. In the former case it is necessary to solve for negative energies and thus imaginary wavevectors are found. As such, the states become exponentially bound to the scatterers. In the latter case the states propagate through the system with real wavevectors.

Then, by keeping all the potential heights to a constant $V$ we vary the width of the first cell, $v$, and so vary the width of the second well as $w=d-v$. We then also set the well widths to both be constant and equal to $d/2$. Then the potential heights are varied as $V_1=W$ and $V_2=U-W$ so that $V_1+V_2=U$ with $U$ as some constant.

\subsection{Negative Potentials with Varying Widths}

So, firstly, we investigate the case of negative scatterers of constant height. In this case we have that $q_k\to iq_k$ and $V_{1,2}\to-V$. This causes the trigonometric functions in the transcendental equation to become hyperbolic. Thus the transcendental equation possesses only two roots for $q_k$.

It must be noted that the choice of $q_k\to iq_k$ over $q_k\to -iq_k$ is a trivial one. Taking $q_k\to iq_k$ yields the bulk well wavefunctions as $C_ne^{-q_kx}+D_ne^{q_kx}$ whilst taking $q_k\to-iq_k$ yields $C_ne^{q_kx}+D_ne^{-q_kx}$. In other words, $C_n$ and $D_n$ swap roles, which makes no matter to $r(k)$ and $\theta_{\cal Z}$ as has been pointed out.

\begin{figure}
\begin{minipage}{.5\linewidth}
\begin{overpic}[width=\linewidth]{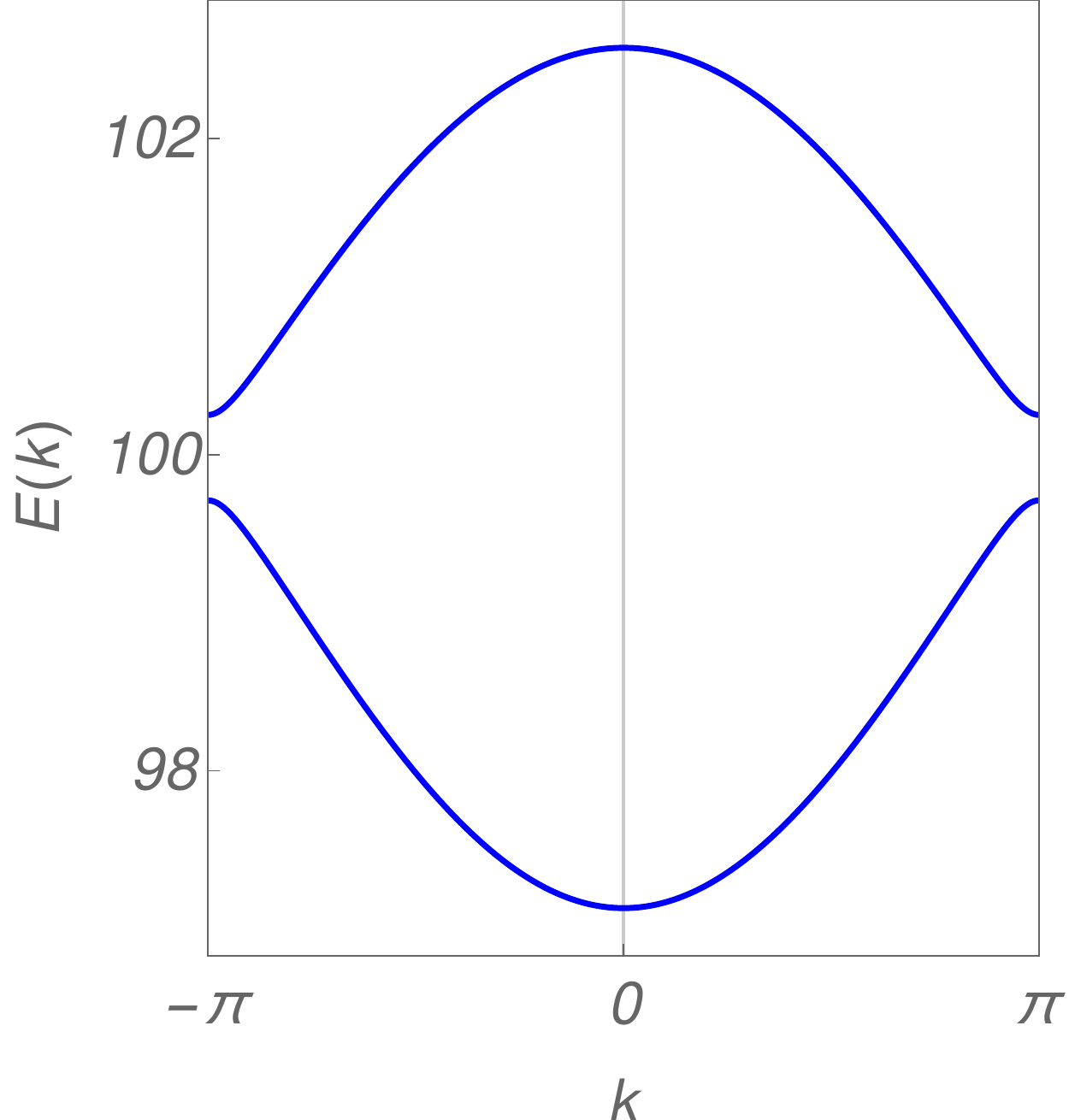}\put(82,92){(a)}
\end{overpic}
\begin{overpic}[width=\linewidth]{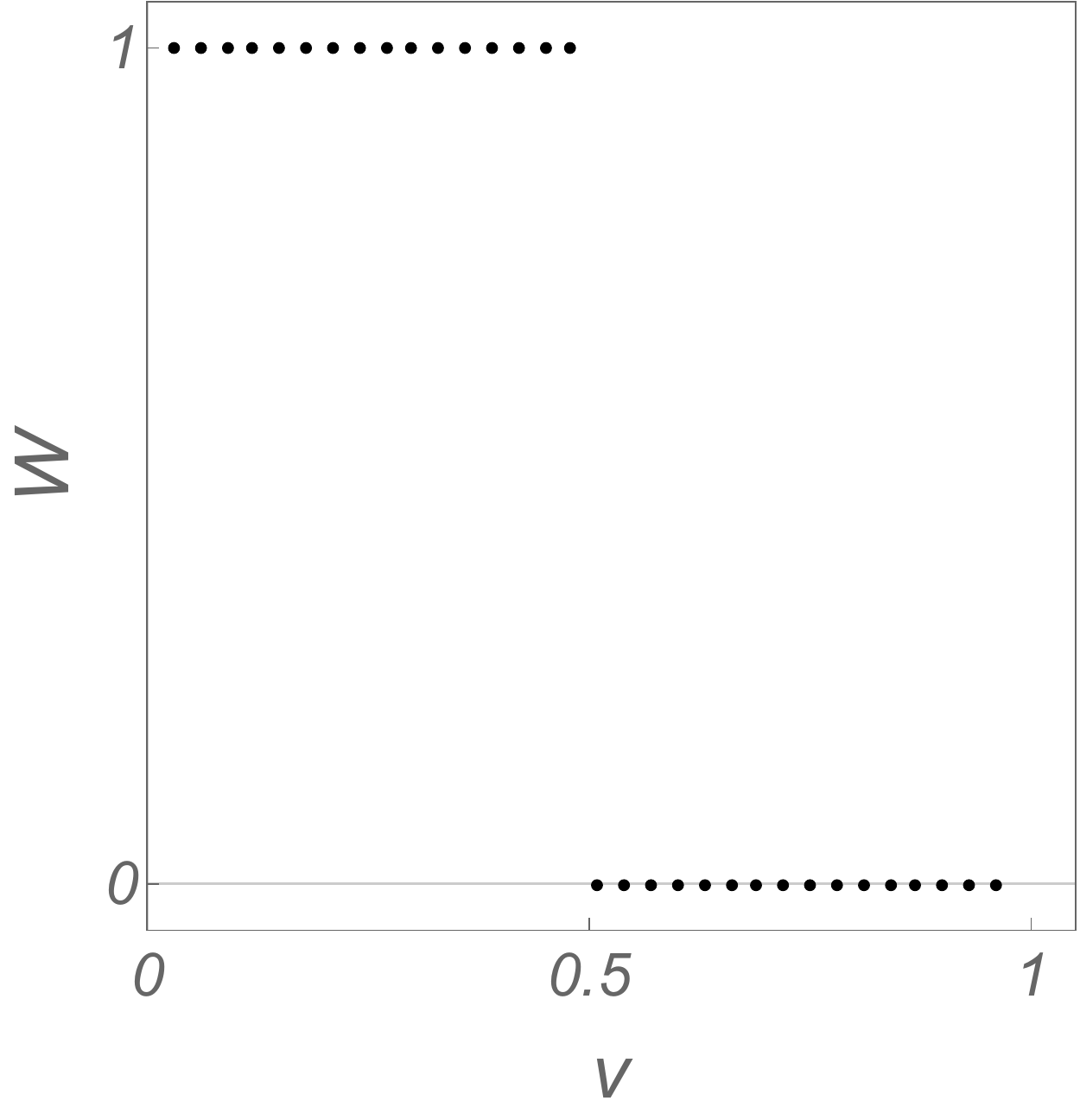}\put(85,92){(c)}
\end{overpic}
\end{minipage}%
\begin{minipage}{.5\linewidth}
\begin{overpic}[width=\linewidth]{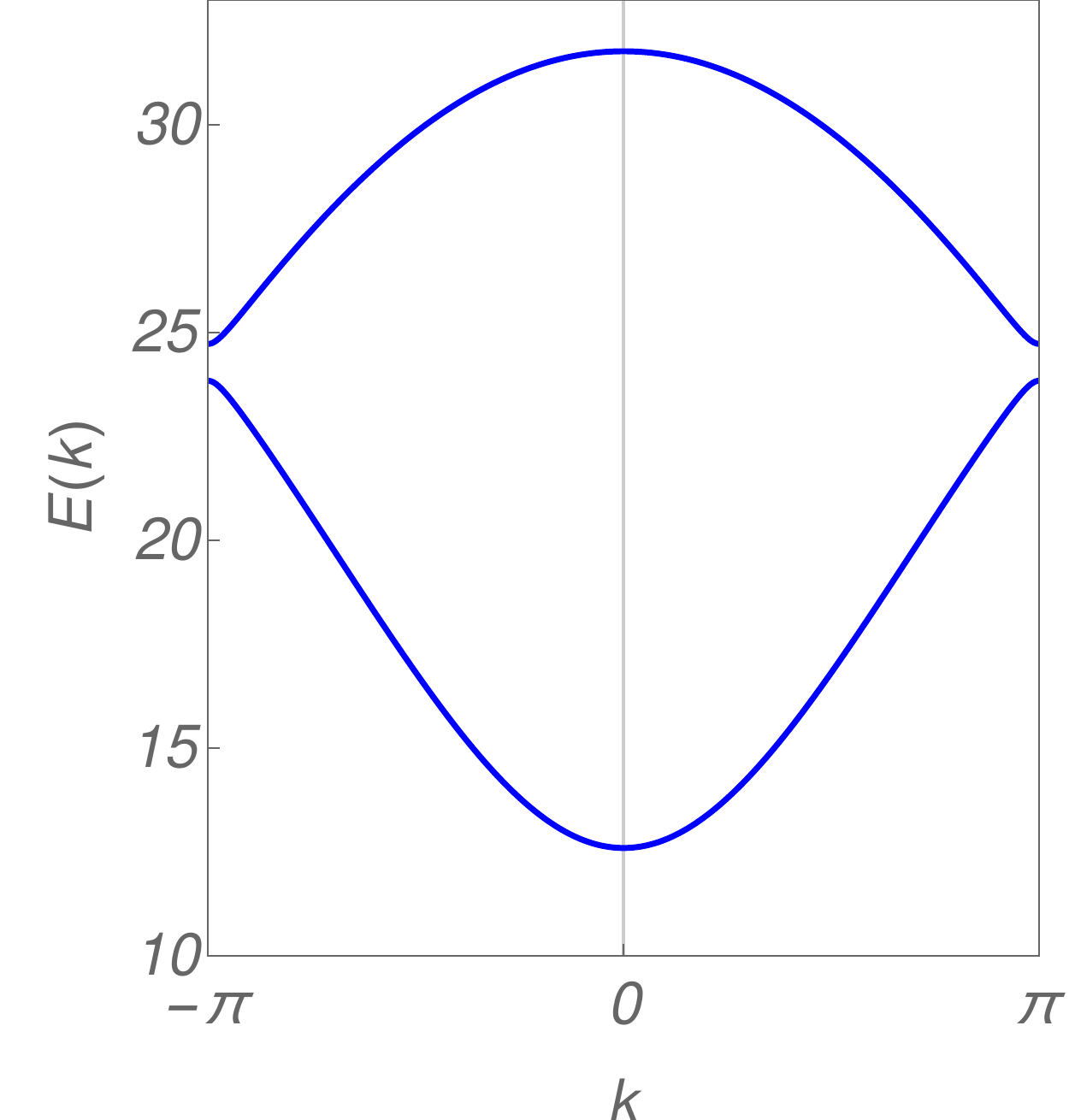}\put(82,92){(b)}
\end{overpic}
\begin{overpic}[width=\linewidth]{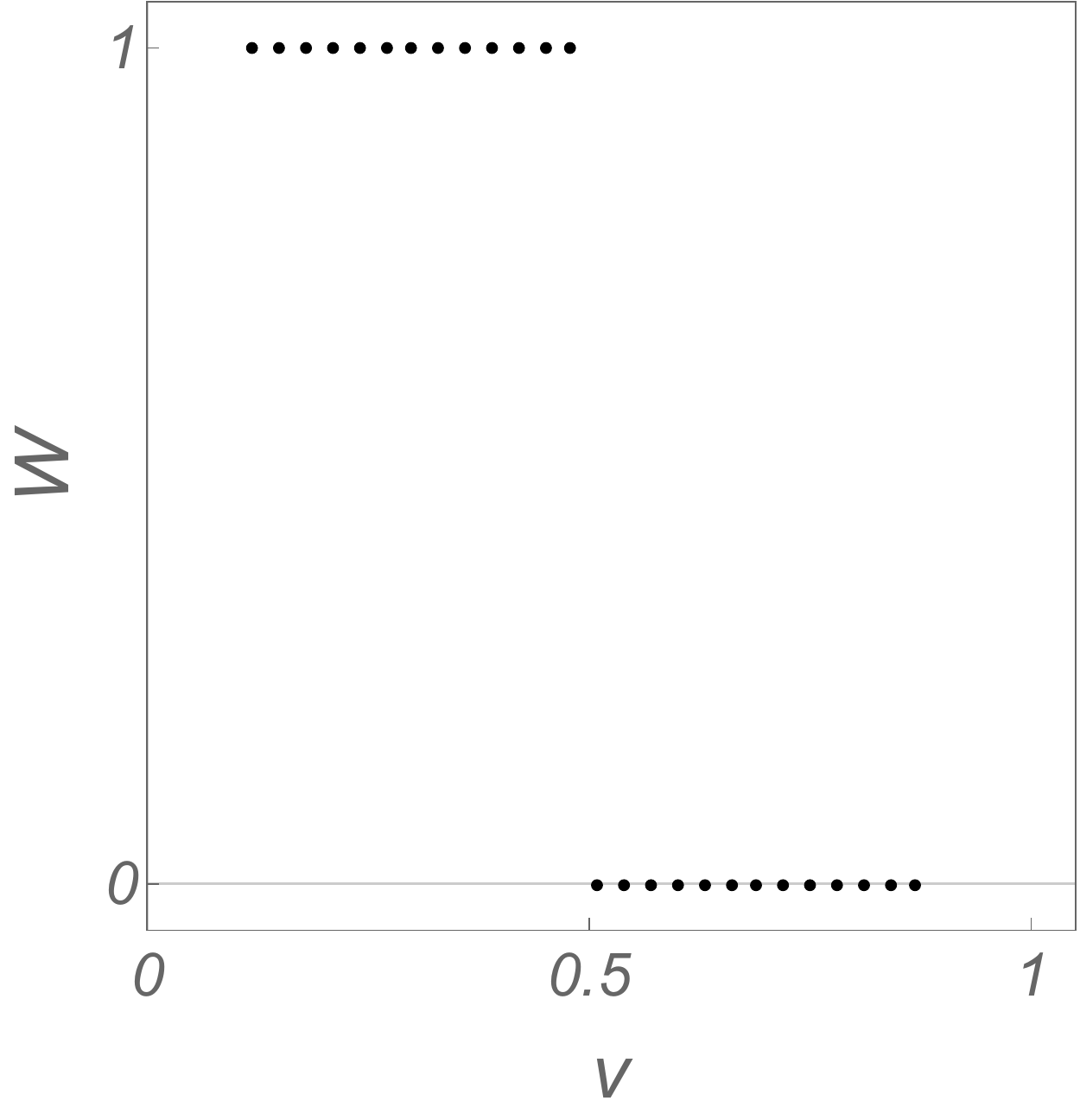}\put(85,92){(d)}
\end{overpic}
\end{minipage}%
\caption{(Colour on-line) Here are presented the quasi-momentum, $q_k$, spectra over the Brillouin zone (top) and the topological invariant (bottom) for the negative BP K-P model with constant potential heights. Panel (a): $|V|=10$ and $v=.51$, panel (b): $|V|=5$ and $v=.51$, panel (c): invariant of the top band, and panel (d): invariant of the bottom band, both for $V=5$. Note that, ${\cal W}_r$ is unaffected by the value of $V$ so long as both bands are present, {\it i.e.} that $V$ is not so small that the lower band is destroyed.}
\label{fig:negw}
\end{figure}

In Figs.~\ref{fig:negw}(a,b) are shown the quasi-momentum spectra for two of these very cases. On the left is shown the case with $|V|=10$ and on the right that of $|V|=5$. The former is seen to be symmetric, exactly akin to the SSH model, about the mid-gap point of $q_k=|V|$ whilst the latter is clearly not.

\begin{figure}
\centering
\begin{minipage}{.5\linewidth}
\begin{overpic}[width=\linewidth]{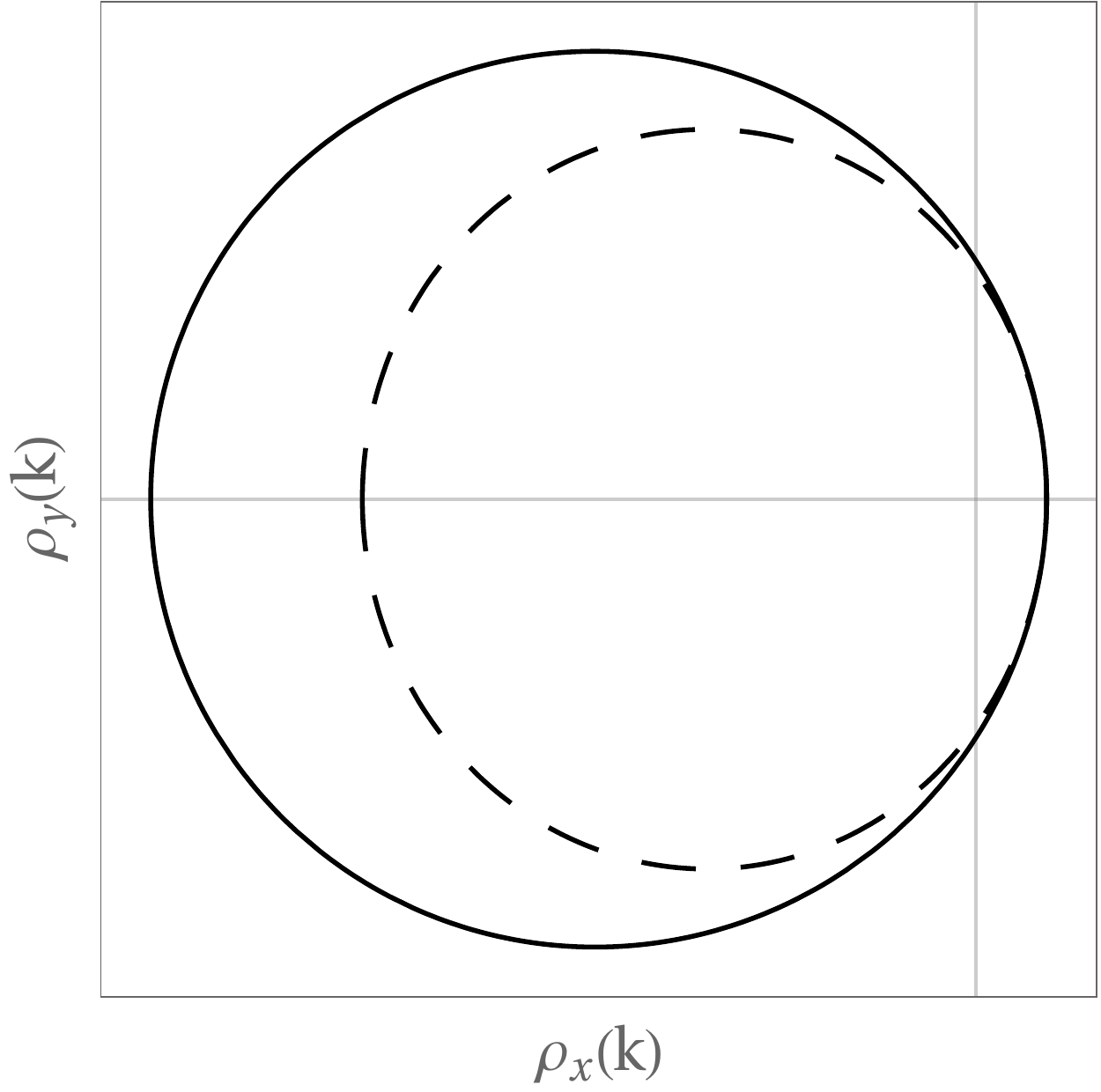}\put(88,15){(a)}
\end{overpic}
\begin{overpic}[width=\linewidth]{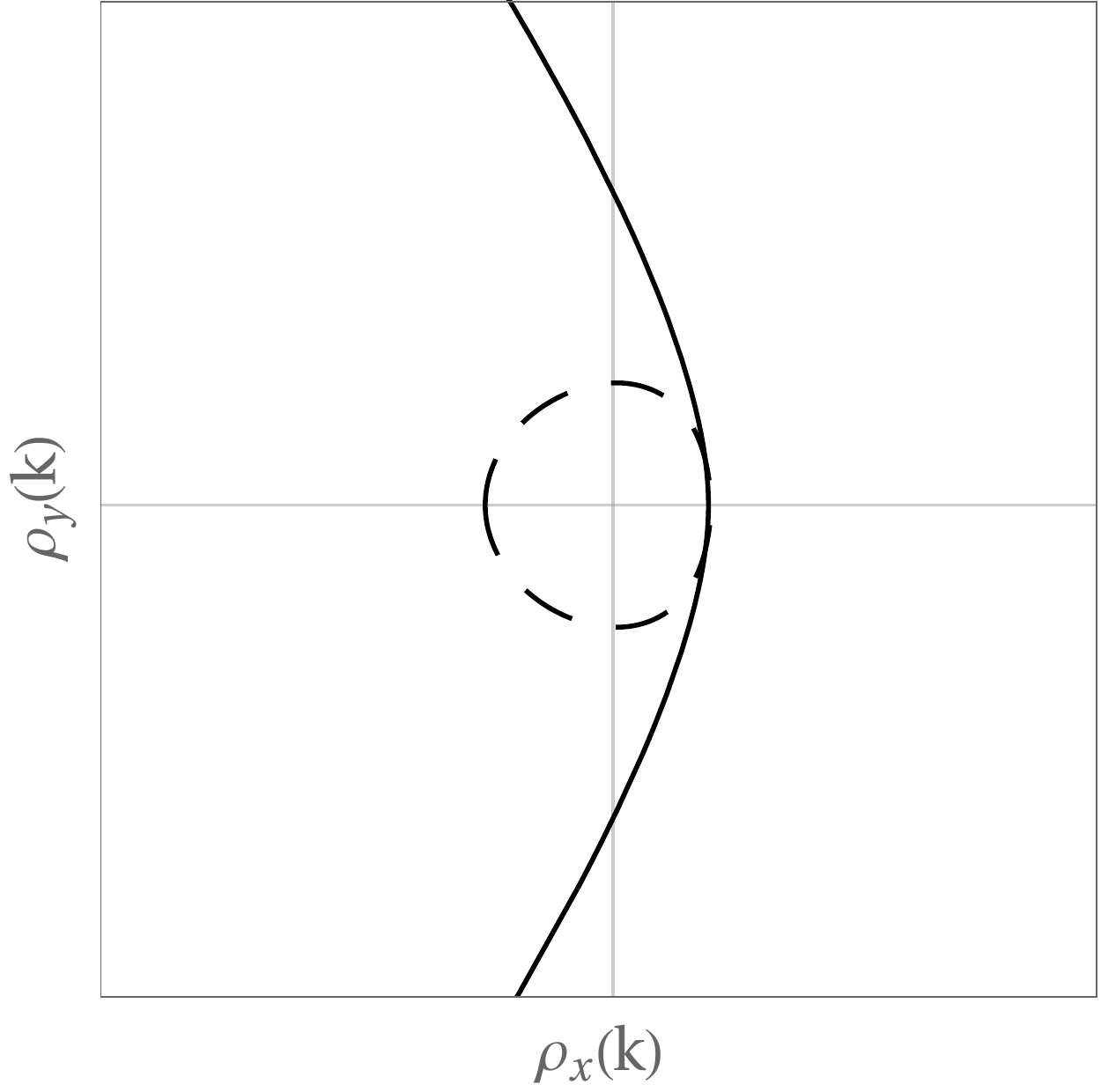}\put(87,15){(c)}
\end{overpic}
\end{minipage}%
\begin{minipage}{.5\linewidth}
\begin{overpic}[width=\linewidth]{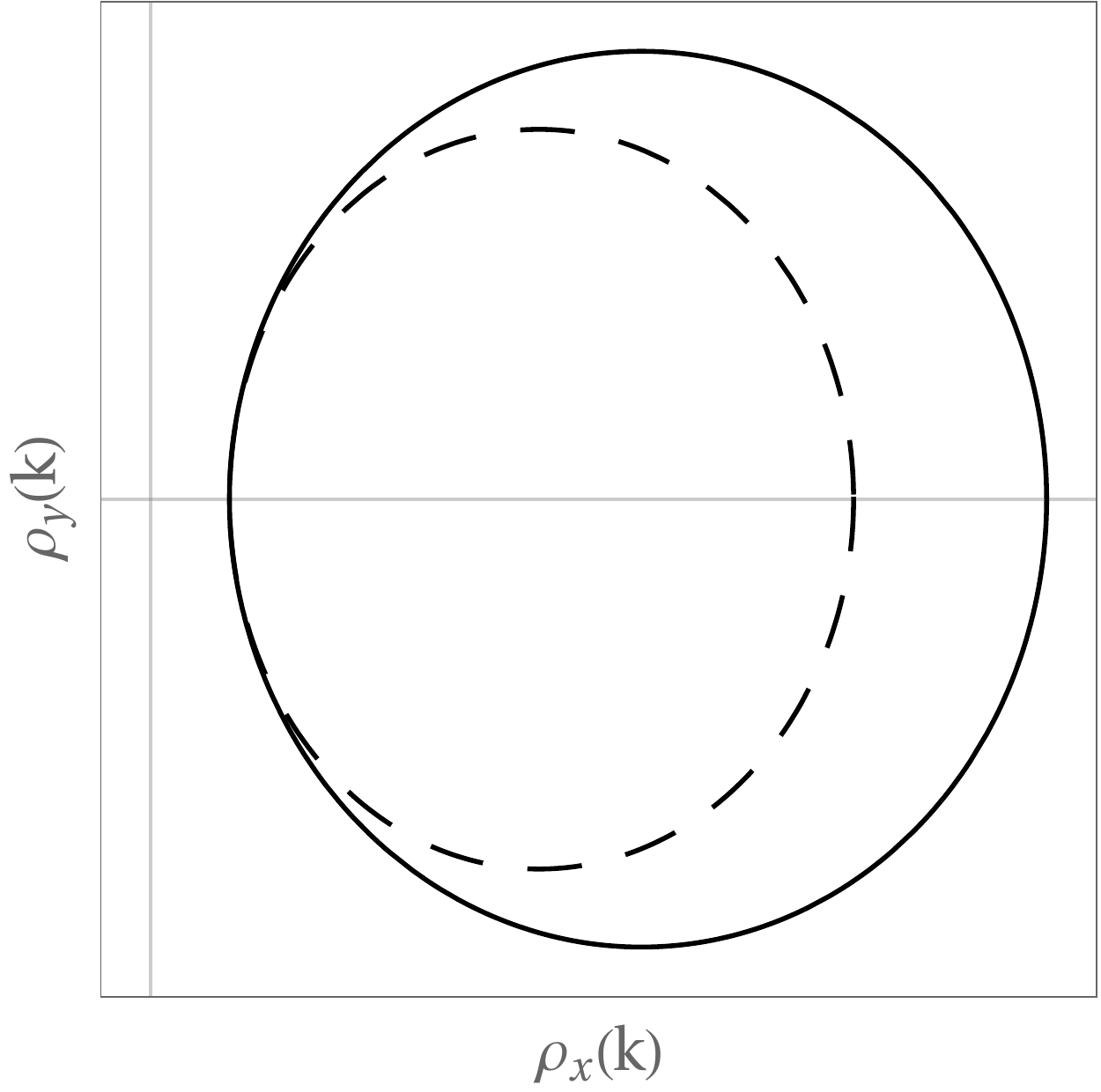}\put(88,15){(b)}
\end{overpic}
\begin{overpic}[width=\linewidth]{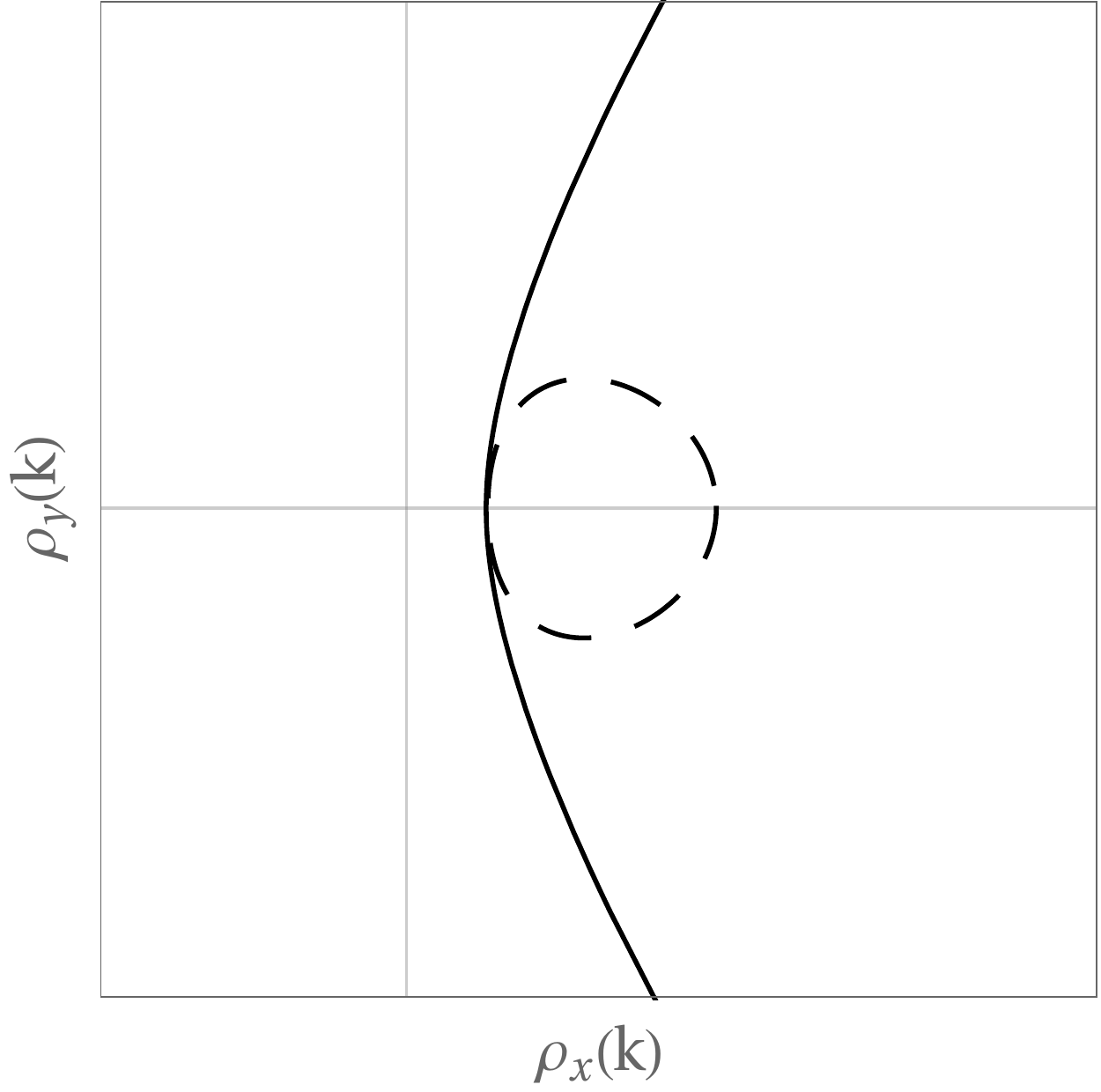}\put(87,15){(d)}
\end{overpic}
\end{minipage}%
\caption{The windings of the reflection coefficients before and after the topological transition from $v<w$ to $v>w$ in the negative scatterer case for varying $|V|$ and different $v$. Dashed/thick corresponds with the lower/upper bands since $q_k$ is band dependent. Panel (a): $|V|=10$ and $v=.49$, panel (b): $|V|=10$ and $v=.51$, panel (c): $|V|=5$ and $v=.49$, and panel (d): $|V|=5$ and $v=.51$.}
\label{fig:negKPrcoeffsW}
\end{figure}

If we take the reflection coefficient of the unit cell as in Eq.~(\ref{eqn:rcoeff}) with $q_k\to iq_k$ and $V_{1,2}\to-V$ then we see that it may be decomposed into real and imaginary parts as $\rho(k)=\rho_x(k)+i\rho_y(k)$, where:
\begin{equation}\label{eqn:Wrho}
\rho(k)=e^{-q_kd}\left[\frac{V^2e^{q_k(2v-d)}+q_k^2e^{ikd}}{(V-q_k)^2}\right].
\end{equation}
(The Greek symbol labelling is used here to differentiate this case from the positive scatterer case in which Roman symbols will be used.) Then, using the quasi-momenta, which are functions of $k$, as found using the transcendental equation we may parametrically plot $\rho(k)$ as a function of $k$, as shown in Fig.~\ref{fig:negKPrcoeffsW}.

The thick/dashed circles in the plots are the windings of $\rho(k)$ using the $q_k$ as obtained from the lower/upper bands, as may be readily deduced. The important observation is that the winding number, {\it i.e.} how many times $\rho(k)$ encompasses the origin, is one on the left and zero on the right. This is entirely analogous to the SSH model in which $h(k)=v+we^{-ik}$ winds once when $v<w$ but never when $v>w$. This is indeed the topological invariant that is sought in order to characterise any topologically protected edge states within the system and observe the resultant bulk-boundary correspondence. $\tau(k)$, on the other hand, shows the opposite winding behaviour as $\rho(k)$ of $0\rightarrow-1$ upon the transition and so this invariant does not correspond with protection of edge states. (The negative sign appears due to the $e^{-ikd}$ term in $\tau(k)$.)

Furthermore, this topological behaviour remains in the second case of $|V|=5$ regardless of the asymmetry of the bands as may be seen in the bottom of Fig.~\ref{fig:negKPrcoeffsW}. The bound states localise to the scatterers and the odd and even scatterers behave as distinct sublattices. This is an indication of the ability for the system to retain its topological character while being adiabatically deformed. Indeed the system ought to retain it since there is no sublattice dependent potential present, which would break chiral symmetry. Such a term in the SSH model manifests itself as a $\sigma_z$ contribution to the Hamiltonian and thus the winding of ${\bm d}(k)$ occurs above the $d_x-d_y$ plane, thereby failing to encompass the origin.

As confirmation for the winding number of the reflection coefficient being the topological invariant we consider also the Zak phase of the two bands. In order for ${\cal W}_{\cal Z}=\theta_{\cal Z}/\pi$ to be a good quantum number, as explained earlier, the unit cell must be constructed symmetrically. In this case this requires that the edges of the unit cell occur at $x_0=-d/2$ and $x_3=d/2$ with the potential scatterers at $x_1=-v/2$ and $x_2=v/2$. Then, the invariant of each band as a function of $v$ may be seen in Figs.~\ref{fig:negw}(c,d).

The final piece is to solve the finite system and show that mid-gap energy states exist within the topologically non-trivial regime for both an odd and an even number of scatterers.

\begin{figure}
\centering
\begin{minipage}{.5\linewidth}
\begin{overpic}[width=\linewidth]{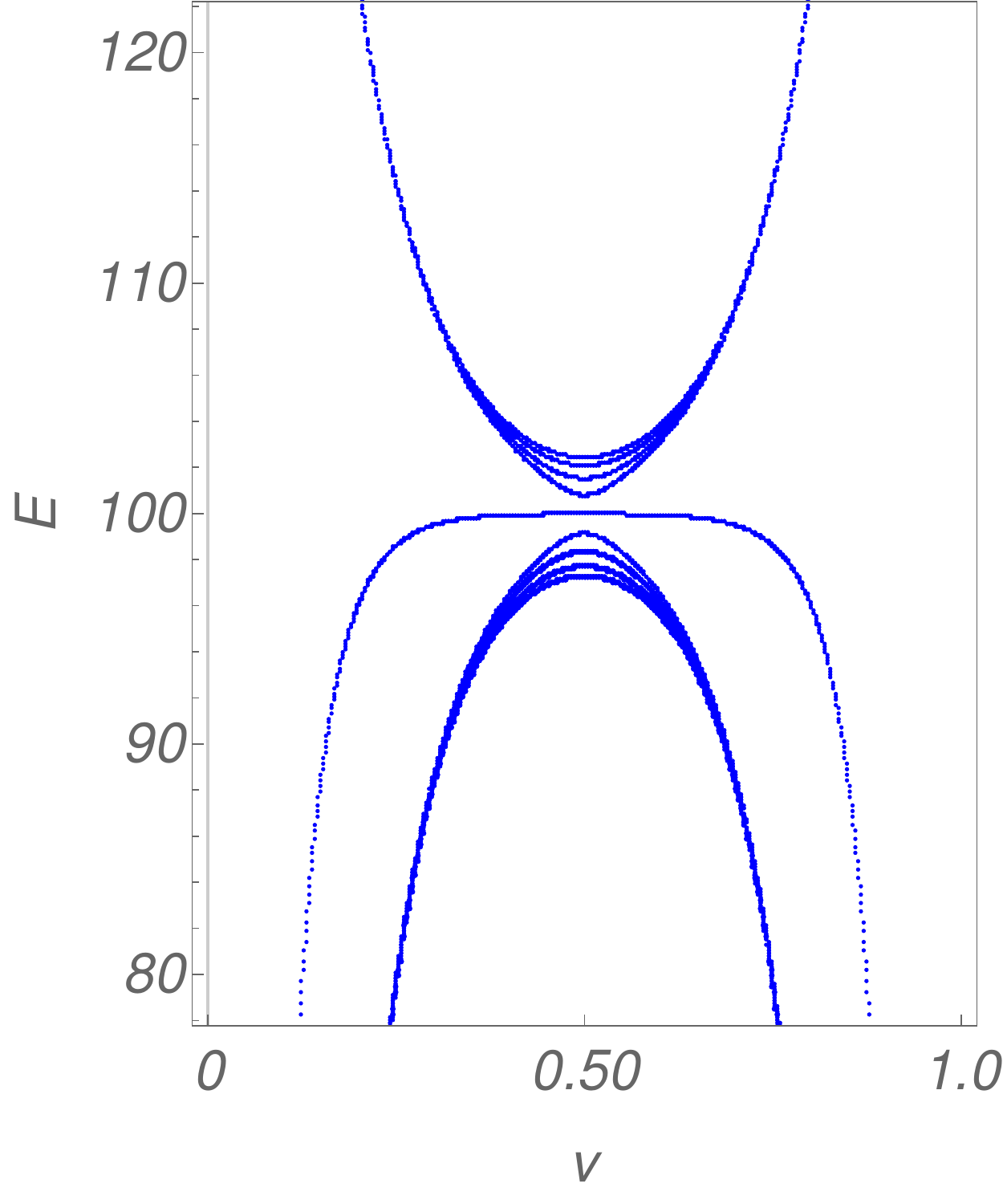}\put(73,18){(a)}
\end{overpic}
\begin{overpic}[width=\linewidth]{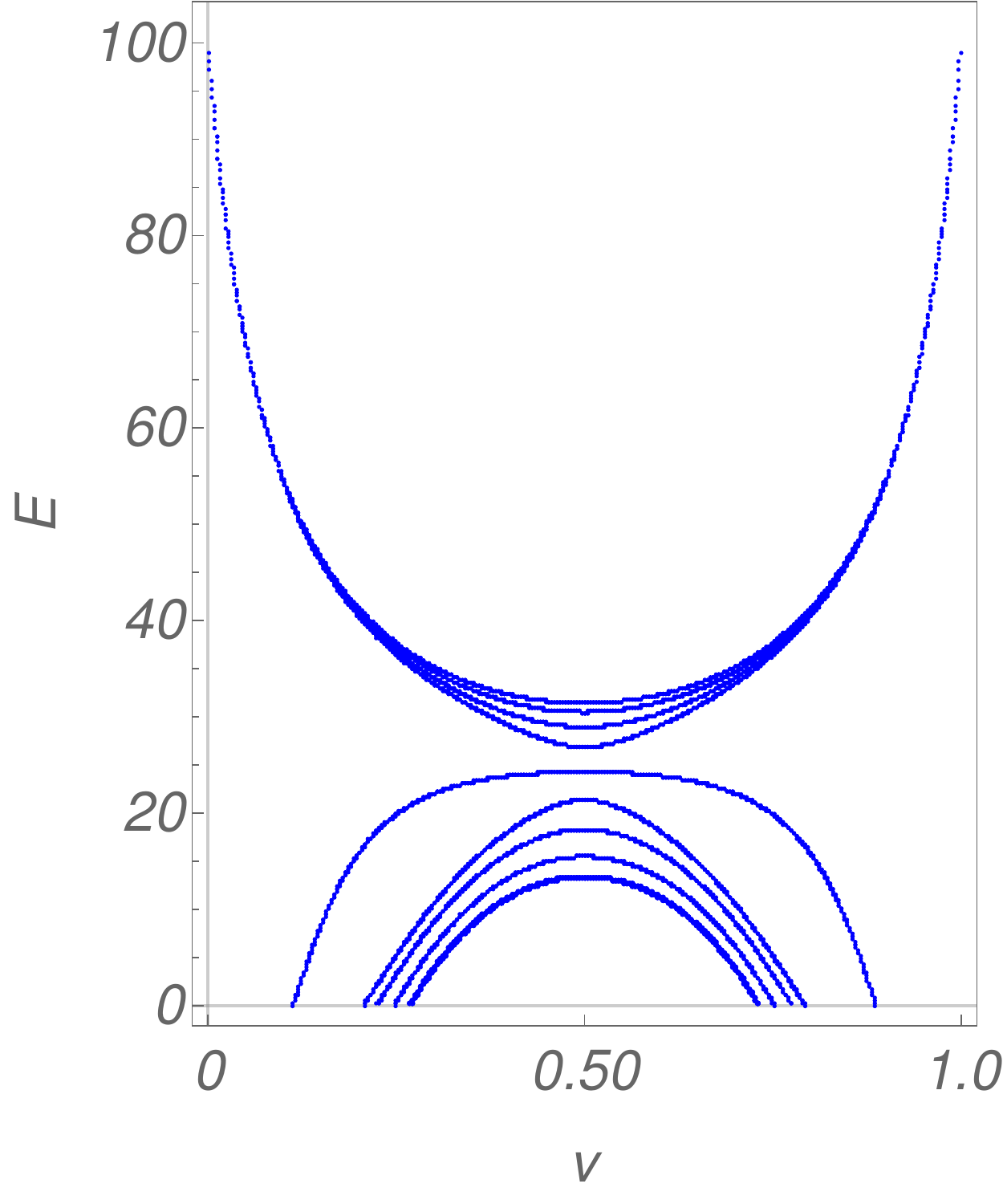}\put(73,18){(c)}
\end{overpic}
\end{minipage}%
\begin{minipage}{.5\linewidth}
\begin{overpic}[width=\linewidth]{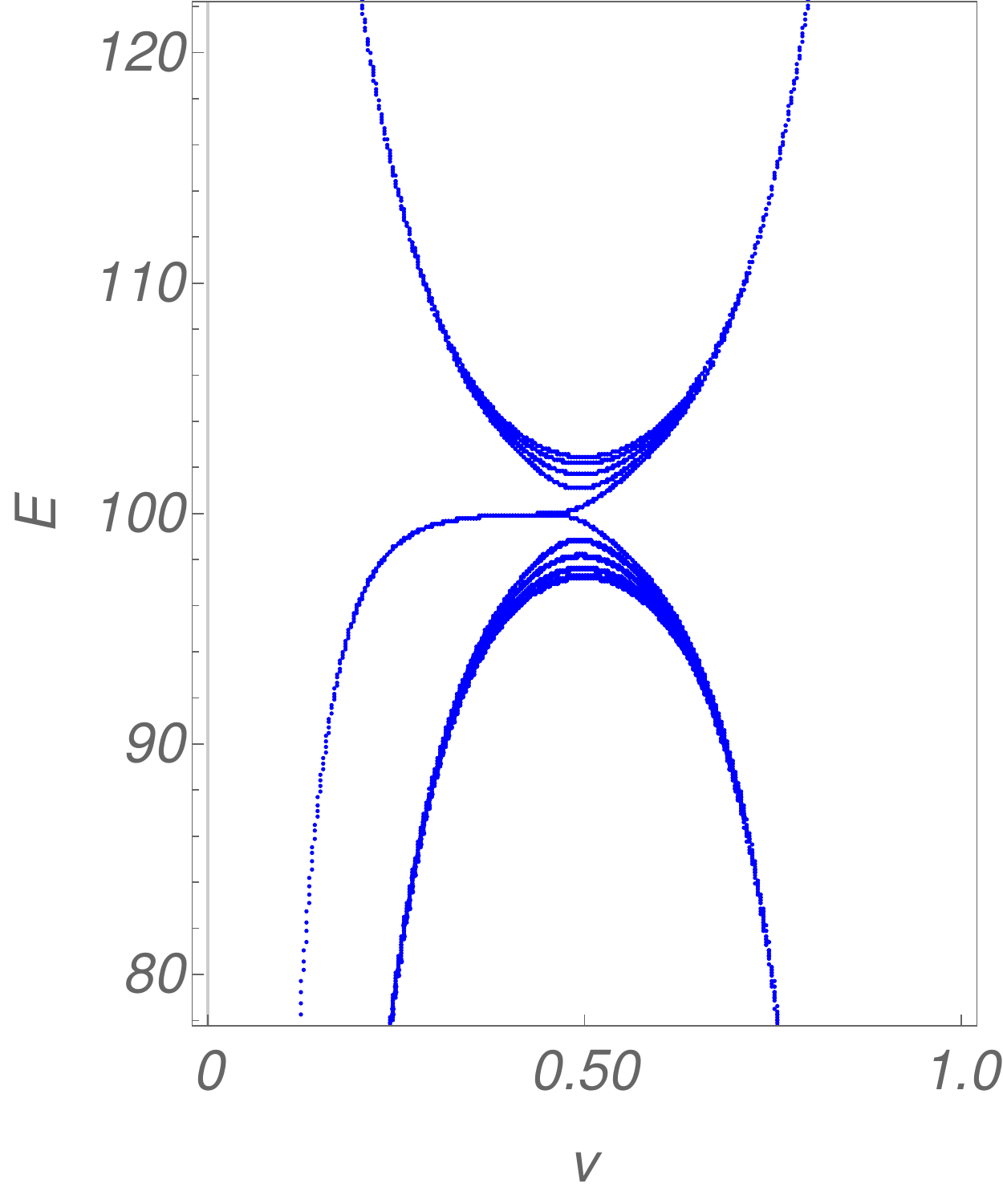}\put(72,18){(b)}
\end{overpic}
\begin{overpic}[width=\linewidth]{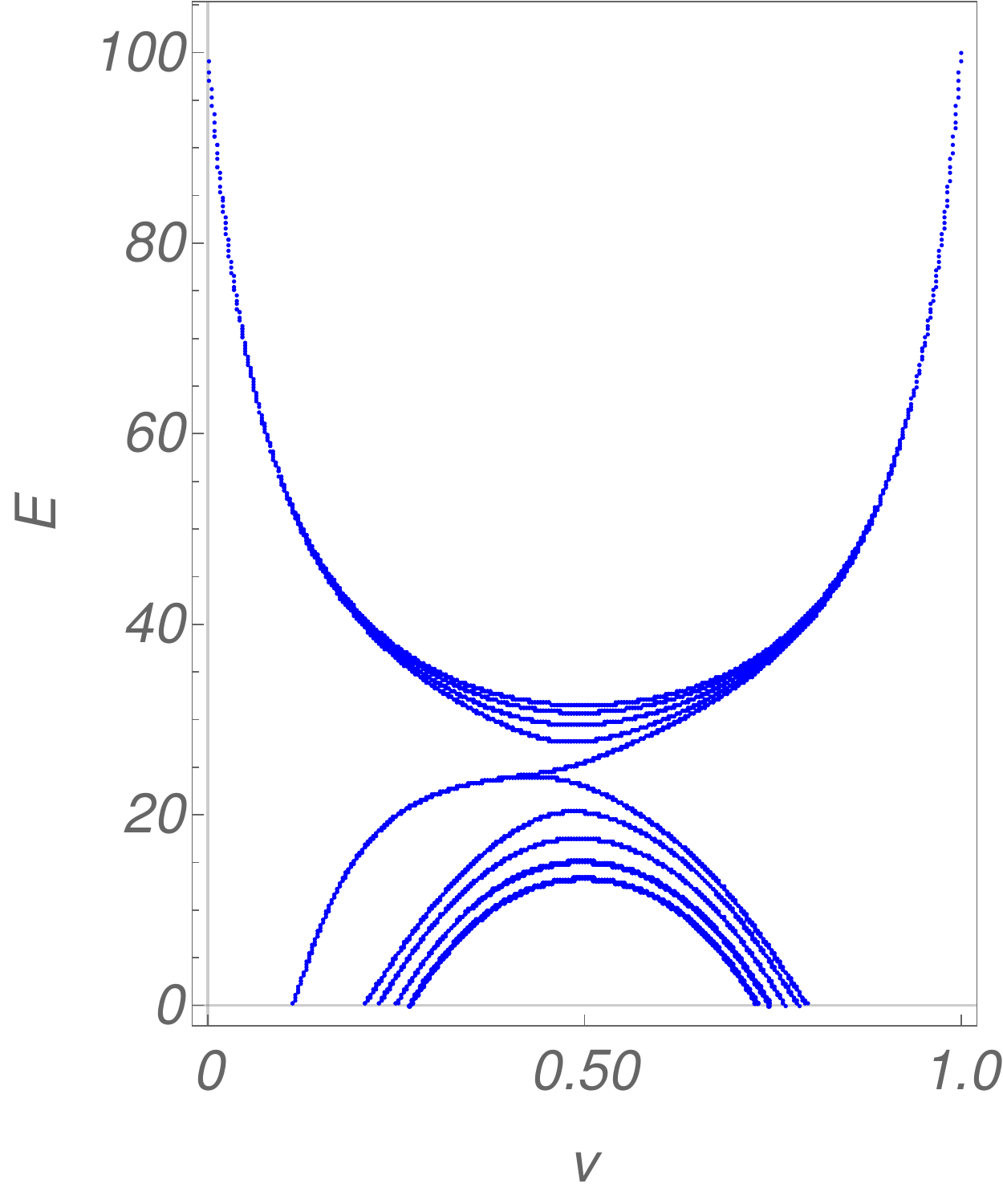}\put(72,18){(d)}
\end{overpic}
\end{minipage}%
\caption{(Colour on-line) The quasi-momentum, $q$, spectra for the finite bipartite Kronig-Penney model with varying $N$ and $|V|$ as a function of $v$ as it is varied from zero to its maximum value of $d=1$. Panel (a): $N=9$ and $|V|=10$, panel (b): $N=10$ and $|V|=10$, panel (c): $N=9$ and $|V|=5$, and panel (d): $N=10$ and $|V|=5$.}
\label{fig:negDKPwEn}
\end{figure}

In the present case, we have scatterers of identical heights and so $V_{\cal E}=V_{\cal O}=-V$. As mentioned, we choose $w=1-v$ and so $L$ is determined from this with the number of scatterers within the chain. Those to be considered will have $N=9$ for the odd case and $N=10$ for the even case. Thus, $L=5$ for $N=9$ and $L=5+v$ for $N=10$. The reasoning for having $L=5+v$ rather than simply $L=5$ is to keep the resultant spectrum symmetric about the line $v=w=d/2$.

The quasi-momentum band spectra are plotted in Fig.~\ref{fig:negDKPwEn}. As may be seen, there are clear `zero' energy mid-gap states of clearly similar nature as one would find in the SSH model. Indeed, if one were to solve the SSH model with hoppings of $v$, $w=1-v$ and a constant potential $V=10$, as is done in Appendix \ref{appB}, one would see energy bands strikingly similar to those as shown here. The states exist precisely in the middle of the gap at the potential height of the scatterers $|V|=10$. In the context of the tight-binding SSH model, this manifests itself as a $V\mathbb{1}_2$ term.

For the odd case, the edge state remains mid gap when both $v<w$ and $v>w$ since it simply migrates from one edge to the other. In other words, since the ends are terminated by $v$ and $w$ the act of sending $v$ to be greater than $w$ simply inverts the mirror symmetry of the chain. As such, the edge state moves accordingly.

For the even case, the edge state only exists within the topologically non-trivial region of $v<w$. This is because the edges terminate with $v$ wells and thus the act of sending $v$ to be greater than $w$ does not preserve the mirror symmetry of the chain. Thus, when $v>w$, no edge states may exist.

Comparing these plots in Fig.~\ref{fig:negDKPwEn} with the top two plots as in Fig.~\ref{fig:SSH} in Appendix \ref{appB}, one sees that, in the even case, the topological transition and coming together of the bulk bands to form a degenerate edge state is a sharp one. By which it is meant that there is very little communication/overlap between the edges of the chain that would cause these bands to remain non-degenerate even within the non-trivial region. Thus the even edge state(s) are very well-defined and localised strongly over the vast majority of the non-trivial region $v<d/2$.

The degenerate nature of the even case edge state indicates the absence of different sublattice-dependent on-site potentials that would otherwise differentiate the two sublattices non-trivially. A property usually associated with symmetric bands, it is seen to remain present in the case of asymmetric bands as may be seen in Fig.~\ref{fig:negDKPwEn}. The upper and lower bands are clearly asymmetric however the edge states remain mid-gap and topological.

\begin{figure}
\centering
\begin{minipage}{.5\linewidth}
\begin{overpic}[width=\linewidth]{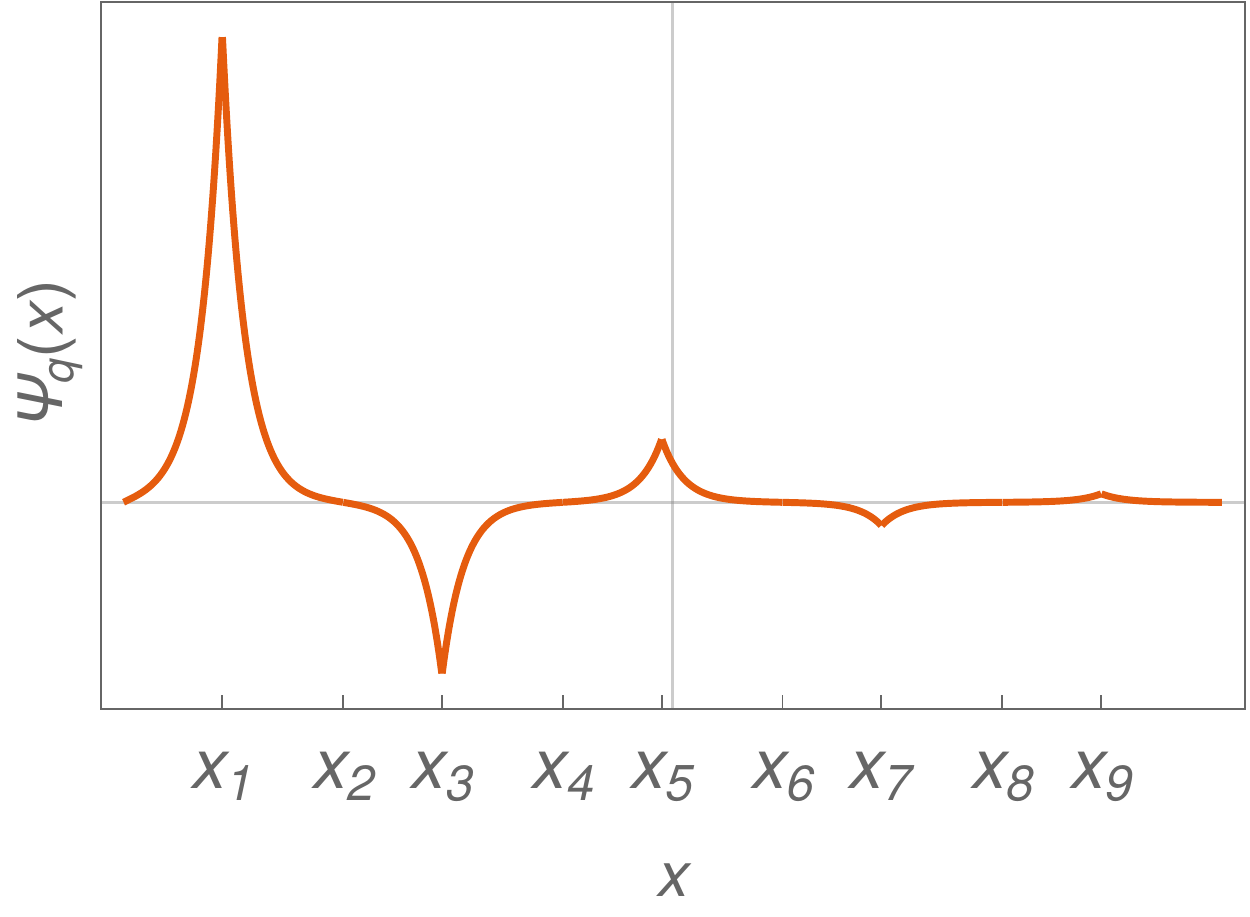}\put(88,20){(a)}
\end{overpic}
\begin{overpic}[width=\linewidth]{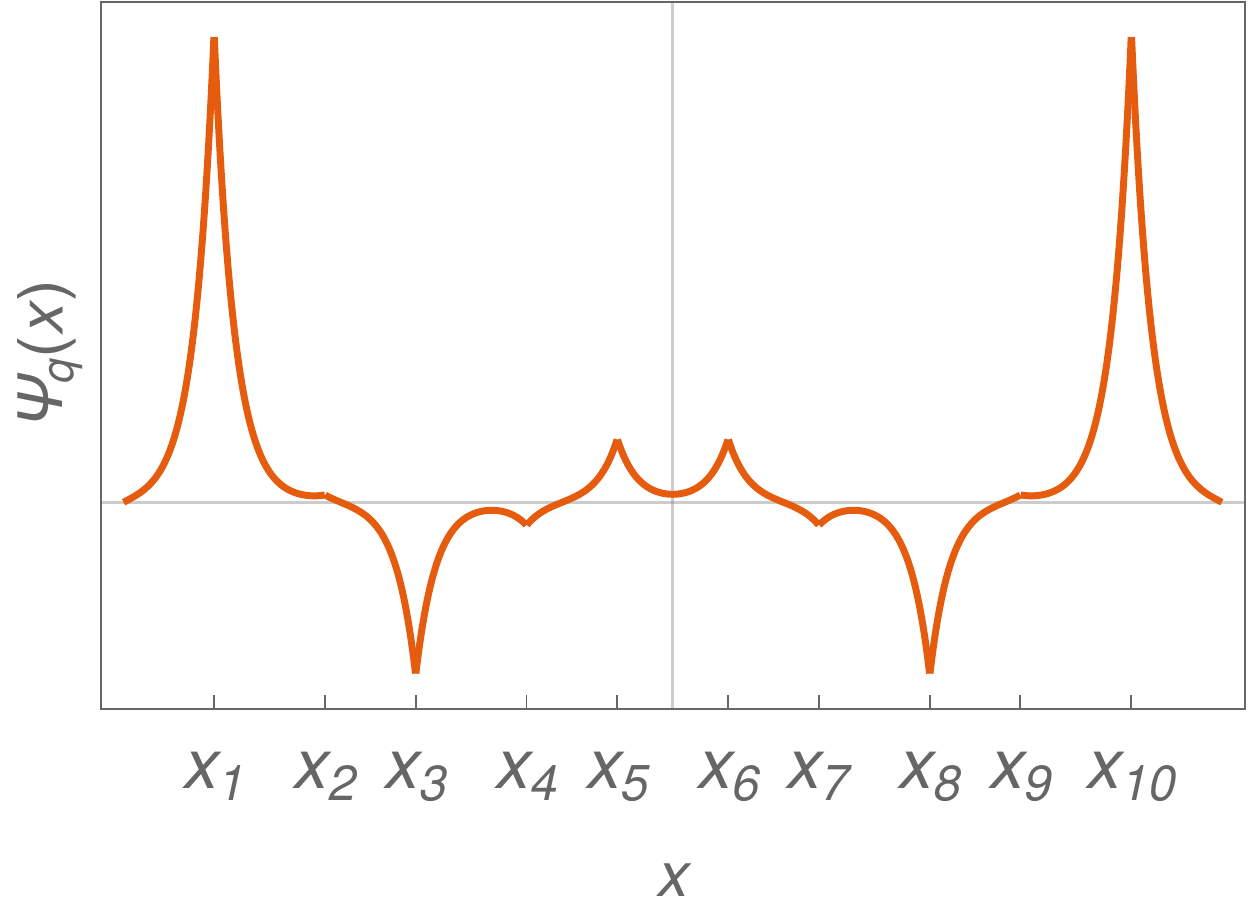}\put(88,20){(c)}
\end{overpic}
\end{minipage}%
\begin{minipage}{.5\linewidth}
\begin{overpic}[width=\linewidth]{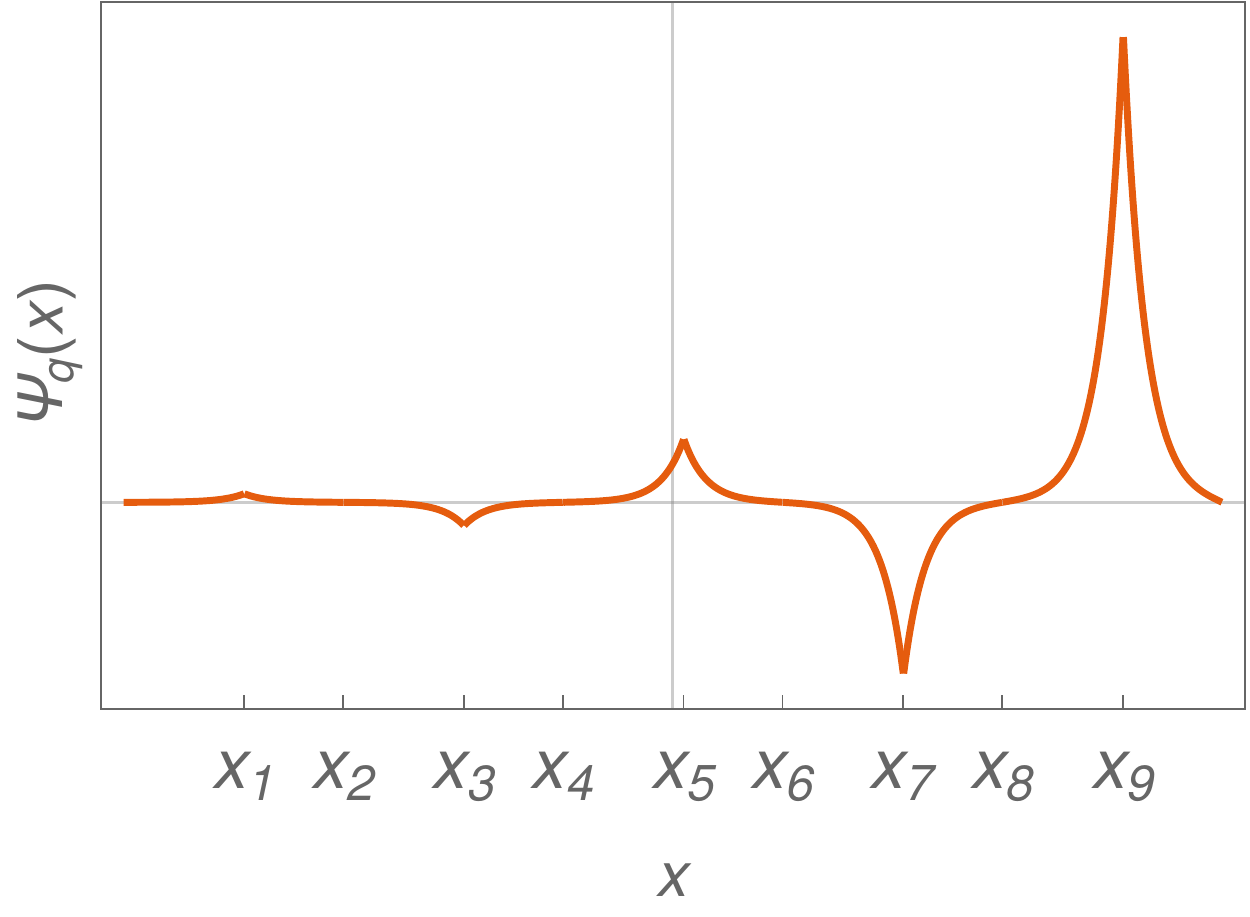}\put(88,20){(b)}
\end{overpic}
\begin{overpic}[width=\linewidth]{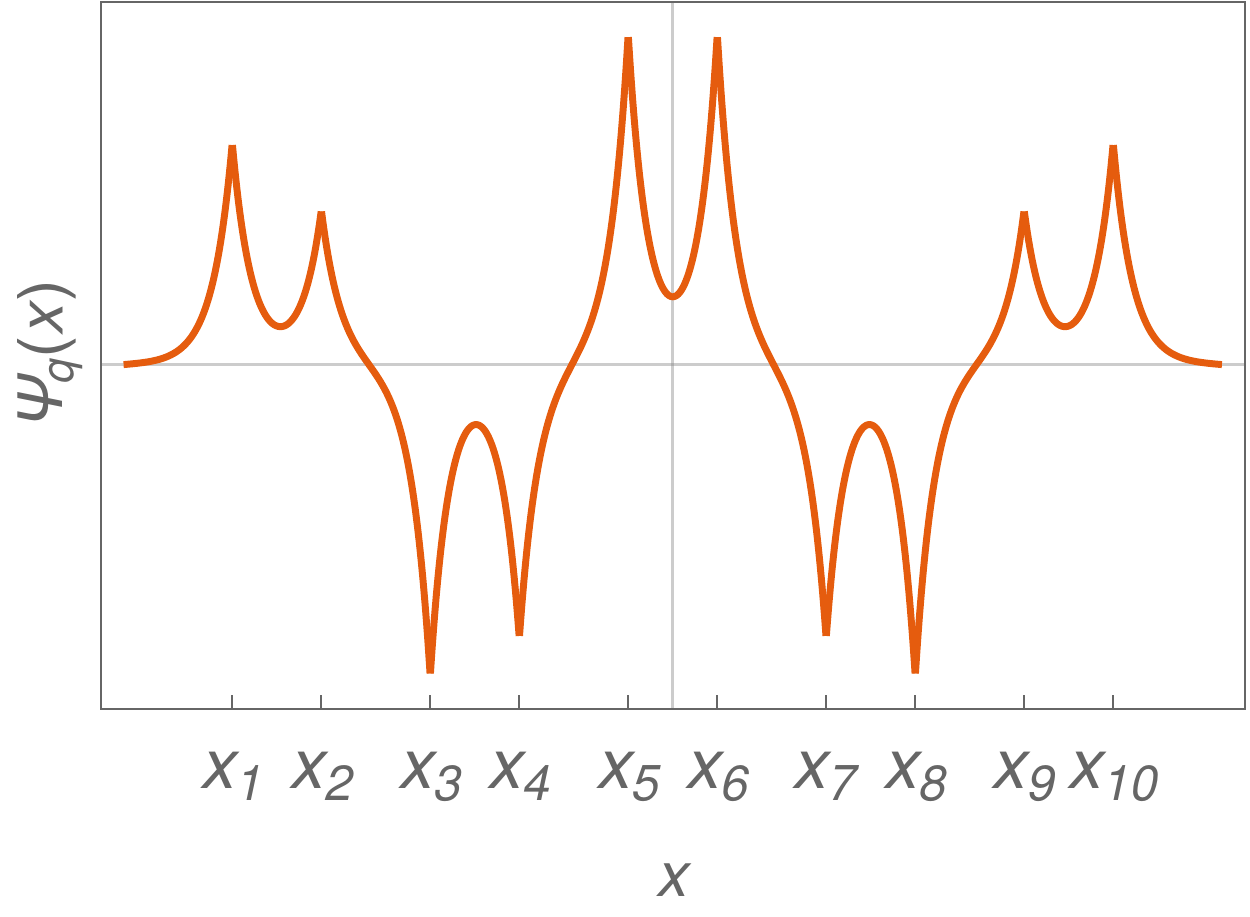}\put(88,20){(d)}
\end{overpic}
\end{minipage}%
\caption{(Colour on-line) The wavefunction of the fifth state over the chains of the finite negative BP K-P model with $|V|=10$ and varying $N$ and $v$. Panel (a): $N=9$ and $v=.45$, panel (b): $N=9$ and $v=.55$, panel (c): $N=9$ and $v=.45$, and panel (d): $N=10$ and $v=.55$.}
\label{fig:negDKPst}
\end{figure}

Finally, when $v$ becomes too small or large, {\it i.e.} $v<0.2$ and $v>0.8$, the edge states cease to be mid-gap in that their energy begins to vary with $v$. However, they do not become bulk states but instead remain exponentially localised to the edge. Furthermore, within the bulk, the lower band is destroyed leaving only the upper band when $v<0.2$ and $v>0.8$. This is as a result of hybridisation between the two scatterers when they are brought too close to each other. The edge state remains highly localised as one of the scatters is brought closer to the hard walls. It is then forced to be an edge state by a trivial localisation. In the bulk, when $v>0.8$ and $v<0.2$, the lower band ceases to exist and so its Zak phase and winding number become undefined whilst the upper band remains retaining its topological character.

Due to the hybridisation of the bulk states between the neighbouring potentials as the $v-w$ difference becomes larger, the topological character of the chain is best seen when $v$ and $w$ are almost, but not exactly, equal to each other. This is consistent (and obvious) when considering the bulk bands, which become flat when $v$ is radically different from $w$.

\begin{figure}
\centering
\begin{minipage}{.5\linewidth}
\begin{overpic}[width=\linewidth]{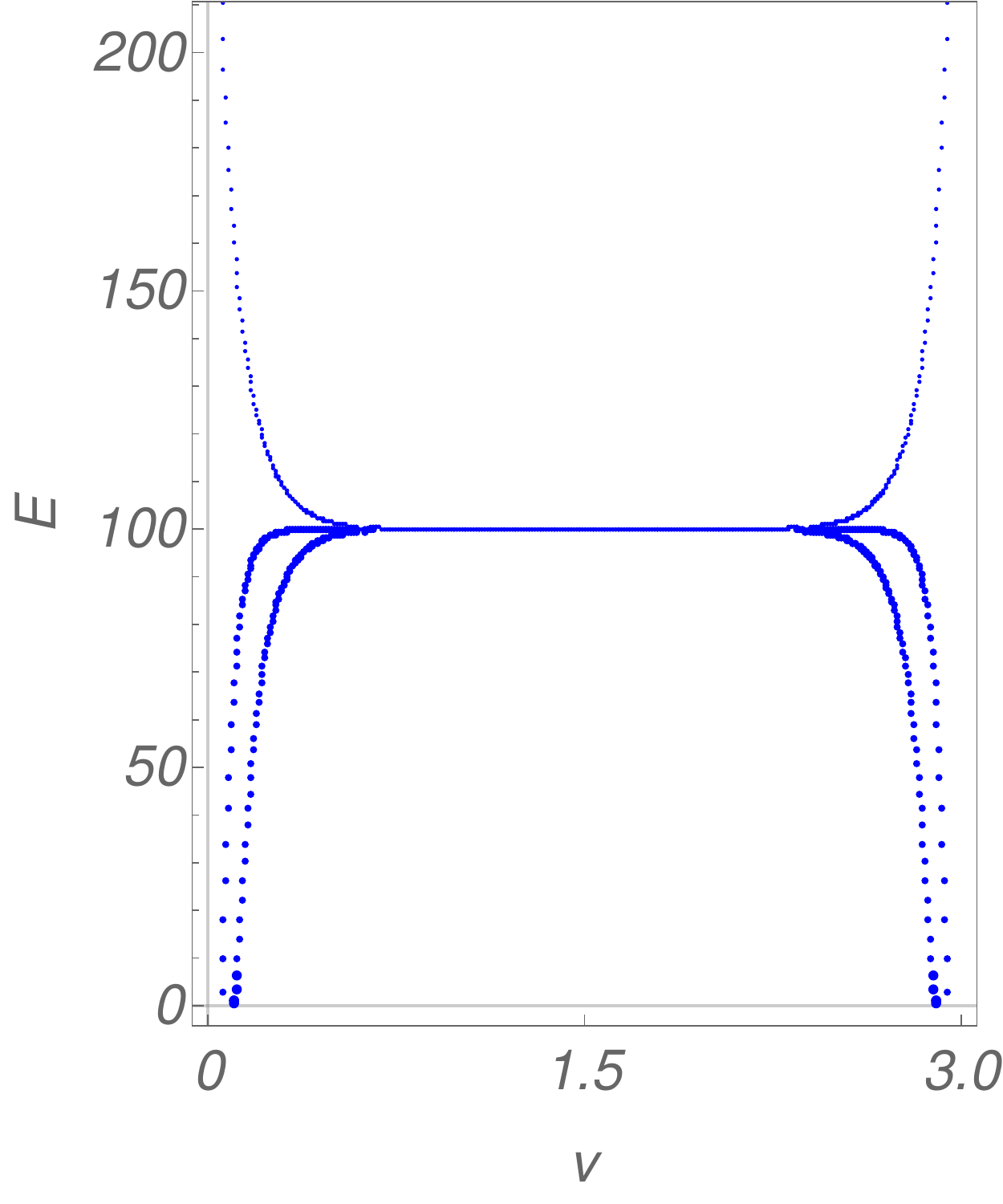}\put(20,93){(a)}
\end{overpic}
\end{minipage}%
\begin{minipage}{.5\linewidth}
\begin{overpic}[width=\linewidth]{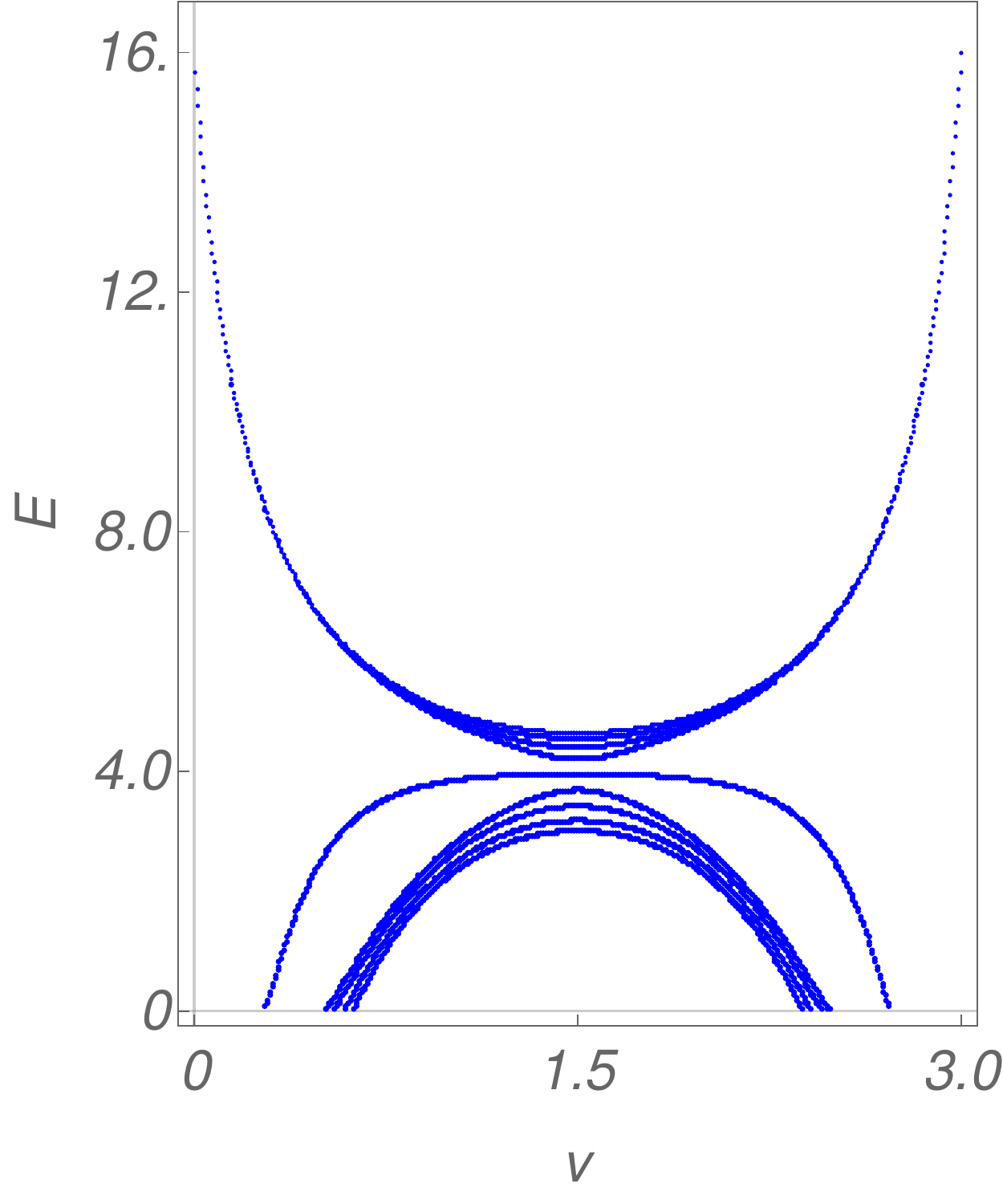}\put(20,93){(b)}
\end{overpic}
\end{minipage}
\caption{(Colour on-line) The quasi-momentum, $q$, spectra for the finite bipartite Kronig-Penney model with $N=9$ and varing $|V|$ as a function of $v$ from zero to its maximum value of $d=3$. Panel (a): $|V|=10$, and panel (b): $|V|=2$.}
\label{fig:negDKPwEx}
\end{figure}

All this conjecture may be seen in the plots of the edge states for the cases of $N=9$ and $N=10$ in Fig.~\ref{fig:negDKPst}. In the odd case, the edge state migrates from one side of the chain to the other upon the transition and in the even case it only exists when $v<w$. In this case, it exists with equal weight at either end thus exhibiting the phenomenon of fractionalisation of charge.\cite{Asboth:2016}

Finally, if the size of the unit cell is taken to be larger then the protected edge states remain. However, they become indiscernible from the bulk bands when $v\sim w$ as the gap between the upper and lower bands shrinks. Indeed, the numerical solution ultimately fails to pick up all the states in such a narrow region of $q$. In order that this gap remain open as $d$ increases, the potential heights must be reduced. This may be seen in Fig.~\ref{fig:negDKPwEx}.

Thus, in order to see topological edge states, a chain of well-spaced scatterers must have small potentials heights whilst a chain of narrowly-spaced scatterers must have large potentials heights.

\subsection{Negative Potentials with Varying Heights}

\begin{figure}
\centering
\begin{minipage}{.5\linewidth}
\begin{overpic}[width=\linewidth]{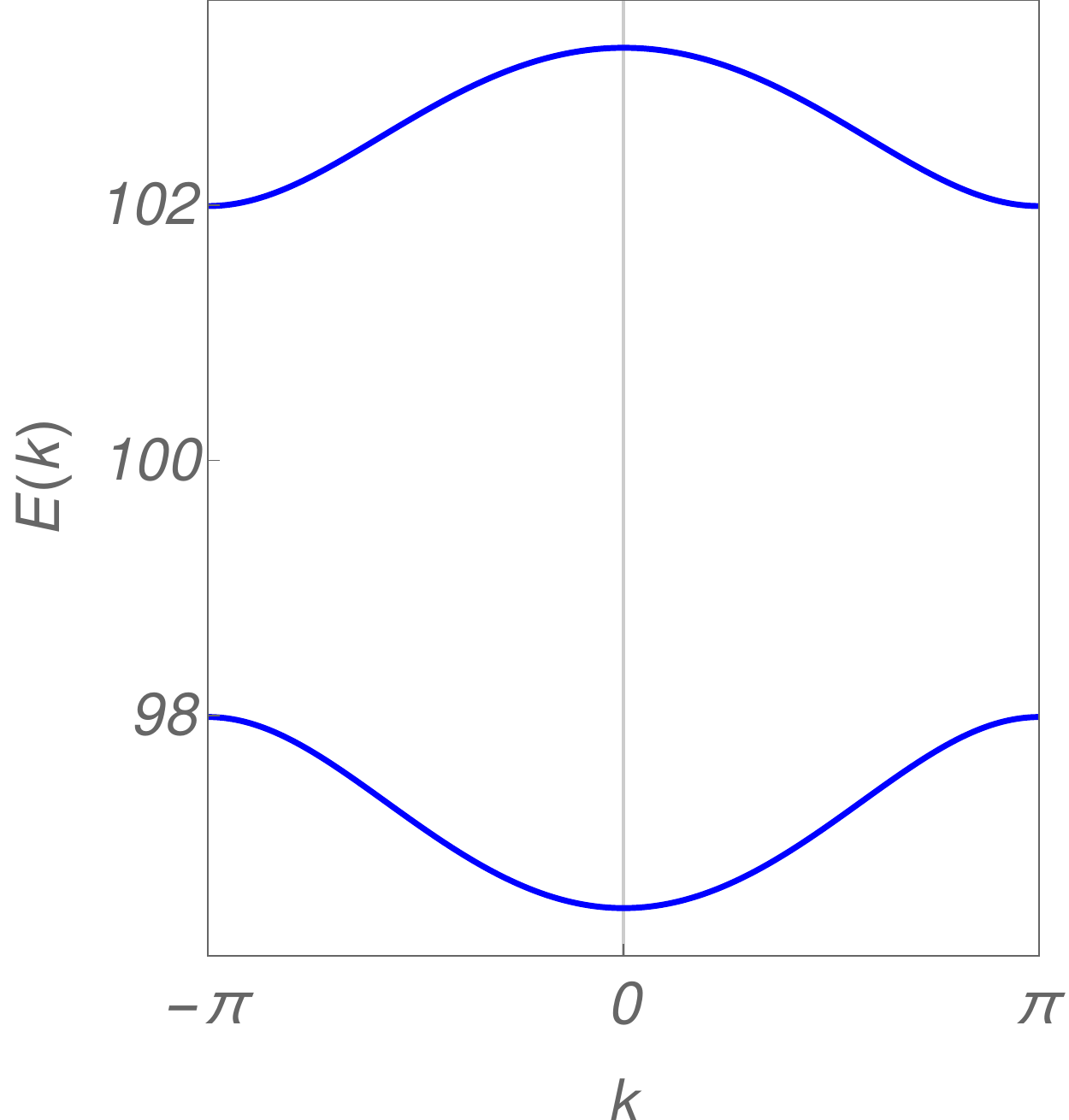}\put(20,18){(a)}
\end{overpic}
\begin{overpic}[width=\linewidth]{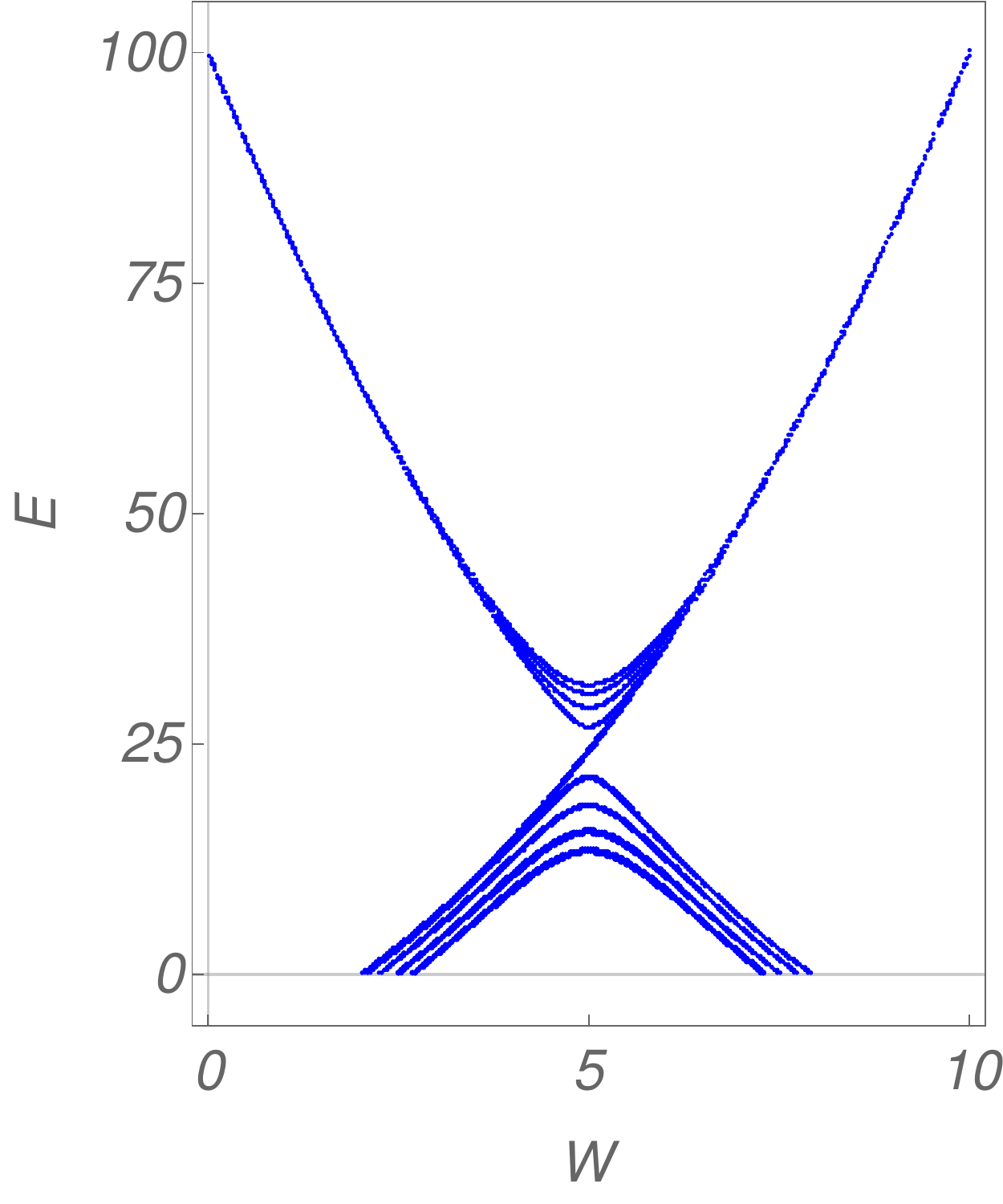}\put(16,22){(c)}
\end{overpic}
\end{minipage}%
\begin{minipage}{.5\linewidth}
\begin{overpic}[width=\linewidth]{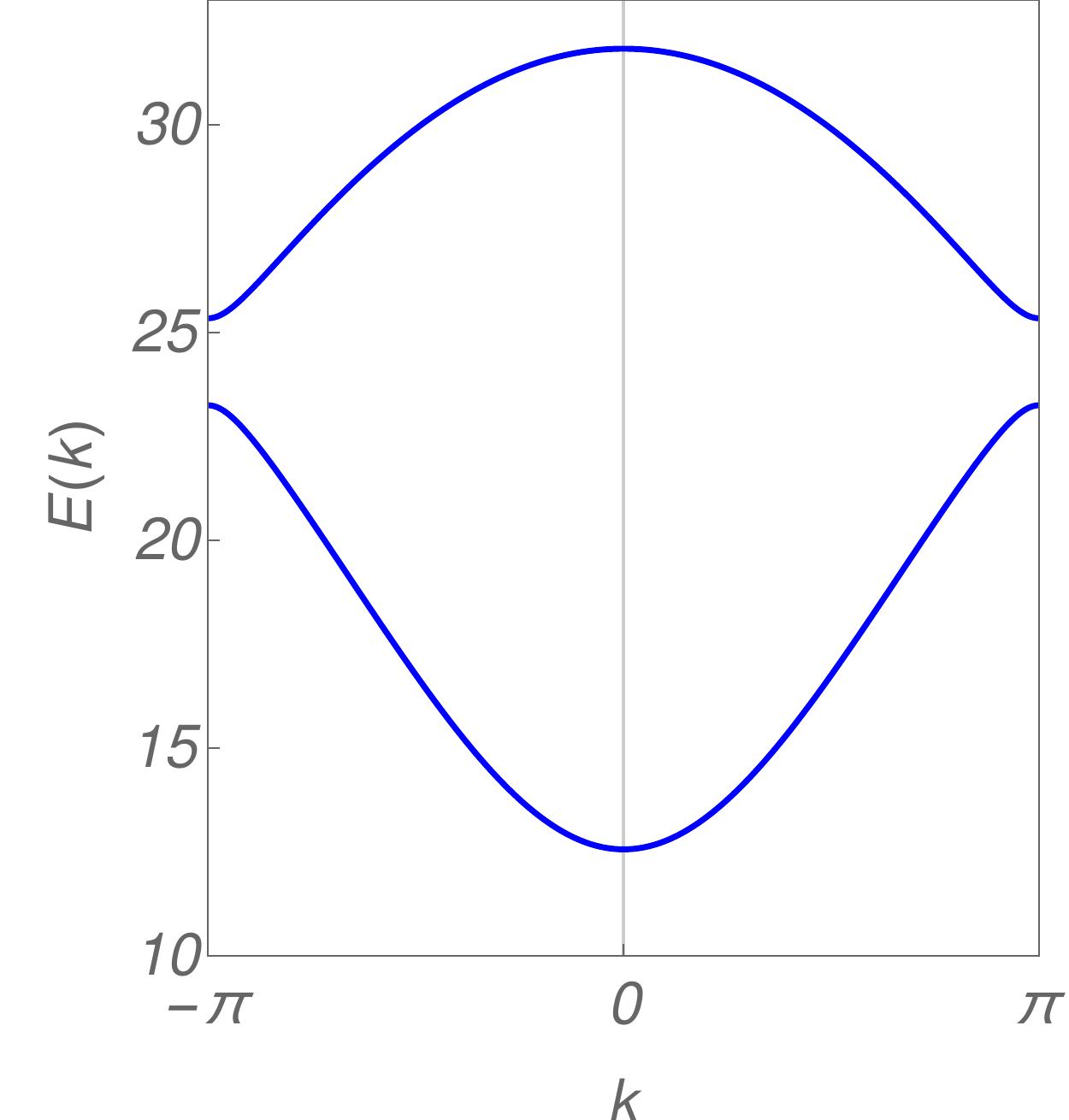}\put(20,18){(b)}
\end{overpic}
\begin{overpic}[width=\linewidth]{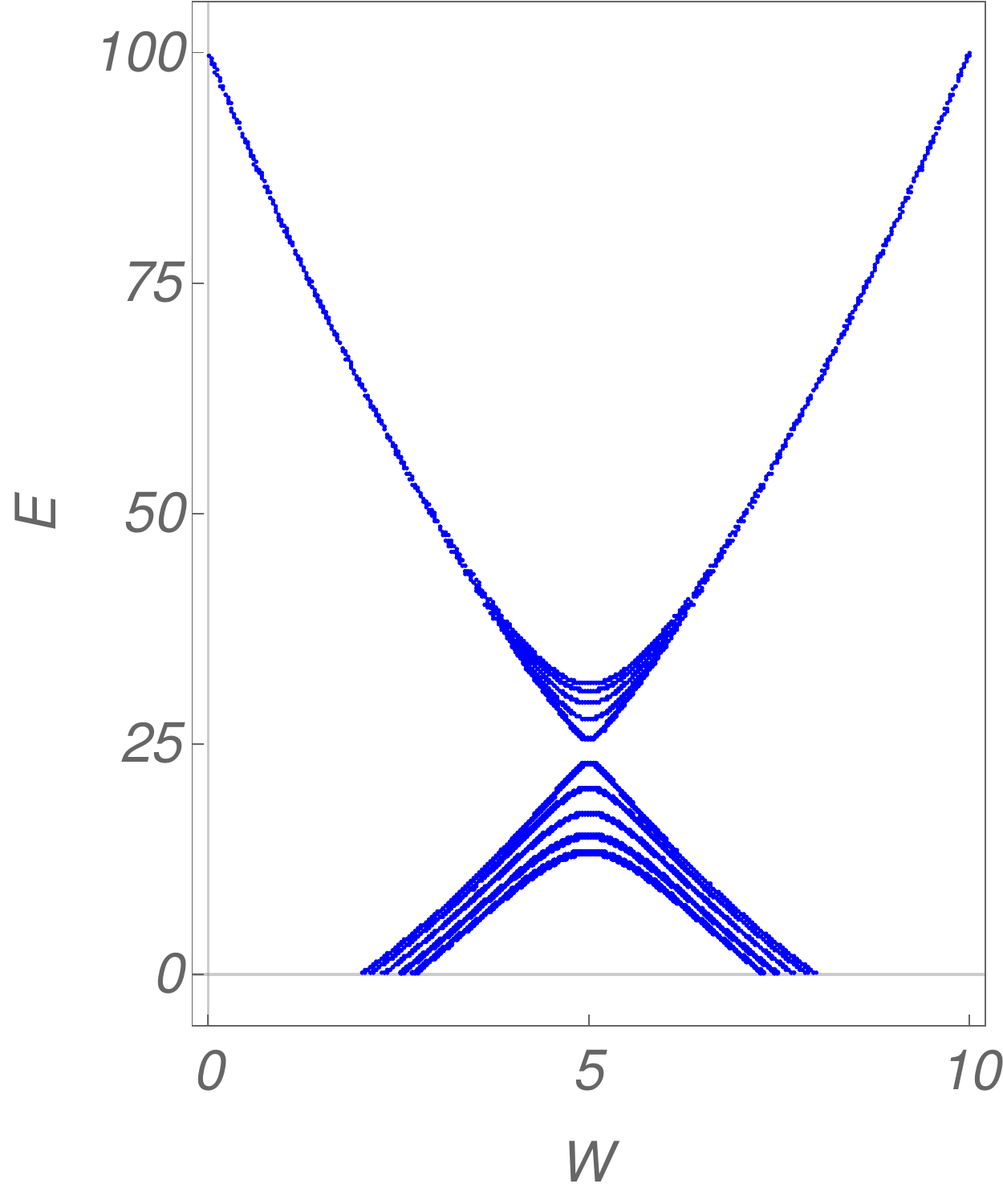}\put(16,22){(d)}
\end{overpic}
\end{minipage}%
\caption{(Colour on-line) The quasi-momentum spectra, $q$, for the negative bipartite Kronig-Penney model with constant widths $v=w=d/2=0.5$ and varying heights. The bulk and finite spectra are plotted in the top (a,b) and bottom (c,d) panels respectively. Panel (a): $U=20$ and $W=9.9$, panel (b): $U=10$ and $W=4.9$, panel (c): $U=10$ and $N=9$, and panel (d): $U=10$ and $N=10$. }
\label{fig:negDKPhEn}
\end{figure}

We now consider the case in which the distances between the potential scatterers are fixed to be equivalent and equal to $d/2$ whilst varying the potential heights. Again, the wavevector is made to be imaginary $q_k\to iq_k$ however the potentials are varied as $V_1\to -W$ and $V_2\to -U+W$. The trigonometric functions within the transcendental equation become hyperbolic once more and so two roots for $q_k$ are expected again.

The bulk spectrum for two cases are shown in Figs.~\ref{fig:negDKPhEn}(a,b). The left has $U=20$ and $W=9.9$ whilst the right has $U=10$ and $W=4.9$. Interestingly, the bands have identical forms to the case of varying widths with the lowest point of the upper band at $\max(|V_1|,|V_2|)$ and the highest point of the lower band at $\min(|V_1|,|V_2|)$.

This comes as no surprise however since both are bipartite unit cells of bound states. The difference then comes as a result of the different physical origin and character of the bipartite-ness and the total absence of chiral symmetry. Indeed, the differing potential heights act as on-site potential terms, which in the SSH model destroys chiral symmetry, and the specification that $v=w=d/2$ causes the hopping probabilities to be identical. As such not only is the topological behaviour destroyed but also the very edge states themselves.

This is reflected by the winding of $\rho(k)$, which does not change across the transition. Note that the winding of $\tau(k)$ does change in this case in the exactly the same way as it did in the previous case. This fact highlights that the winding of the transmission coefficient is not the invariant that corresponds to topological protection of edge states. This may be seen further with the calculation of the Zak phase, which may be found to undergo a transition from $2\pi\rightarrow\pi$ for the upper band and $\pi\rightarrow0$ for the lower band. Since we see no edge states in the finite system, this invariant does not apply to topological protection. (It may yet apply to some other topological effect, perhaps some form of charge pumping, but that is not for us to say.)

The finite system is solved and spectra are shown in Figs.~\ref{fig:negDKPhEn}(c,d). The potential heights are chosen as $|V_{\cal O}|=W$ and $|V_{\cal E}|=U-W$ with $U=10$ and the widths of each well are set to $v=w=d/2$. As may be seen, a state is seen to migrate from the lower band to the upper band in the odd scatterer case as $W$ is increased across the transition point of $W=U/2$. As in the previous case, this is exactly the same result as one would find in the SSH model of the same configuration as is shown in Appendix \ref{appB}.

This can be explained physically by the following mirror symmetry argument. For an even number of scatterers there are equal numbers of small and large potentials. In this $N=10$ case we have five small and five large for all $V_1$ and $V_2$. However, when $N$ is odd, this is not the case and there is a mismatch. For $N=9$ there are five small and four large scatterers when $V_1<V_2$. Thus, when $V_1$ becomes greater then $V_2$ this switches such that we have four small and five large. Thus, one of the five states that existed in the lower band with lower energy corresponding to one of the five small potentials is ejected to join the upper band, which now has higher energy, that also now corresponds to the five large potentials.

\subsection{Positive Potentials with Varying Widths}

\begin{figure}
\centering
\begin{minipage}{.5\linewidth}
\begin{overpic}[width=\linewidth]{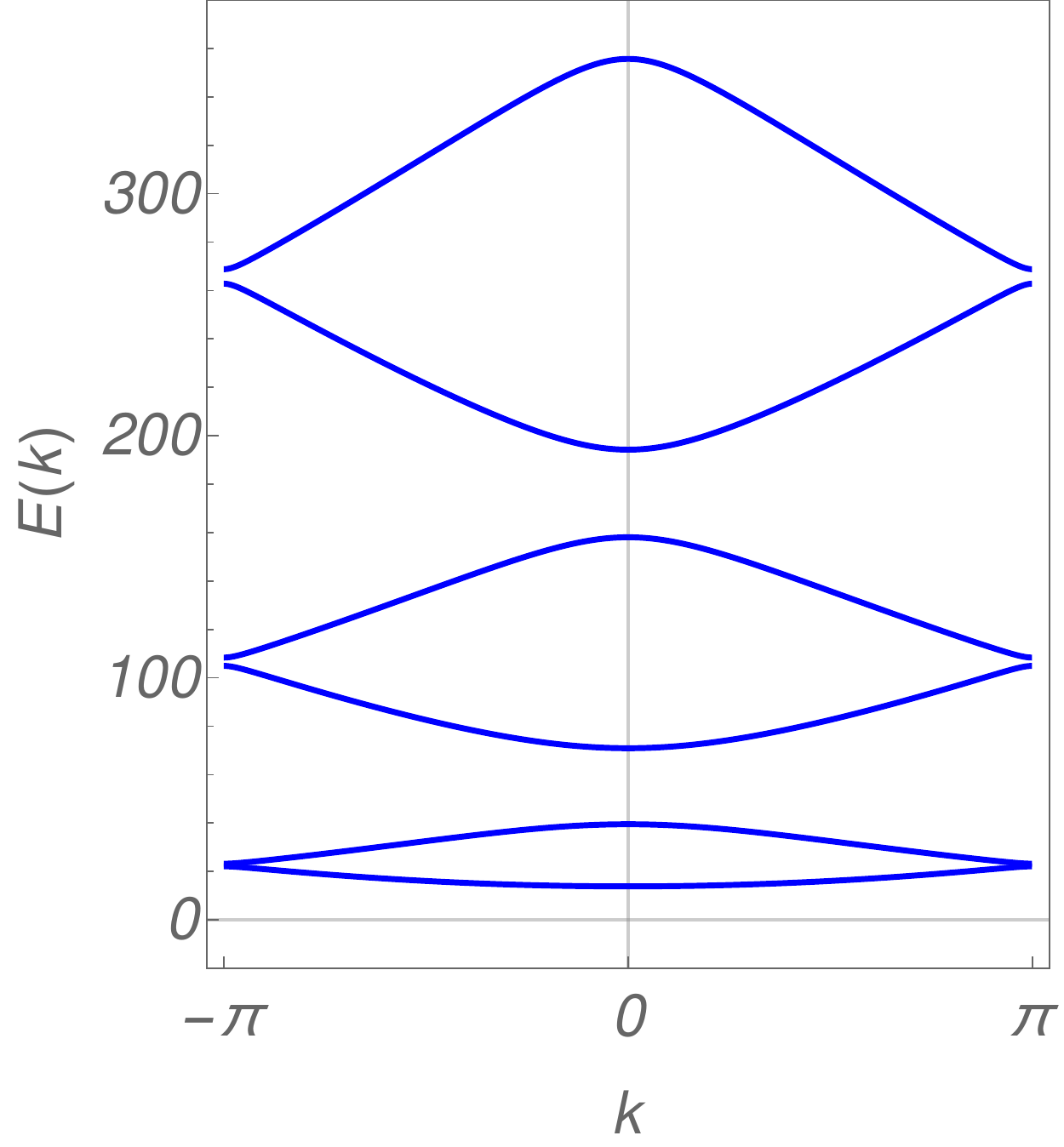}\put(19,29){(a)}
\end{overpic}
\begin{overpic}[width=\linewidth]{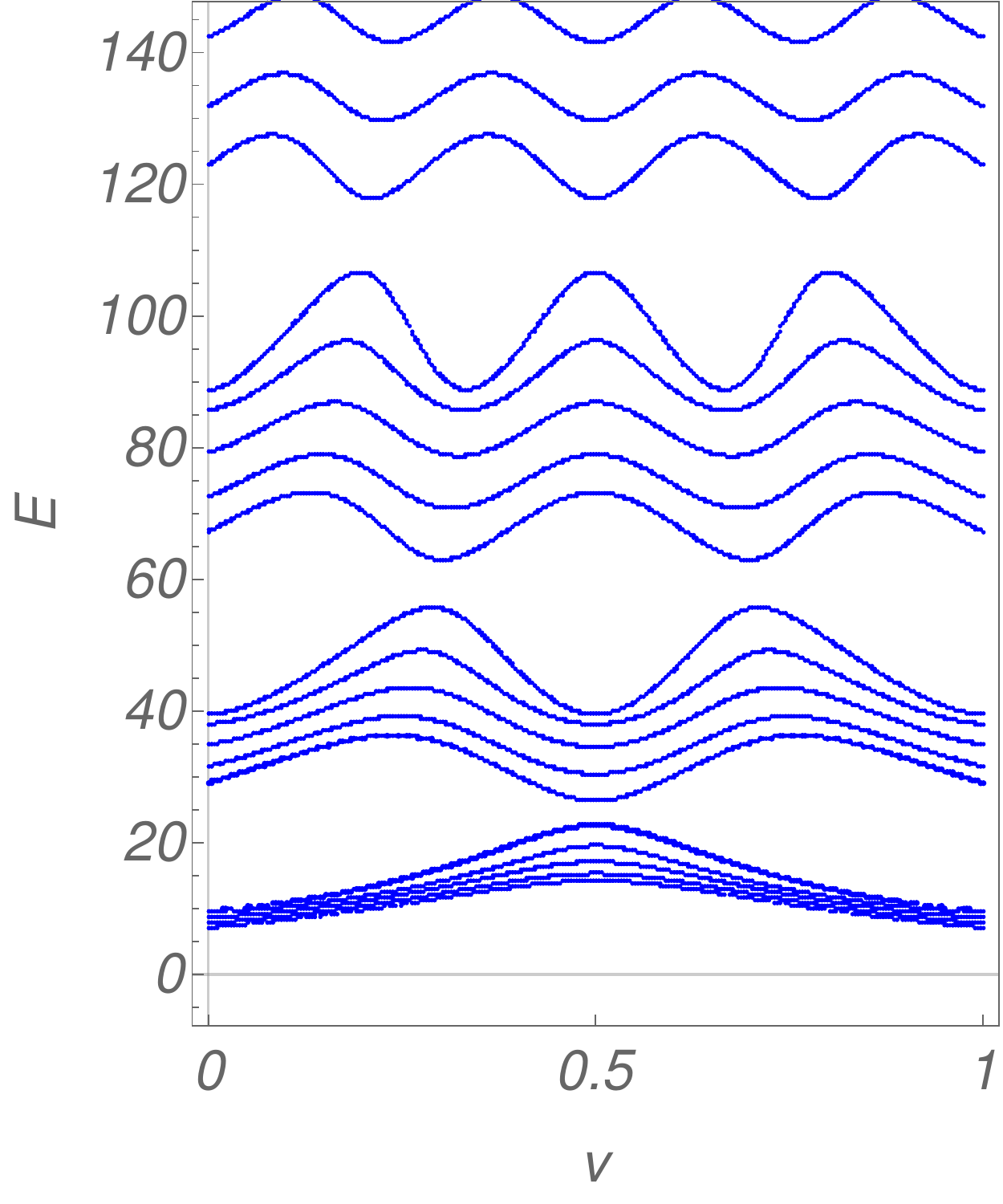}\put(17,28){(c)}
\end{overpic}
\end{minipage}%
\begin{minipage}{.5\linewidth}
\begin{overpic}[width=\linewidth]{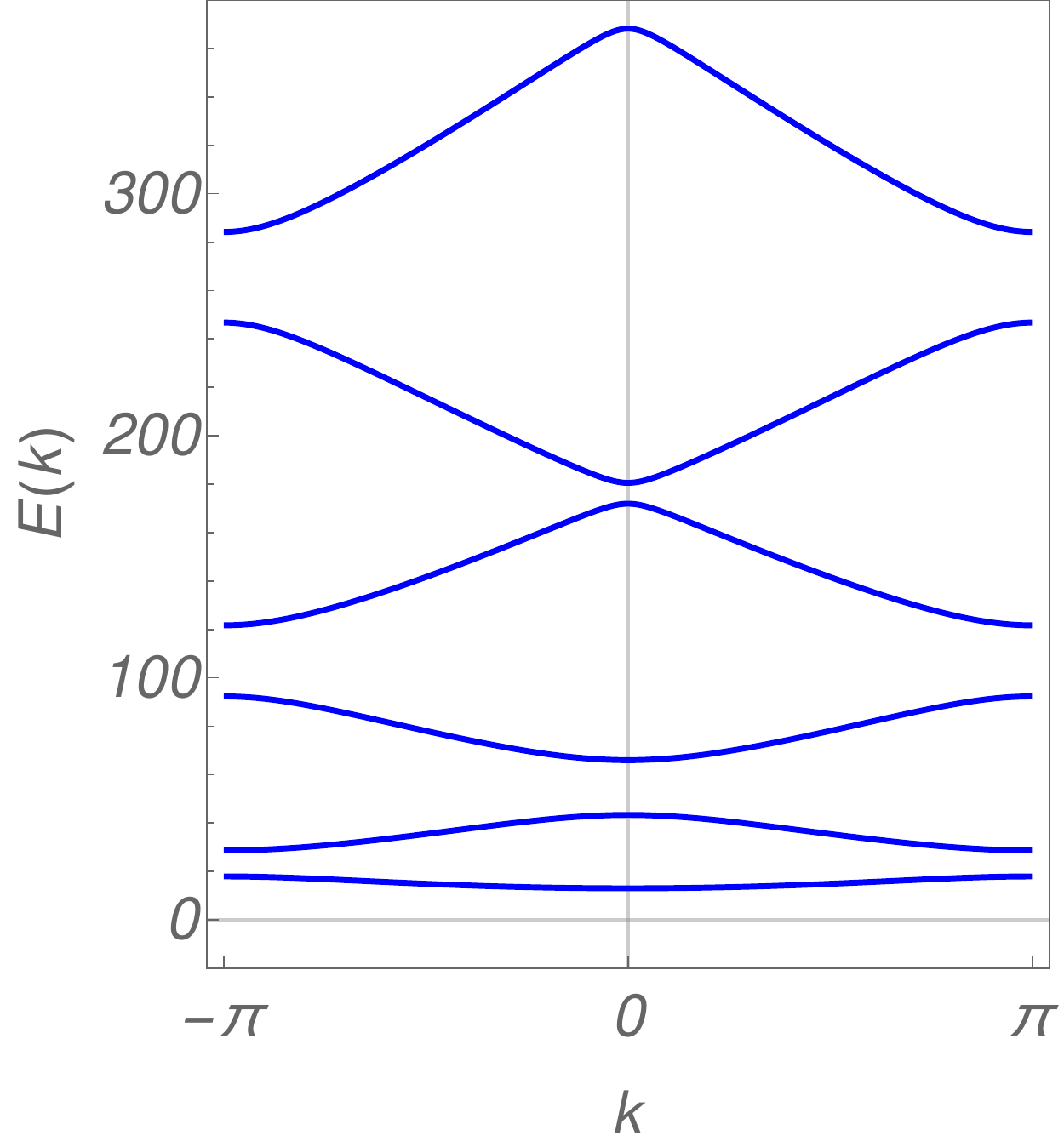}\put(19,29){(b)}
\end{overpic}
\begin{overpic}[width=\linewidth]{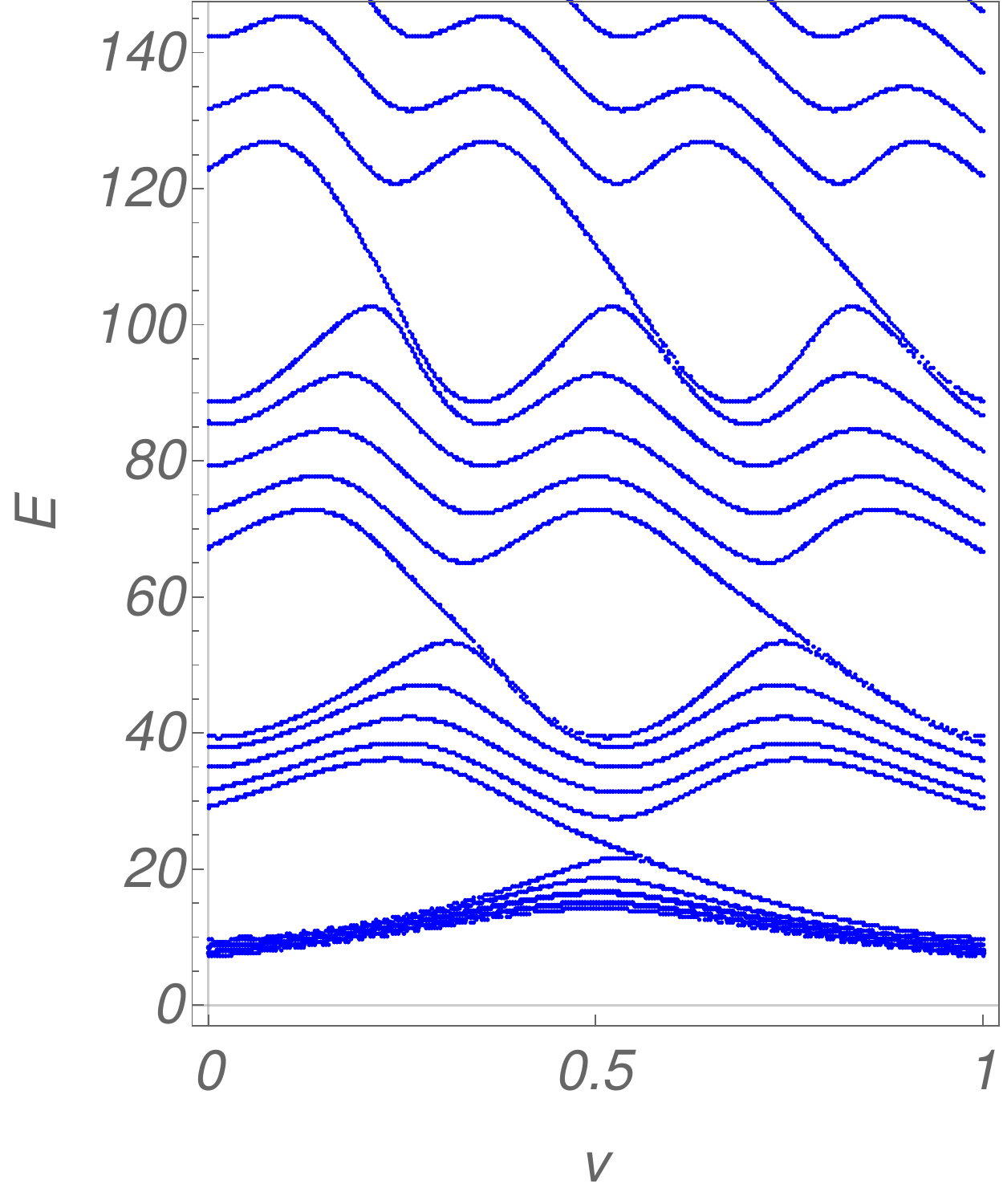}\put(17,27){(d)}
\end{overpic}
\end{minipage}
\caption{(Colour on-line) The quasi-momentum, $q$, spectra for the positive bipartite Kronig-Penney model with constant potential heights of strength $|V|=5$ and unit cell width $d=1$. The bulk and finite spectra are plotted in the top (a,b) and bottom (c,d) panels respectively. Panel (a): $v=.49$, panel (b): $v=.4$, panel (c): $N=9$, and panel (d): $N=10$.}
\label{fig:posKPspectra}
\end{figure}

We turn now to the case of positive potential heights with varying widths. In this case we have that $V_{1,2}\to V$ and real wavevector so we expect to find infinitely many bands due to the sinusoidal functions in the transcendental equation. Indeed, this is the case, as may be seen in the plots in Figs.~\ref{fig:posKPspectra}(a,b).

Interestingly, here, decreasing the height acts to not only lower the energies of the bands but also to effectively close any band gaps. This is because the higher energy bands will simply ignore the effects of the potentials if the heights are much smaller. This is evident on the left-hand plot too since the band gaps that occur for $k=0$ decrease in size as $q_k$ increases.

However, an interesting feature emerges as seen in the right-hand plot. At $k=0$, the fourth and fifth bands appear to be close to touching as opposed to the second and third bands, which are very far apart from each other. At some certain value of $v$ the bands do indeed cross and will reopen if $v$ is changed once again. This would appear to be the same mechanism of a topological transition through band closing and reopening as in the negative scatterer case. However, it is not, as the winding numbers of the reflection coefficient do not undergo a transition from zero to one. It is instead a trivial band closing and so any edge states obtained within the system will not be topological.

\begin{figure}
\centering
\begin{minipage}{.5\linewidth}
\begin{overpic}[width=\linewidth]{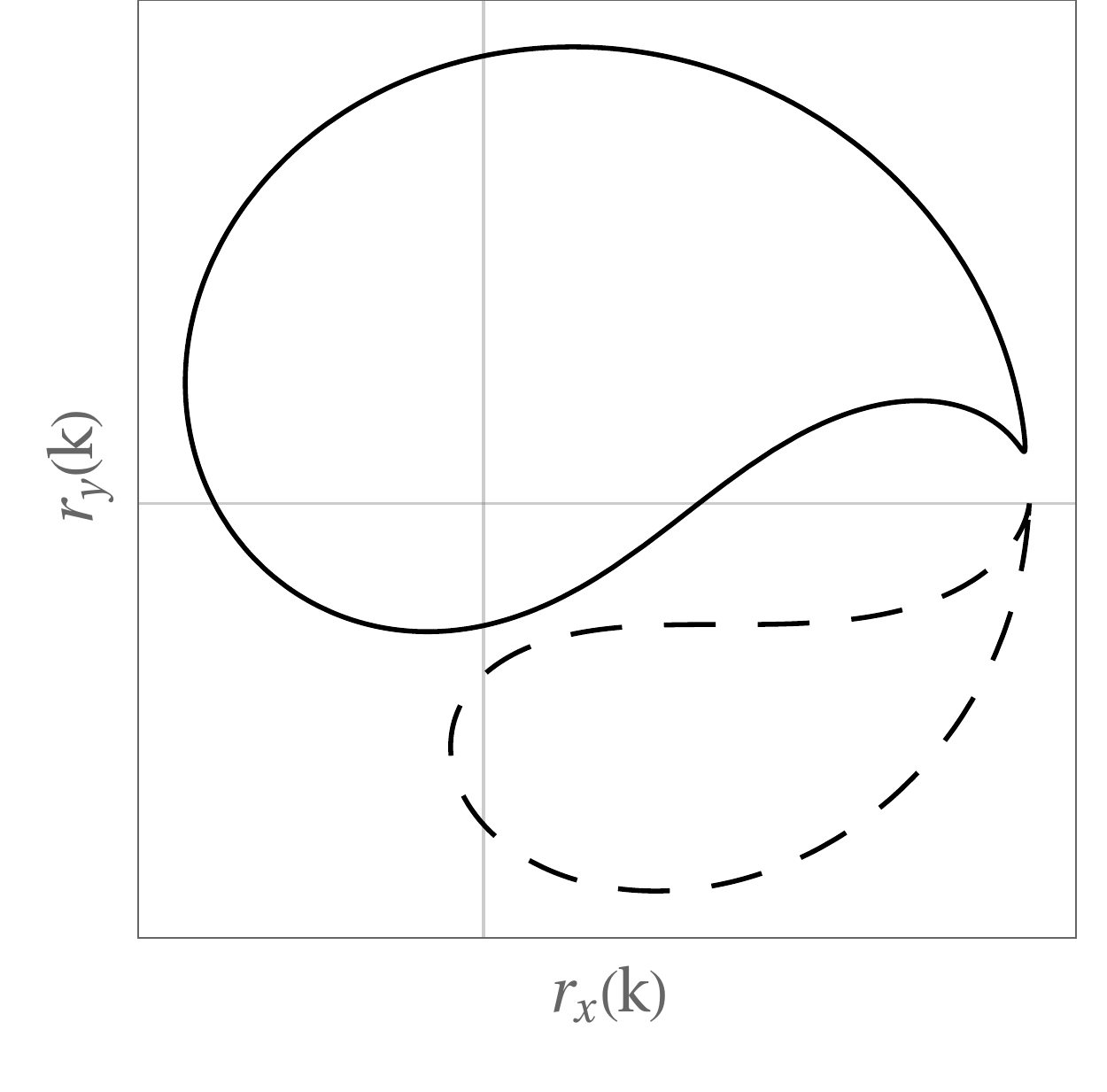}\put(12,16){(a)}
\end{overpic}
\end{minipage}%
\begin{minipage}{.5\linewidth}
\begin{overpic}[width=\linewidth]{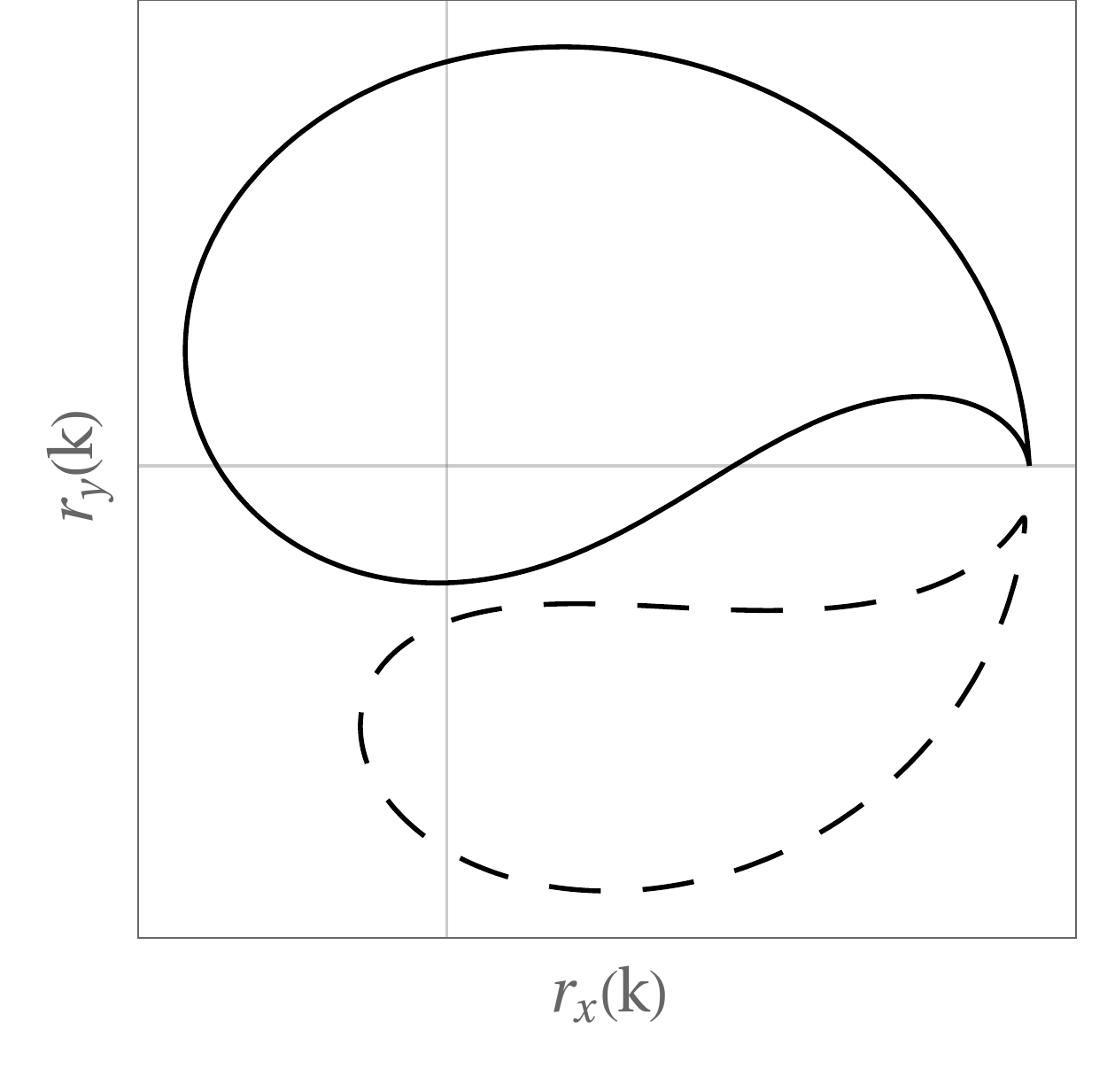}\put(12,16){(b)}
\end{overpic}
\end{minipage}
\caption{The windings of the reflection coefficients for the first two bands of the bulk bipartite Kronig-Penney model with positive scatterers of strength $|V|=5$ and unit cell width $d=1$. Dashed/thick correspond to the first/second bands since $q_k$ is band dependent. Panel (a): $v=.49$, and panel (b): $v=.51$.}
\label{fig:posDKPrcoeffs}
\end{figure}

For this case, the reflection coefficient reads:
\begin{equation}
r(k)=e^{iqd}\left[\frac{V^2e^{iq(d-2v)}-q_k^2e^{ikd}}{(V-iq_k)^2}\right],
\end{equation}
which may be decomposed as $r(k)=r_x(k)+ir_y(k)$. Considering, for clarity, the first two bands wherein a band closing is expected when $v=w=d/2$, plots of their reflection coefficients may be seen before and after this point in Fig.~\ref{fig:posDKPrcoeffs}. As may be seen, the windings are zero and one before the transition, panel (a), and zero and one afterwards, panel (b), indicating that the transition is not topological. The same may be applied to all the other band closings, both at $k=0$ and $k=\pm\pi/d$, of the bulk spectrum to see the same result; the absence of change in the winding number upon the transition. In this case, the transmission coefficient passes through the origin of the $r_x-r_y$ plane so its winding is undefined/unquantised.

This same behaviour may be observed in the Zak phase. When calculated, all the bands possess the same phase of $\theta_{\cal Z}=0$ for all $v$. There are indeed no topological transitions at any of the band closing points and so the edge states are not topologically protected and the underlying topology of the system is trivial.

Again, this is confirmed when the finite solution is solved. Two chains consisting of $N=9$ and $N=10$ scatterers are shown in Figs.~\ref{fig:posKPspectra}(c,d) wherein edge states can be seen that vary in energy with $v$. In the odd case, single edge states exist within the band gaps but not at a constant mid-gap energy. Indeed they are seen to move in $q$ as $v$ is varied and to track the bulk states quite closely. In the even case, states continually detach from an upper band and come together to form a ``nearly''-degenerate edge state that again tracks the bulk closely before joining the lower band.

As may be seen on close inspection of the edge states between the third and fourth bands of Fig.~\ref{fig:posKPspectra}(d), the edge states are not perfectly degenerate. This indicates that chiral symmetry is absent from the system since it would imply that the edge states are not simultaneously eigenstates of both the Hamiltonian and the chiral operator.\cite{Asboth:2016} Since the wells all have the same potential environment, this lifting of the degeneracy must stem from the long-range interactions of the local well wavefunctions over the lattice. This same behaviour may be observed in the lower band edge states by zooming in on the band spectrum.

\begin{figure}
\centering
\begin{minipage}{.5\linewidth}
\begin{overpic}[width=\linewidth]{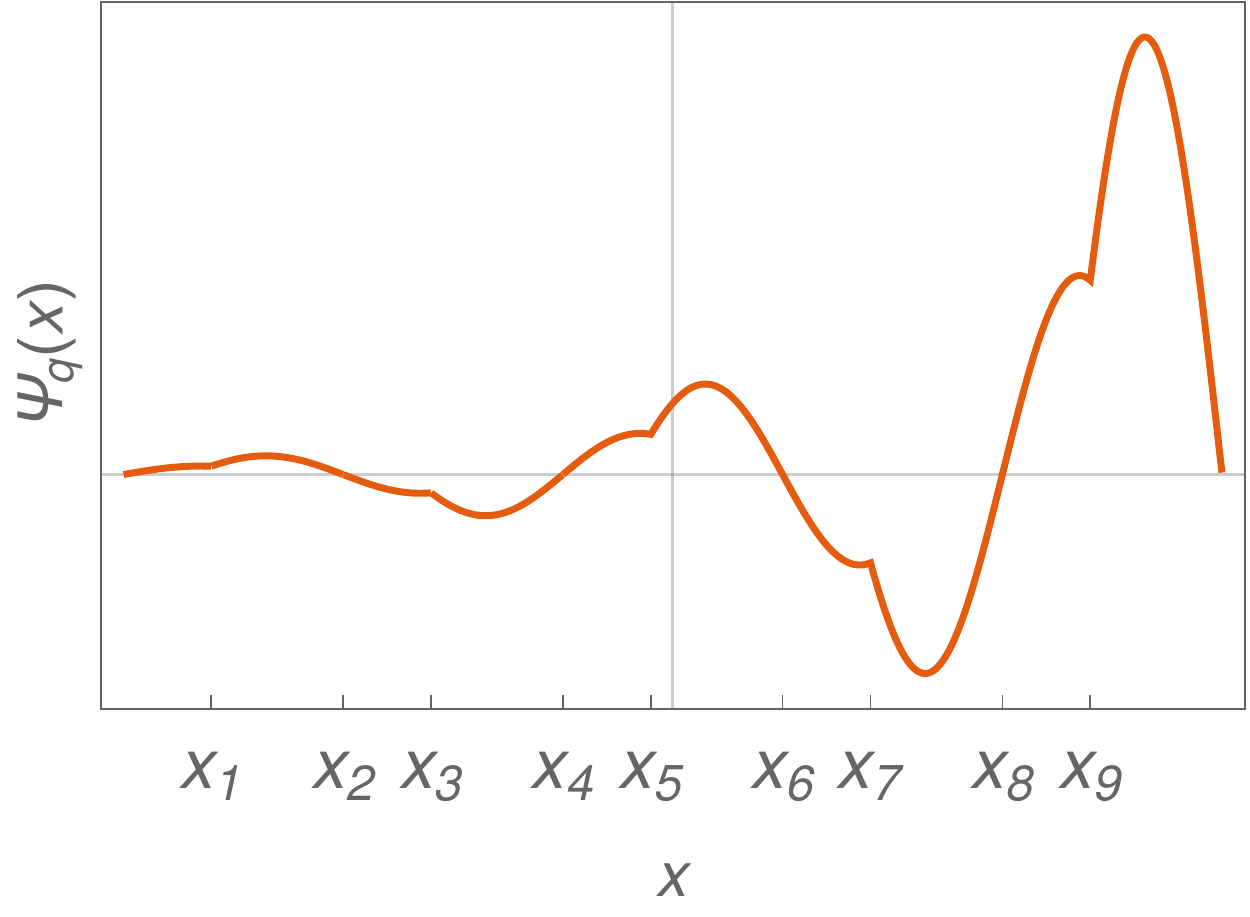}\put(9,65){(a)}
\end{overpic}
\begin{overpic}[width=\linewidth]{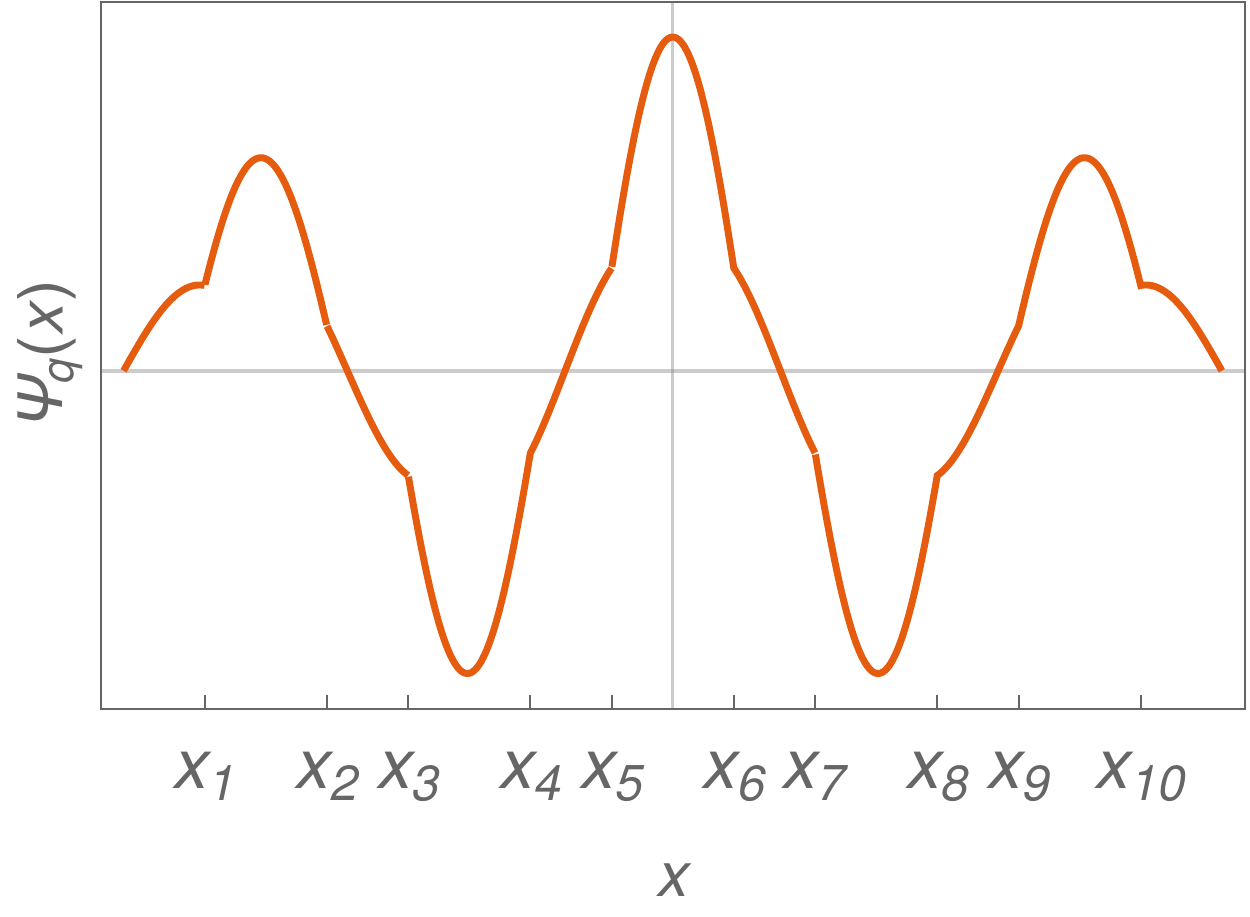}\put(9,20){(c)}
\end{overpic}
\end{minipage}%
\begin{minipage}{.5\linewidth}
\begin{overpic}[width=\linewidth]{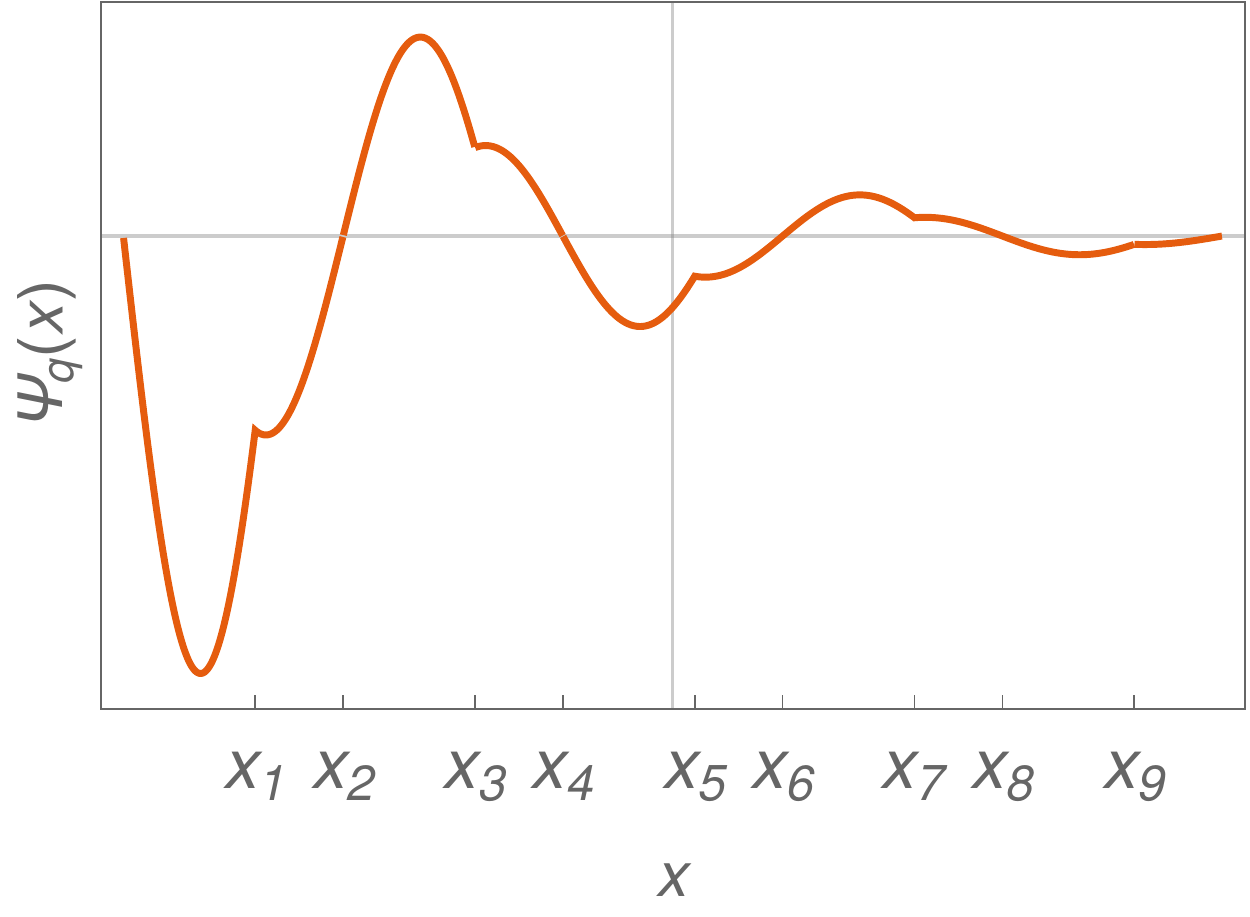}\put(9,65){(b)}
\end{overpic}
\begin{overpic}[width=\linewidth]{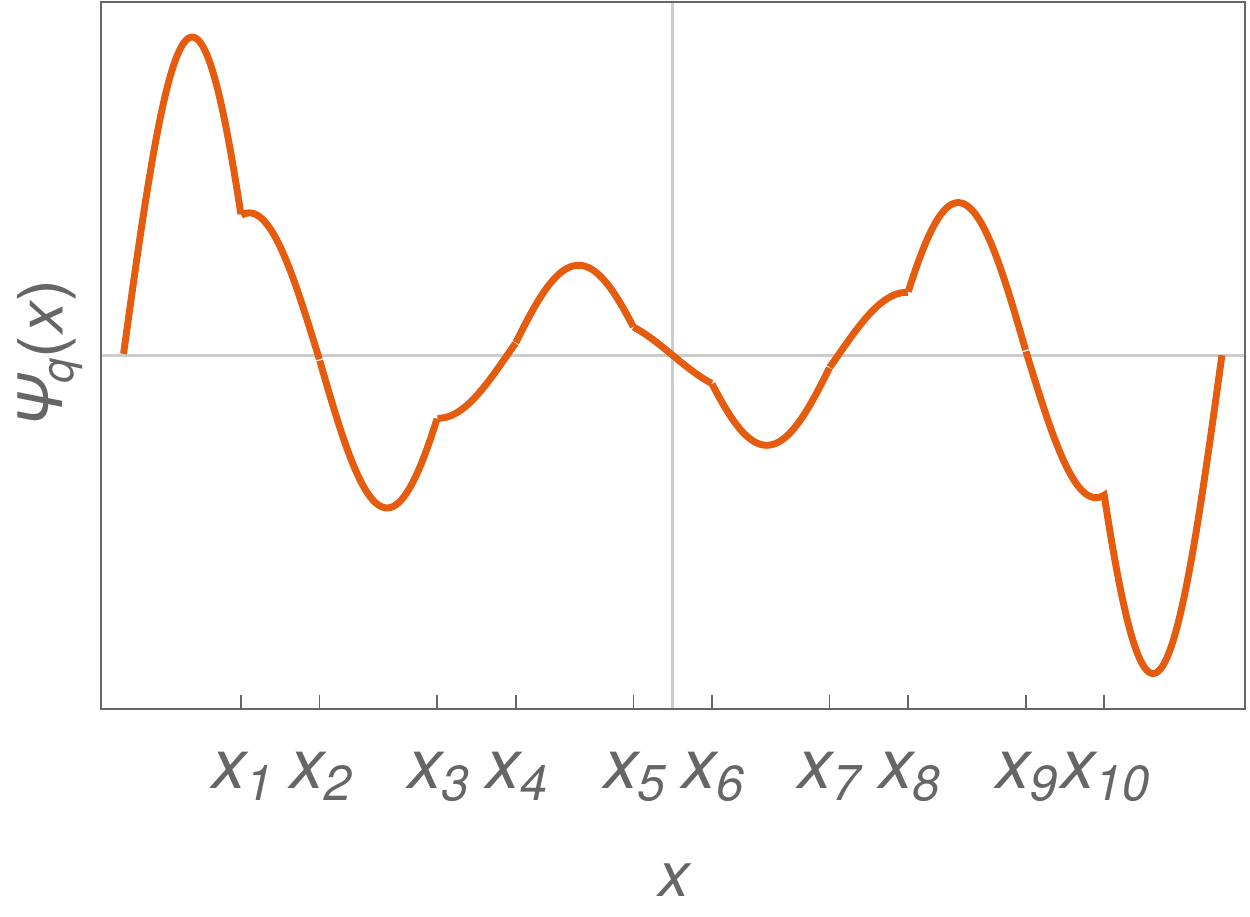}\put(9,20){(d)}
\end{overpic}
\end{minipage}%
\caption{(Colour on-line) The wavefunctions of the fifth state over the chain of the finite bipartite Kronig-Penney model with $|V|=10$ before and after the topological transition at $v=d/2$. Panel (a): $N=9$ and $v=.4$, panel (b): $N=9$ and $v=.6$, panel (c): $N=10$ and $v=.4$, and panel (d): $N=10$ and $v=.6$.}
\label{fig:posDKPst}
\end{figure}

For higher bands, there are multiple detachments within each mid-gap and the value of $v$ for which these occur are $v_c=md/(n+1)$ where $n$ is the number of band closings of the lower band and $m$ is an integer that can take values less than $n$. As such, the closings between the second and third bands occur at $v_c=d/3$ and $v_c=2d/3$. Interestingly, the even edge state in fact exists in the opposite region of the phase space in comparison the negative case. Here they exist when $v>v_c$ whereas, in the negative potential case, the edge state existed for $v<v_c=d/2$.

These detaching states that are initially mid-gap eventually end up as bulk states as $v$ is increased from zero to one, in a similar way to the cases of varying heights. In other words, a state migrates from one band to another, in this case from the upper to lower band. The same mirror symmetry argument as used there may be applied here to explain why. In this case, for an even number of scatterers there are an odd number of wells. Thus, when $v<v_c$ there are fewer narrow wells than wider wells and when $v>v_c$ this inverts hence bands are ejected from upper bands.

This is the self-same argument that explains why the protected edge state of the odd chain in the first considered case ($A$) migrated across the chain upon the transition. However, there a topological invariant was identified with its existence, whereas here there is no such distinction. As always, this is the crucial step to the establishing of bulk-boundary correspondence: finding an invariant in the bulk and mapping it to an edge state.

Plotting the states across the chain shows clearly the lack of topological protection. As shown in Fig.~\ref{fig:posDKPst}, the states can be seen to exponentially localise to the edges however they are not solely confined to a single sublattice, which would indicate chiral symmetry breaking, protection-destroying, next-next-etc.-nearest-neighbour interactions. This has to do with the fact that, when chiral symmetry is present, the topologically protected edge states of the SSH model are confined to exist solely on a single sublattice due to being eigenstates of the chiral operator.\cite{Asboth:2016}

This may be explained physically as being a result of the long-range interaction between the local well wavefunctions. In the negative heights case, the localisation of the wavefunctions to the scatterers ensured that the interactions between neighbouring sites were of nearest-neighbour type. So too, in the simple SSH model, are next-nearest-neighbour interactions ignored. This is indeed not the case here since the local well wavefunctions, which are not exponentially localised, may interact strongly with each other over the chain.

Moreover, in the odd case, upon the transition of $v<v_c\to v>v_c$, the wavefunction of the edge state migrates across the chain but has its sign flipped in the process. The same effect is seen in the even edge state where its weight has opposite signs at either end of the chain. This is distinctly not the case in the topological SSH model and so further shows the non-topological nature of these edge states. Indeed, this behaviour may be mimicked in the SSH model by including next-nearest neighbour interactions. It must be noted, however, that there are cases of the SSH model in which certain next-nearest neighbour interactions in fact preserve the topological protection.\cite{Li:2014,Gonzalez:2018}

\begin{figure}
\centering
\begin{minipage}{.5\linewidth}
\begin{overpic}[width=\linewidth]{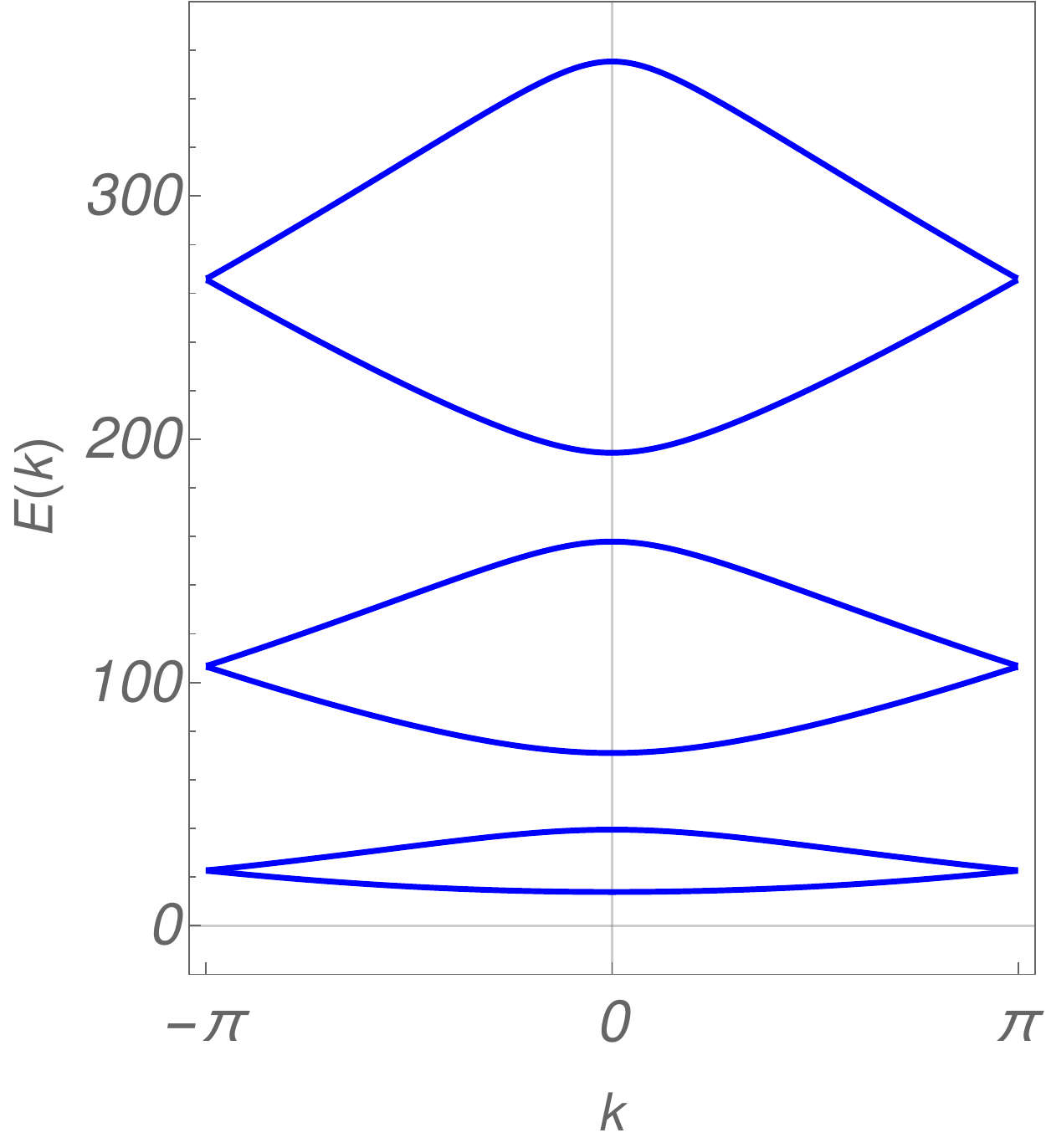}\put(18,52){(a)}
\end{overpic}
\begin{overpic}[width=\linewidth]{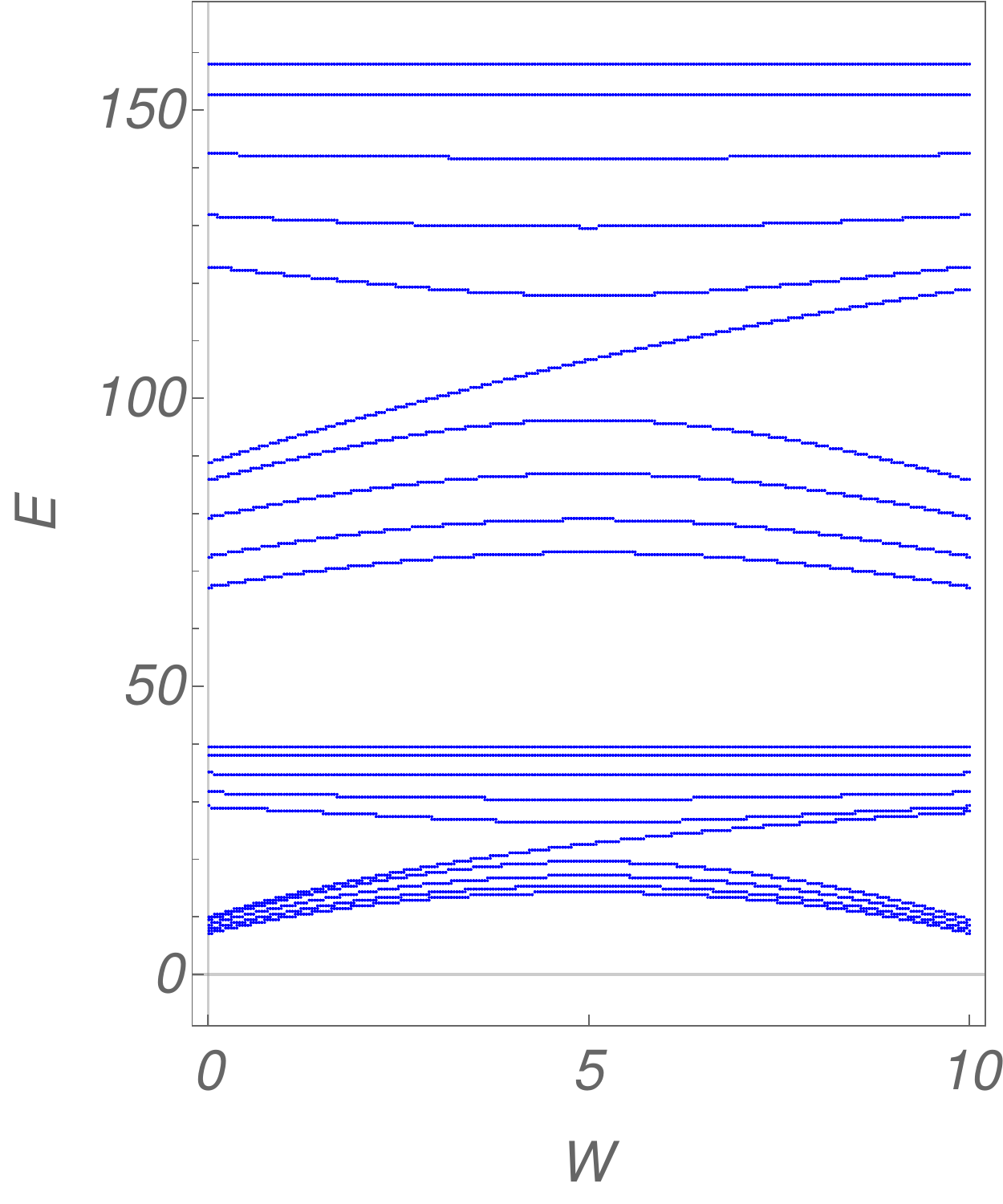}\put(17,42){(c)}
\end{overpic}
\end{minipage}%
\begin{minipage}{.5\linewidth}
\begin{overpic}[width=\linewidth]{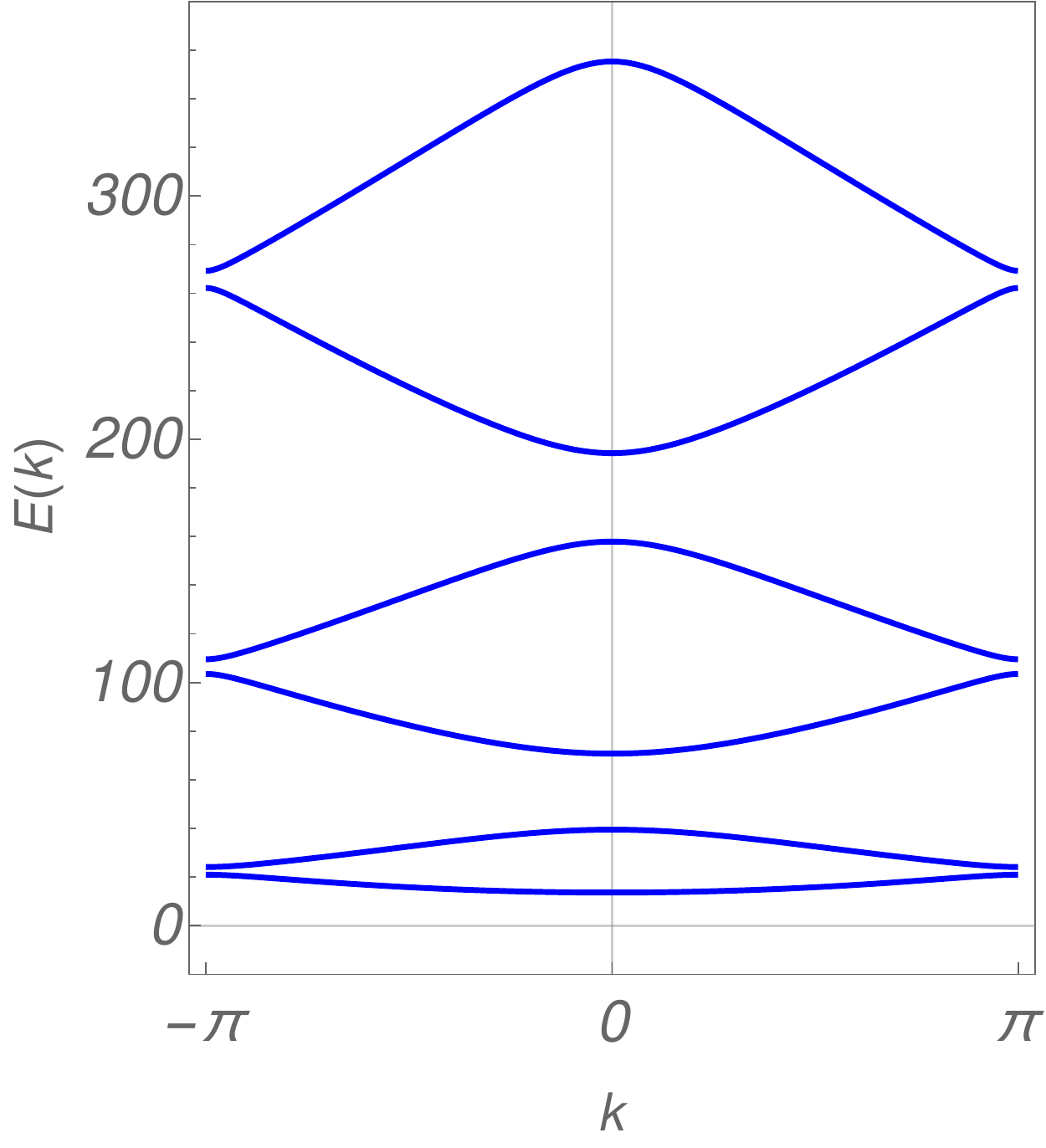}\put(18,52){(b)}
\end{overpic}
\begin{overpic}[width=\linewidth]{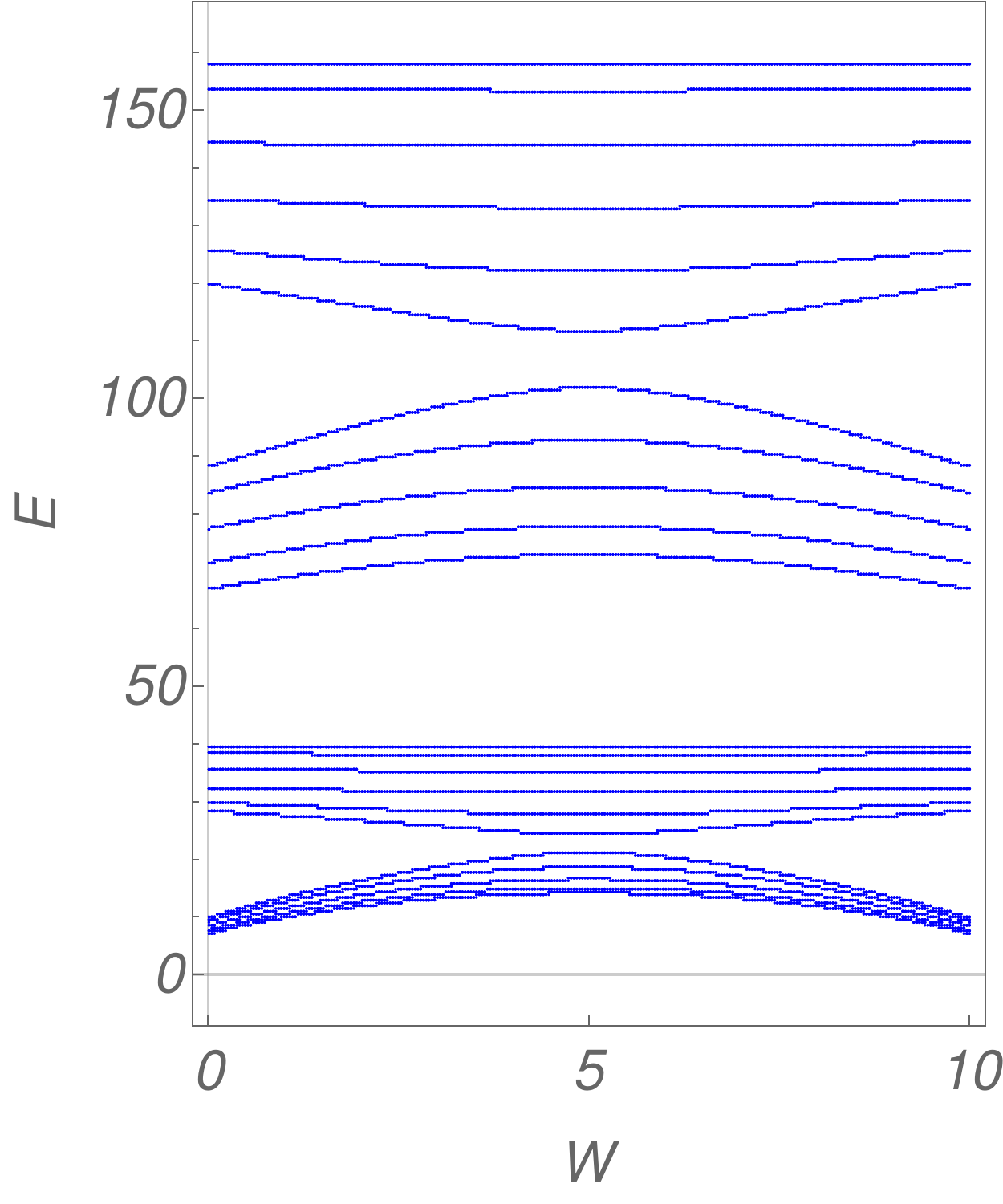}\put(17,42){(d)}
\end{overpic}
\end{minipage}%
\caption{(Colour on-line) The quasi-momentum spectra for the positive bipartite Kronig-Penney model with constant widths $v=w=d/2=0.5$ and varying heights. The bulk and finite spectra are plotted in the top (a,b) and bottom (c,d) respectively. Panel (a): $U=10$ and $W=4.9$, panel (b): $U=10$ and $W=4$, panel (c): $U=10$ and $N=9$, and panel (d): $U=10$ and $N=10$. As opposed to the previous case, modulating $W$ only modulates the gaps at the edges of the Brillouin Zone. This may be seen by comparing Fig.~\ref{fig:posDKPhEn}(b) with Fig.~\ref{fig:posKPspectra}(b).}
\label{fig:posDKPhEn}
\end{figure}
\begin{figure}
\begin{minipage}{.5\linewidth}
\begin{overpic}[width=\linewidth]{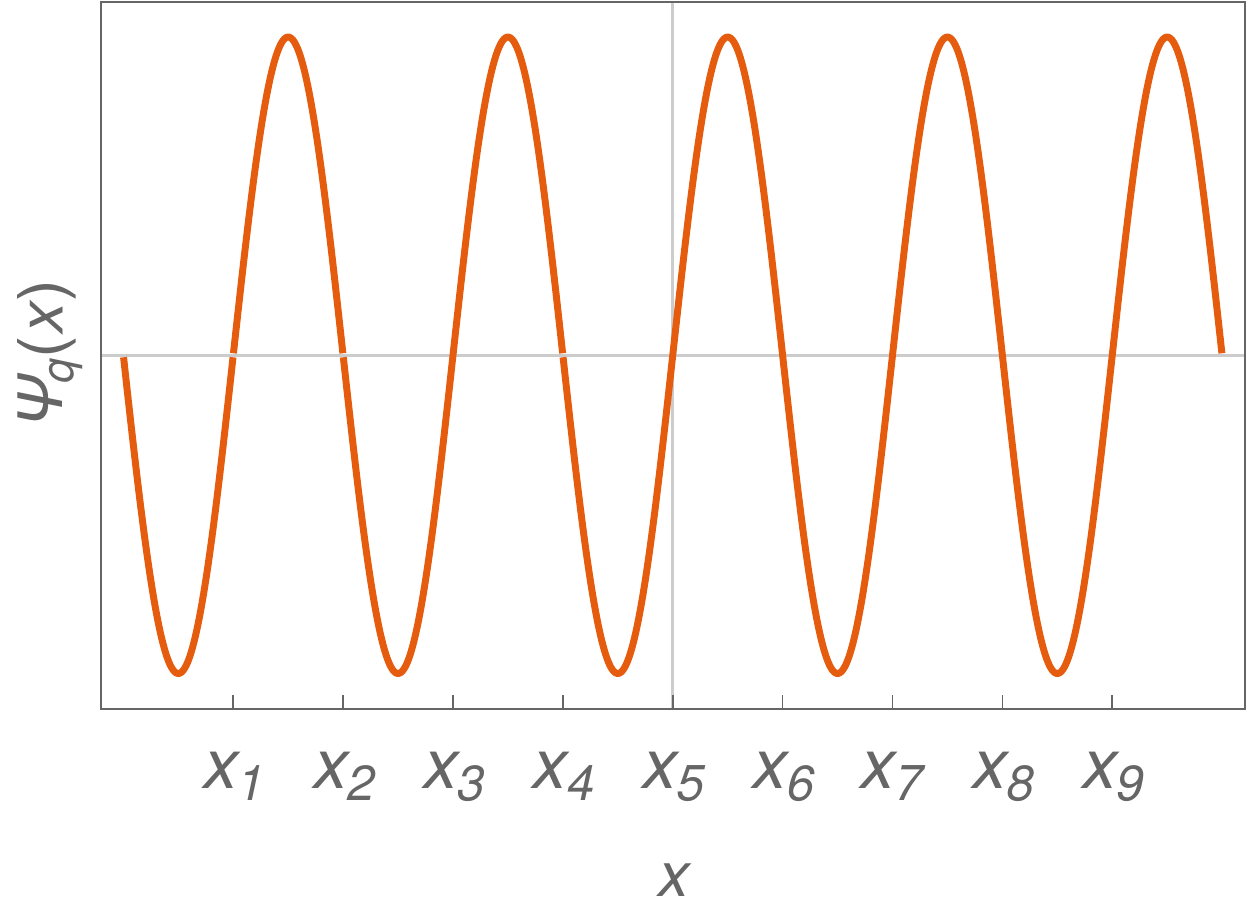}\put(89,20){(a)}
\end{overpic}
\end{minipage}%
\begin{minipage}{.5\linewidth}
\begin{overpic}[width=\linewidth]{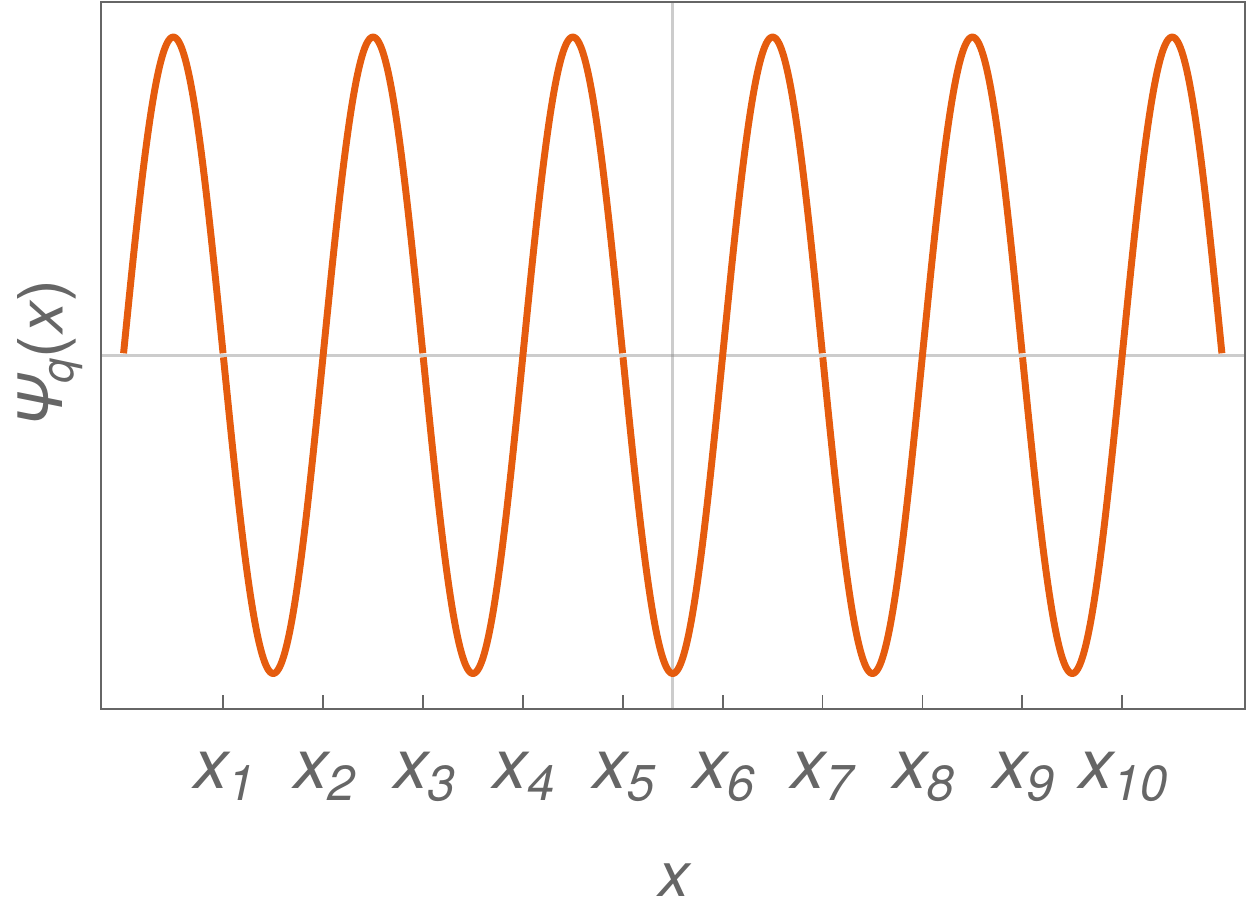}\put(89,20){(b)}
\end{overpic}
\end{minipage}
\caption{(Colour on-line) The probability densities of the $N+1$-th state over the chain of the finite positive bipartite Kronig-Penney model with $U=10$ and $W=4$. Panel (a): $N=9$, and panel (b): $N=10$.}
\label{fig:posDKPst3}
\end{figure}

\subsection{Positive Potentials with Varying Heights}

Now we turn to, what would initially appear to be, the strongest candidate of all the considered systems for the hosting of topological edge states. In the positive heights case, the wavevector is real and so the local wavefunctions exist within the wells. All the wells are identical in this case since $V_0=0$ and $v=w=d/2$. Thus, with analogy to the SSH model, the scatterer heights $V_1=W$ and $V_2=U-W$ would take the place of the hopping integrals $v$, $w$.

The bulk solution is solved and the band spectra are shown in Figs.~\ref{fig:posDKPhEn}(a,b). They may be seen to be almost identical to the previous positive case. The difference, however, is that the band gaps situated along $k=0$ never close as a result of the modulation of $W$ as opposed to the previous case. As a result, the finite solution will show band closings and potential band crossings only when $V_1=V_2$, {\it i.e.} $W=U/2$, whereat the gaps at $k=\pm\pi/d$ close.

The windings of the reflection and transmission coefficients may be calculated and plotted, however they behave in much the same way as in cases $B$ and $C$; that is, there is no topological transition. This may be seen to be consistent through a calculation of the Zak phase, which reveals that it is in fact unquantised in this system. Therefore, this system has no topological features.

As confirmation, the finite system is solved and the spectra are shown in Figs.~\ref{fig:posDKPhEn}(c,d). As may be seen, the only band crossings occur when $W=U/2$. A state may be seen to migrate between bands to the upper band only in the odd number of scatterer case by the same mirror symmetry argument as used throughout the previous cases.

The striking characteristic of these band spectra is the appearance of entirely flat bands. As it turns out, these are the $mN+1$ bands where $m$ is an integer and their characters over the chain are shown in Fig.~\ref{fig:posDKPst3}. As may be seen, they are effectively entirely localised with equal weight within each well with zero overlap between the wells. Indeed, this is the very reason for their flatness. 

\section{Summary and Conclusions}

To summarise, the simplified Kronig-Penney model with Dirac potential scatterers has been extended so as to become bipartite. This may be done by alternately modulating the distances between the scatterers or the heights of the scatterers themselves. The cases of negative and positive potential heights were investigated wherein the wavevector was imaginary and real respectively.

The solution proceeded within the scattering, rather than tight-binding, formalism wherein the boundary conditions at each Dirac scatterer were solved to obtain a unit cell scattering matrix. Such a formalism is different from the lattice-scattering-matrix approach of Ref.~\onlinecite{Fulga:2012}, which relies on an underlying tight-binding model. In our case we directly solve the Schr\"odinger equation for a bipartite Kronig-Penney chain of scatterers and obtain, albeit numerically, the exact wavefunction, without the need of constructing an effective lattice model.

Topologically protected mid-gap states were seen to be present in only one of the cases considered; that of negative heights and varying widths. This behaviour was found to be characterised by the winding of the reflection coefficient, ${\cal W}_r$, over the Brillouin zone, which was confirmed by the calculation of the Zak phase, $\theta_{\cal Z}$.

A summary of these integer values for both $\theta_{\cal Z}$ and ${\cal W}_r$ may be seen in the table in Fig.~\ref{fig:table}. Although, in case $B$, the Zak phase shows not only quantised behaviour but also step-wise integer change over the transition, no edge states were found within the finite system. As such, this integer change over the transition may be an indication of a different topological effect than the protection of edge states against adiabatic deformations.

The three other cases were seen to lack such protected states, however were not wholly uninteresting as a result. In the cases of varying heights, the mirror symmetry of the entire chain meant that a state migrated between bands for an odd number of scatterers, however, there were no band crossing states in the even case. Finally, no edge states were observed within the system.

Furthermore, the positive heights and varying widths case was seen to possess strikingly similar characteristics to the negative heights and varying widths case. Namely that edge states of almost identical character exist between the bulk bands. They were, however, seen to not possess topological protection since their wavefunctions were not confined solely to a single sublattice and the even states were seen to be non-degenerate. Points that were backed up by the bulk solution and both indications of the absence of chiral symmetry and topological protection. Finally, in all of the cases $B$, $C$ and $D$ the variation of the potential may in fact be seen to give rise to charge pumping. In such a case, the Zak phase that quantises the number of pumped states is defined over the parameter space of the variational parameter. In cases $B$ and $D$, this is $W$ whilst in case $C$ it is $v$. In all cases, a further (and very complicated) integral of the wavefunctions over this parameter will yield a quantised Zak phase corresponding to the number of pumped states between bands.\cite{Gasparian:2005,Wang:2013}

\begin{figure}
\centering
\begin{tabular}{c|c|c|c|c|c|c|}
\cline{2-7}
& \multicolumn{2}{c|}{$A$} & \multicolumn{2}{c|}{$B$} & $C$ & $D$
\\
\cline{2-7}
& Ub & Lb & Ub & Lb & All bands & All bands
\\
\hline
\multicolumn{1}{|c|}{$\theta_{\cal Z}$} & $\pi,0$ & $\pi,0$ & $2\pi,\pi$ & $\pi,0$ & 0,0 & Unquantised
\\
\multicolumn{1}{|c|}{${\cal W}_{\rm r}$} & $1,0$ & $1,0$ & $1,1$ & $1,1$ & $1,1$ or $0,0$ & $1,1$ or $0,0$
\\
\multicolumn{1}{|c|}{${\cal W}_{\rm t}$} & $0,-1$ & $0,-1$ & $0,-1$ & $0,-1$ & Unquantised & $0,-1$
\\
\hline
\end{tabular}
\caption{A table that summarises the different invariants for each of the considered cases. In each case, $x,y$ denotes the invariant in the topologically non-trivial ($x$) and trivial ($y$) regions for the bands. In cases $A$ and $C$ these are $v<w$ and $v>w$ respectively, whilst in cases $B$ and $D$ these are $V_1<V_2$ and $V_1>V_2$ respectively. Ub and Lb stand for Upper and Lower bands respectively. The only guaranteed topologically protected edge states thus occur in case $A$ wherein the invariants $\theta_{\cal Z}/\pi$ and ${\cal W}_{\rm r}$ conform. This is confirmed by the appearance of zero-energy mid-gap edge states. Thus the invariant that encodes the number of topologically protected edge states is the winding number of the reflection coefficient.}
\label{fig:table}
\end{figure}

An interesting point still remains with respect to the third case. Mid-gap edge states are still present between each pair of bands for all $v$ in the odd case and for $v>v_c$ in the even case when two bulk states come together to form a ``nearly''-degenerate edge state. Since the wells are all identical there is no on-site potential term so the topological protection is destroyed as a result of the long-range interactions.  This is a small but crucial effect so, even though these edge states are seen to not possess topological protection, they bare distinct resemblances to the edge states in the negative potential case.

Within the SSH model, the hopping parameters, $v$ and $w$, over the chain are determined from the overlap of the atomic wavefunctions that are used within the tight-binding approximation. This same mechanism is at play here in that the realisation of the bipartite nature of the chain is in toying with the overlap between wells/sublattices by changing the amount that the wavefunctions overlap through modulation of the distance. The sublattices are given different environments due to the different overlap between the wells.

It comes as no surprise then that the edge states bare a resemblance to SSH edge states. However, the whole story is not complete without solving for the invariant within the bulk and so establishing a bulk-boundary correspondence. Only then can the edge states be correctly characterised as topological. Nevertheless, it is interesting that the two cases of varying widths with positive/negative potential heights show superficial similarities that in fact differ when the bulk solution is considered. This is a testament to bulk-boundary correspondence and its relationship with topological protection.

This is not the case when the potential heights are modulated. In those cases, the environments are made different by changing the nature of the `atomic' wavefunctions, in this K-P case these are the well wavefunctions $\psi_{q,i}(x)$. Thus, the chains cannot host states as are found in the SSH model as the bipartite-ness is inherently and fundamentally different.

This point exemplifies the fact that the topological character of a lattice lies within the microscopic detail of the states and the interactions between them. Thus, it should be impervious to the choice of theory/approximation. If the topology is inherent in the lattice then it will show up regardless of the theoretical approach. As such, the tight-binding model is not integral to the realisation of topologically protected edge states; it may also be identified in the scattering formalism herein.

This second requirement was also seen to apply in the fourth and final case wherein there is no on-site potential, since the wells are identical, but the overlap between the wells extended far beyond the nearest neighbour. As such, no protected states were seen.

In all cases, the topological behaviour was identified with the behaviours of not only the winding of the reflection coefficient but also the Zak phase such that bulk-boundary correspondence could be identified from which the evident edge states were able to be correctly identified as topological. Furthermore, the absence of protected states was confirmed by the behaviour of these quantities in the other three cases.

%As an aside, recent work has shown that, in the tight-binding models, if an imaginary on-site potential term is present, which would manifest itself physically as a gain/loss mechanism within the system, then topologically protected edge states still exist only with an imaginary energy, {\it i.e.} a temporal decay.\cite{Yuce:2015,Yuce1:2018,Yuce2:2018}

Within the context of tight-binding models, the time-independent Schr{\"o}dinger equation is solved as an Hermitian matrix eigenvalue equation. There, the topological character of the matrix Hamiltonians may be categorised by the symmetries that they possess. In order that topologically protected edge states exist, the Hamiltonian must possess the following three symmetries: particle-hole, time-reversal and chiral. Since we have considered Schr{\"o}dinger particles here, the first two symmetries have been trivially satisfied whilst the third was shown to be satisfied only in the first of the four cases.

In the present case, we solve the time-independent Schr{\"o}dinger equation through the scattering formalism whereby the boundary conditions at the potentials are solved. This generates an eigenvalue equation involving a unitary/non-Hermitian matrix (corresponding to positive/negative energy solutions) that, nevertheless, encodes the topological character of the system. This is because, as is shown in Ref.~\onlinecite{Essin:2011}, the topological invariant may be found within the bulk Green's function, which itself may be used to build the unit-cell (bulk) scattering matrix as seen here.

%One might expect then that the scattering matrix $S(k)$ here should possess similar symmetries as the Hamiltonians of tight-binding models or belong to similar Lie groups with an identical topological class; that of the unit circle $S_1$ and as such $\mathbb{Z}$. This, however, is analysis that goes beyond the scope of this paper but would be interesting nonetheless.

Having established the result for a chain of Dirac scatterers, which is a much simplified Kronig-Penney model, this analysis may be readily applied to a system of finite barriers or wells. Such systems are readily encountered in relation to the electromagnetic interaction in the presence of diffraction gratings. In the limit of narrow and tall barriers/wells the same behaviour is expected however as this restriction is relaxed interesting results may be obtained otherwise. 

\section{Acknowledgements}
T.B.S. acknowledges the support of the EPSRC through a PhD studentship grant. A.P. and T.B.S. acknowledge support from the Royal Society International Exchange grant IES\textbackslash R3\textbackslash 170252.

\bibliographystyle{apsrev4-1}
\bibliography{bibliography}

\onecolumngrid
\appendix
\section{Calculation of the Bulk Scattering Matrix}\label{appA}

Taking the boundary conditions that apply for the localised wavefunction within the unit cell as specified in Eq.~(\ref{eqn:bulkBCs}) with the wavefunctions as defined in Eq.~(\ref{eqn:bulkWFs}), the scattering equations across each Dirac potential may be found as:
\begin{align}
\begin{pmatrix}\label{eqn:APPs1}
D_1(k)
\\
C_2(k)
\end{pmatrix}&=S_1(k)
\begin{pmatrix}
C_1(k)
\\
D_2(k)
\end{pmatrix},
\\
\begin{pmatrix}
D_2(k)
\\
C_3(k)
\end{pmatrix}&=S_2(k)
\begin{pmatrix}
C_2(k)
\\
D_3(k)
\end{pmatrix},
\end{align}
where the matrices $S_i(k)$ are found as:
\begin{align}
S_i(k)&=\frac{1}{V_i+iq_k}
\begin{pmatrix}
-V_ie^{2iq_kx_i} & iq_k
\\
iq_k & -V_ie^{-2iq_kx_i}
\end{pmatrix}.
\end{align}
Then, using $\psi_{3,k}(x+d)=\psi_{1,k}(x)e^{ikd}$ to see that $\{C_3,D_3\}=\{C_1e^{-iqd},D_1e^{iqd}\}e^{ikd}$ the second scattering equation may be manipulated to become:
\begin{equation} \label{eqn:APPs2}
\begin{pmatrix}
C_1(k)
\\
D_2(k)
\end{pmatrix}=\tilde{S}_2(k)
\begin{pmatrix}
D_1(k)
\\
C_2(k)
\end{pmatrix},
\end{equation}
where:
\begin{equation}
\tilde{S}_2(k)=\frac{1}{V_2+iq_k}
\begin{pmatrix}
-V_2e^{2iq_k(d-x_2)} & iq_ke^{i(q_k-k)d}
\\
iq_ke^{i(q_k+k)d} & -V_2e^{2iq_kx_2}
\end{pmatrix}.
\end{equation}
Thus, the scattering matrix equation is found simply by substituting Eq.~(\ref{eqn:APPs2}) into Eq.~(\ref{eqn:APPs1}) and so achieving:
\begin{equation}
\begin{pmatrix}
D_1(k)
\\
C_2(k)
\end{pmatrix}=S(k)
\begin{pmatrix}
D_1(k)
\\
C_2(k)
\end{pmatrix},
\end{equation}
where $S(k)=S_1(k)\tilde{S}_2(k)$ is the matrix as quoted in Eq.~(\ref{eqn:Smat}) whose entries are as in Eqs.~(\ref{eqn:rcoeff},\ref{eqn:tcoeff},\ref{eqn:ph}).

As was mentioned in the text, once this scattering matrix equation is solved for $D_1(k)$ and $C_2(k)$, these may be substituted into Eq.~(\ref{eqn:APPs2}) in order to find $C_1(k)$ and $D_2(k)$ whilst $C_3(k)$ and $D_3(k)$ are found using the Bloch condition. Thus, the unit cell wavefunction is entirely determined with the unit cell scattering matrix solely.

\section{Relevant Solutions to the SSH Model}\label{appB}

In the SSH model, the time-independent Schr{\"o}dinger equation is solved as $H\ket{\psi}=E\ket{\psi}$ in the basis of lattice sites. A general finite SSH model Hamiltonian, with the first (left-hand side) lattice site belonging to the $A$ sublattice, takes the following form:\cite{Asboth:2016}
\begin{equation}
H=\sum_{i=1}^N\left[V_i\hat{a}^\dagger_i\hat{a}_i+W_i\hat{b}^\dagger_i\hat{b}_i
+v_i(\hat{a}^\dagger_i\hat{b}_i+\hat{b}^\dagger_i\hat{a}_i)\right]
+\sum_{i=1}^{N-1}w_i(\hat{a}^\dagger_{i+1}\hat{b}_i
+w_i\hat{b}^\dagger_i\hat{a}_{i+1}),
\end{equation}
where $H.C.$ signifies to take the Hermitian Conjugate and the summations are over the unit cells. The first sum involves hoppings within the unit cell whilst the second sum involves hoppings between unit cells. As a simple example, for a system of $2$ unit cells, and thus $4$ lattice sites for the even case and $5$ lattice sites for the odd case, the following single-particle matrix Hamiltonians are generated:
\begin{equation}\label{eqn:SSHhams}
{\cal H}_{\cal E}=
\begin{pmatrix}
V_1&v_1&0&0
\\
v_1&W_1&w_1&0
\\
0&w_1&V_2&v_2
\\
0&0&v_2&W_2
\end{pmatrix},
\quad
{\cal H}_{\cal O}=
\begin{pmatrix}
V_1&v_1&0&0&0
\\
v_1&W_1&w_1&0&0
\\
0&w_1&V_2&v_2&0
\\
0&0&v_2&W_2&w_2
\\
0&0&0&w_2&V_3
\end{pmatrix}.
\end{equation}
To emulate the first system considered within this paper, we take $v_i=v,~w_i=1-v,~V_i=W_i=10,~\forall i$, whilst for the second system, we take $v_i=w_i=0.5,~V_i=t,~W_i=10-t,~\forall i$.

In the thermodynamic limit, $N\to\infty$, wherein periodic boundary conditions may be imposed, the Hamiltonian becomes two-dimensional in the unit cell basis. Then, it takes the form below:
\begin{equation}
{\cal H}(k)=
\begin{pmatrix}
V&v+we^{-ik}
\\
v+we^{ik}&W
\end{pmatrix},
\end{equation}
where $h(k)=v+we^{-ik}$ is the quantity that exhibits the appropriate topological winding. When $V=W$, the resultant $V\sigma_0$ term is an arbitrary energy shift since it does not enter as a $\sigma_z$ term. This is not the case when $V\neq W$ since we may always say that $V=X+Y$ and $W=X-Y$ with appropriate $X$ and $Y$. Then chiral symmetry is broken by the entering of a $\sigma_z$ term, which is seen in the finite band spectra. The migratory state is due to the mirror/reflection symmetry inherent in the odd finite chain.

\begin{figure}
\centering
\begin{minipage}{.5\linewidth}
\begin{overpic}[width=\linewidth]{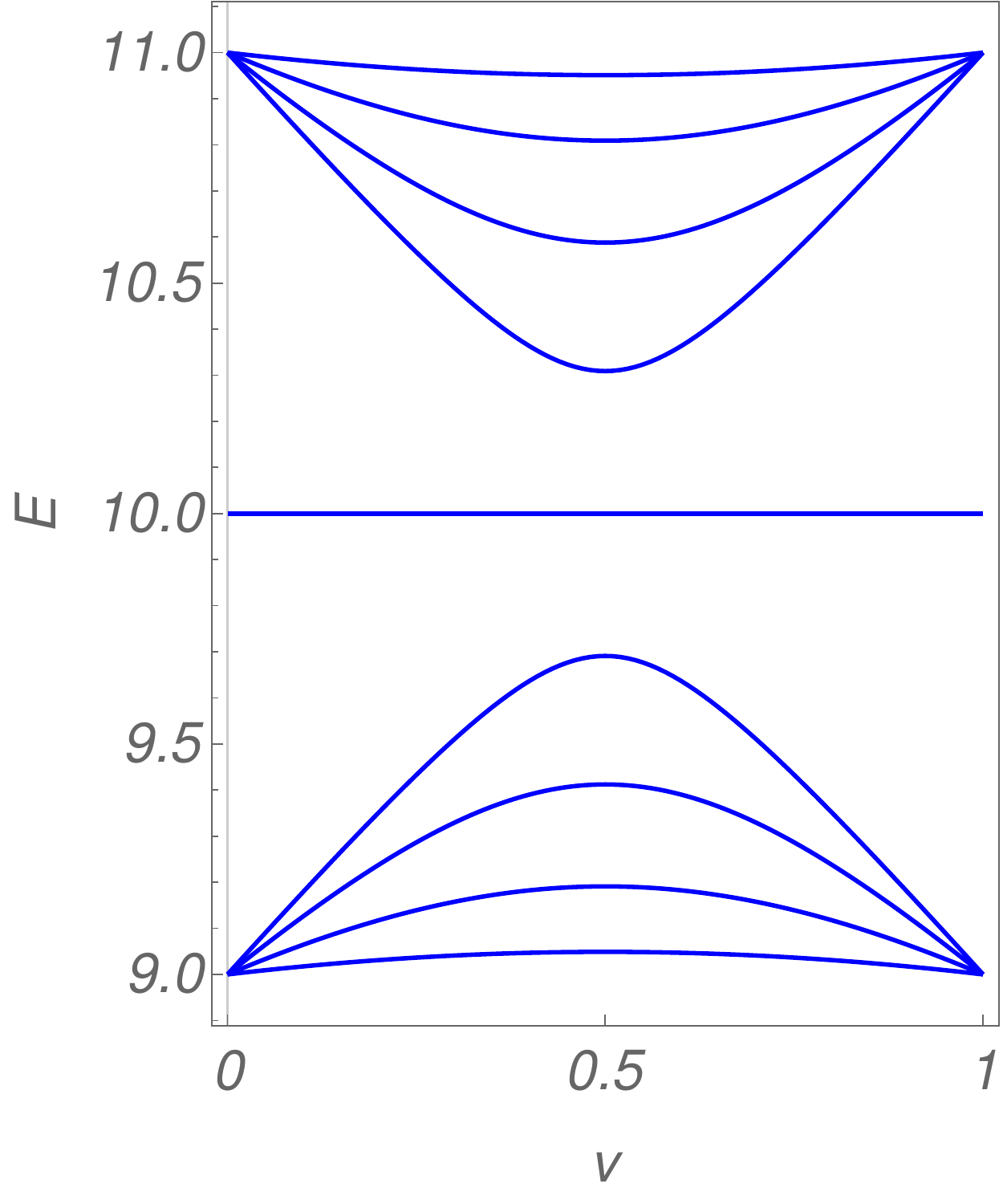}
\end{overpic}
\begin{overpic}[width=\linewidth]{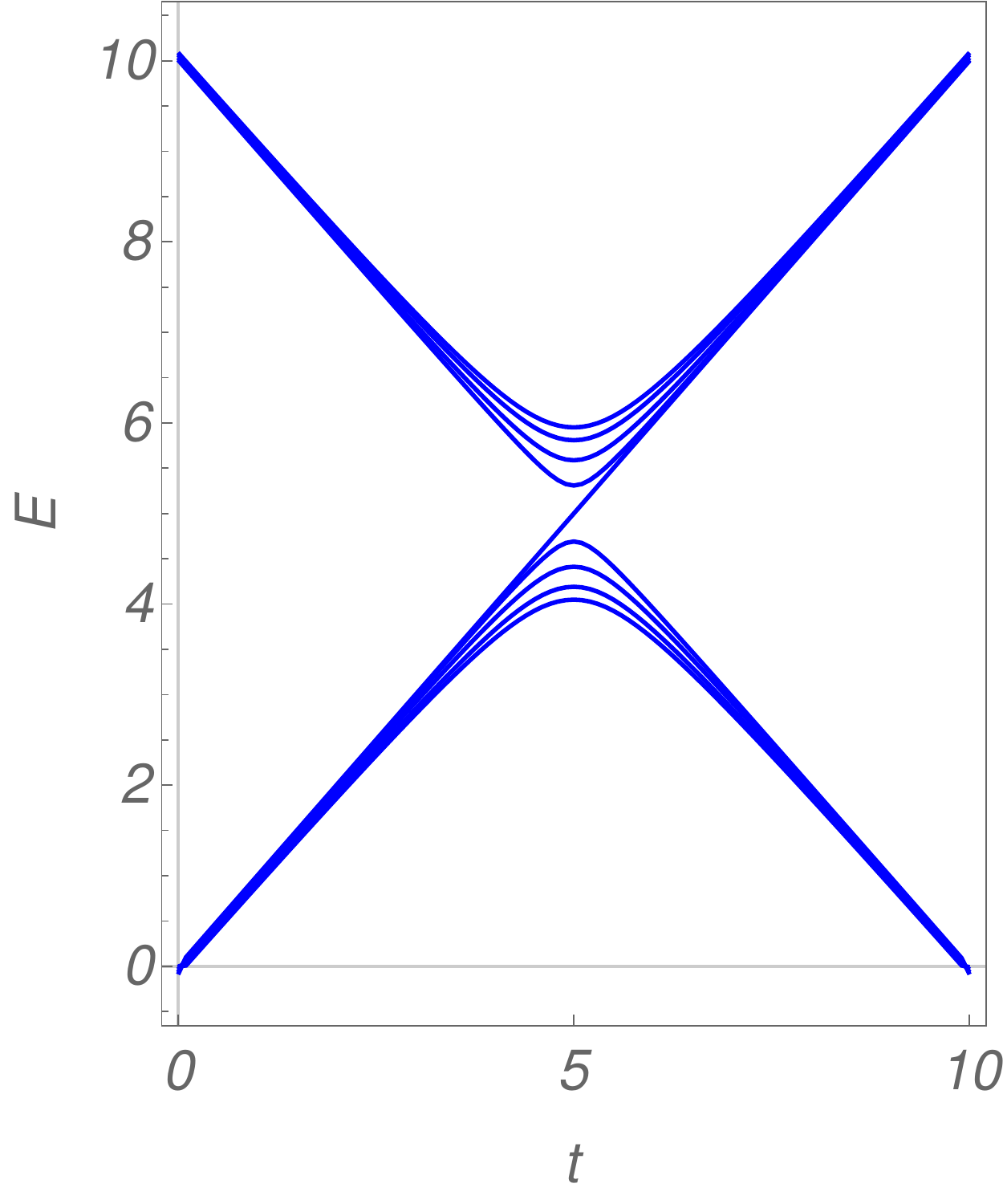}
\end{overpic}
\end{minipage}%
\begin{minipage}{.5\linewidth}
\begin{overpic}[width=\linewidth]{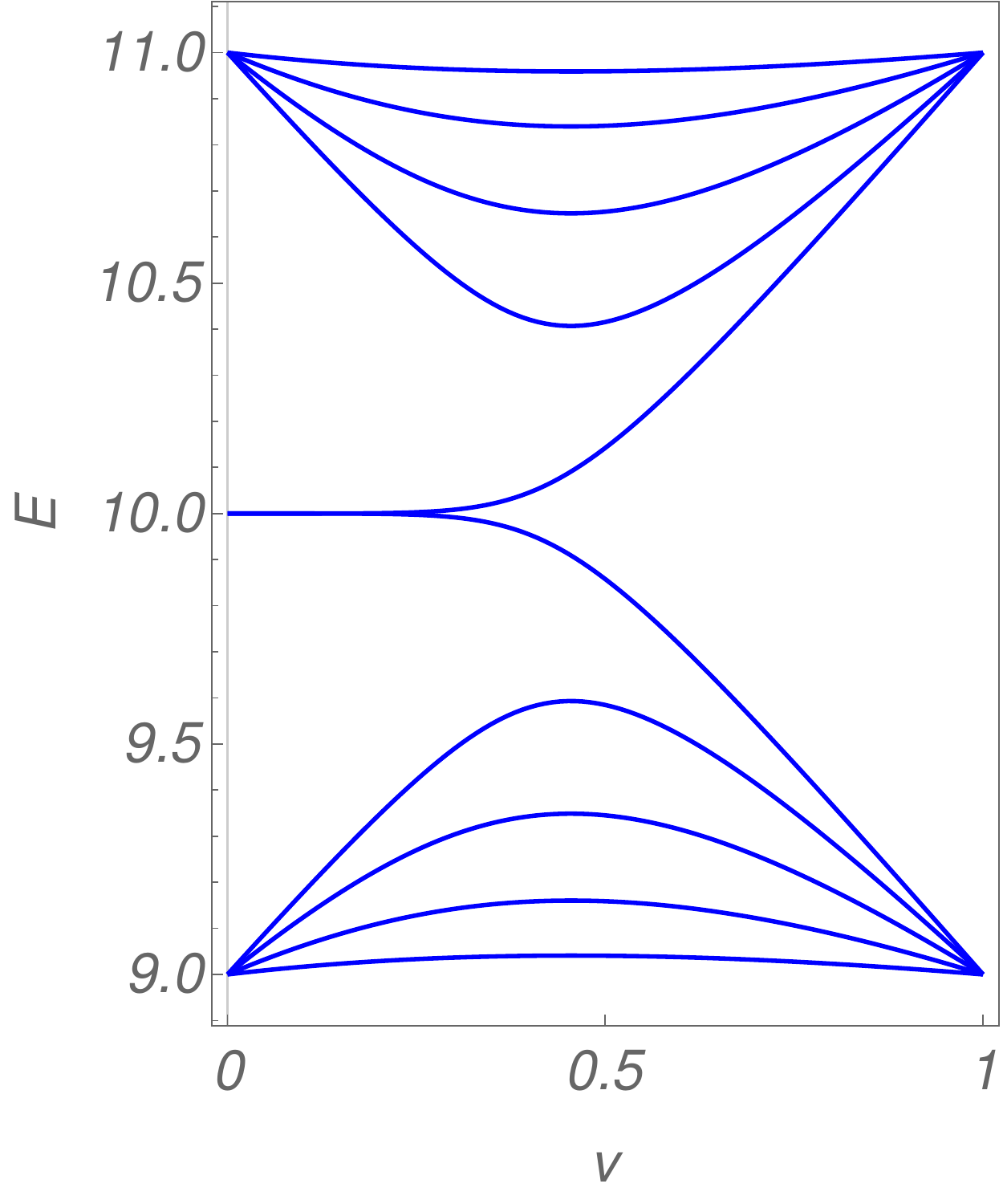}
\end{overpic}
\begin{overpic}[width=\linewidth]{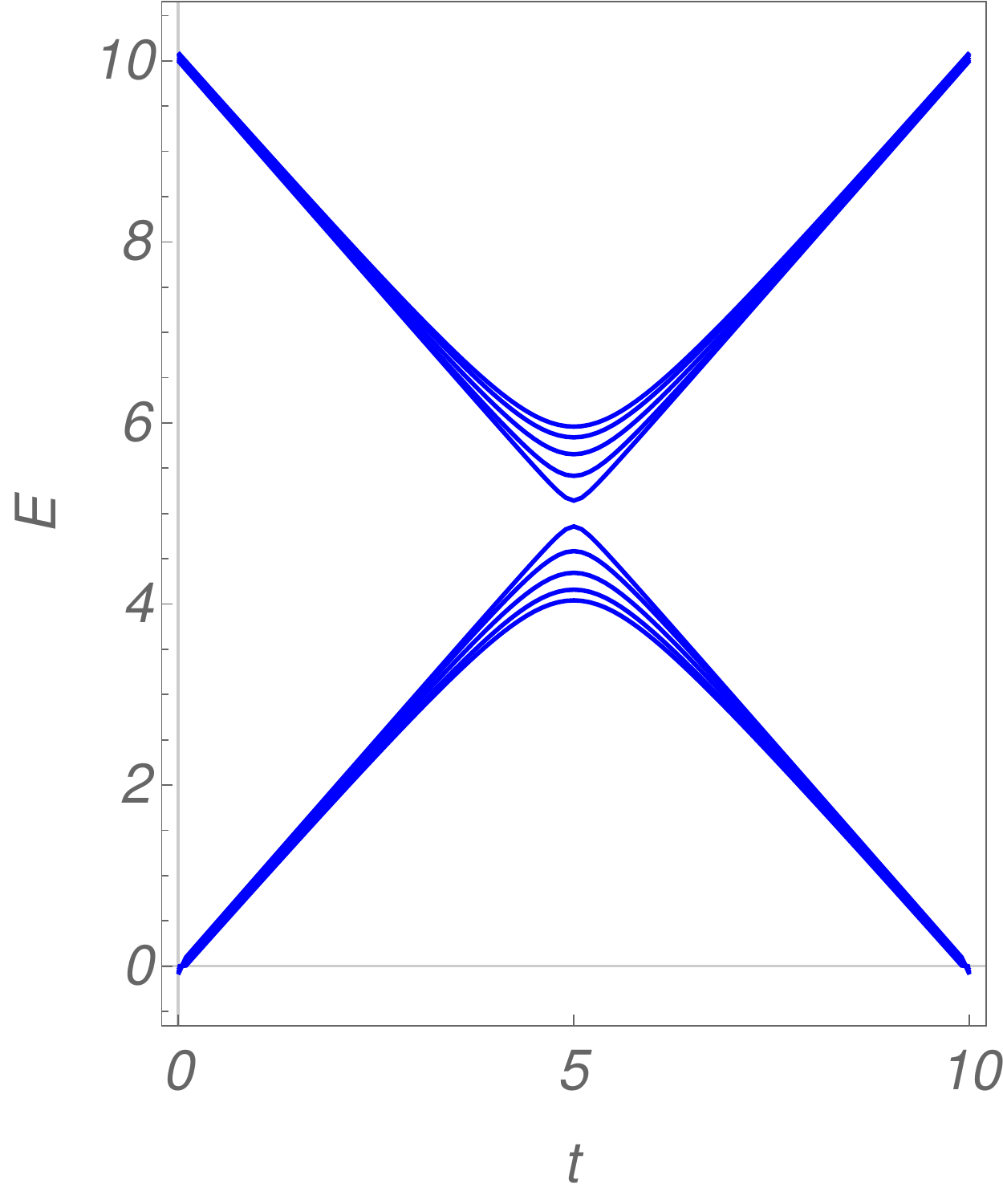}
\end{overpic}
\end{minipage}%
\caption{The band spectra for various SSH models. Top left and top right include only nearest-neighbour hoppings of $v$ and $1-v$ and a constant, lattice independent, on-site potential $V=10$. The bottom left and bottom right include identical hoppings $v=w=.5$ and alternating on-site potentials of $t$ and $U-t$ with $U=10$. Left side is the $N=9$ case and right side is the $N=10$ case. They bare a striking resemblance to the plots as shown in the first two cases considered within this paper.}
\label{fig:SSH}
\end{figure}

\end{document}